\documentclass{emulateapj}

\pdfoutput=1

\usepackage{epsfig}
\usepackage{epstopdf}
\usepackage{wrapfig}
\usepackage{graphicx}
\usepackage{rotating}
\usepackage{color}
\usepackage{subfigure}

\RequirePackage{natbib}

\newcommand{\subscript}[1]{_{\mathrm{#1}}}

\shorttitle{Reverse Shock Dust Destruction}
\shortauthors{Silvia, Smith, \& Shull}
\slugcomment{Accepted by the Astrophysical Journal}

\begin{document}

\title{Numerical Simulations of Supernova Dust Destruction.\\
	I.  Cloud-crushing and Post-processed Grain Sputtering}
	
\author{Devin W. Silvia, Britton D. Smith, and J. Michael Shull}
\affil{CASA, Department of Astrophysical and Planetary Sciences, University of Colorado, UCB 389, Boulder, CO 80309; devin.silvia@colorado.edu; britton.smith@colorado.edu; michael.shull@colorado.edu}

\begin{abstract}
We investigate through hydrodynamic simulations the destruction of newly-formed dust grains by sputtering in the reverse shocks of supernova remnants.  Using an idealized setup of a planar shock impacting a dense, spherical clump, we implant a population of Lagrangian particles into the clump to represent a distribution of dust grains in size and composition. We then post-process the simulation output to calculate the grain sputtering for a variety of species and size distributions. We explore the parameter space appropriate for this problem by altering the over-density of the ejecta clumps and the speed of the reverse shocks. Since radiative cooling could lower the temperature of the medium in which the dust is embedded and potentially protect the dust by slowing or halting grain sputtering, we study the effects of different cooling methods over the time scale of the simulations.  In general, our results indicate that grains with radii less than 0.1 $\mu$m are sputtered to much smaller radii and often destroyed completely, while larger grains survive their interaction with the reverse shock. We also find that, for high ejecta densities, the percentage of dust that survives is strongly dependent on the relative velocity between the clump and the reverse shock, causing up to 50\% more destruction for the highest velocity shocks.  The fraction of dust destroyed varies widely across grain species, ranging from total destruction of Al$_2$O$_3$ grains to minimal destruction of Fe grains (only 20\% destruction in the most extreme cases).  C and SiO$\subscript{2}$ grains show moderate to strong sputtering as well, with 38\% and 80\% mass loss.  The survival rate of grains formed by early supernovae is crucial in determining whether or not they can act as the ``dust factories" needed to explain high-redshift dust.
\end{abstract}


\section{Introduction}
The presence of dust grains in the interstellar medium (ISM) of galaxies can have a wide range of impacts on the evolution of the universe, both at present and at high redshift.  As refractory elements become depleted from the gas phase and locked into grains, they form a considerable reservoir of coolants in the solid-state phase.  The surfaces of these grains are also believed to act as formation sites for molecular hydrogen (H$_2$), which not only serves as an additional coolant but can provide the initial conditions for a rich cloud chemistry.  Grains are capable of reprocessing the radiation field of the ISM by providing a critical transfer mechanism for turning ultraviolet/optical starlight into far-infrared and submillimeter emission seen in interstellar molecular clouds and high-redshift galaxies.  The earliest dust grains may also have influenced the low-metallicity gas by modifying the thermodynamic conditions in the high-redshift universe and altering the star formation process and the initial mass function (IMF) for the earliest second-generation stars \citep{Omukai:2005ly, Clark:2008gf}.

One paradox arises from observations of dust at high redshift.  Several observations of damped Ly$\alpha$ systems indicate the presence of dust grains \citep{Pettini:1994rt, Ledoux:2002ys}, and high-redshift ($z > 6$) quasars seen in the Sloan Digital Sky Survey (SDSS) along with galaxies in the Hubble Deep Field show 1.2 mm thermal emission from dust \citep{Smail:1997gf, Hughes:1998ve, Bertoldi:2003fr}.  These observations imply a total dust mass of $\sim$10$^8$ $M\subscript{\odot}$ formed within the first Gyr after the Big Bang, which is difficult to explain using traditional models of grain formation in evolved low-mass stars \citep{Dunne:2003zr, Dwek:2007lr} and transportation of dust into the ISM through stellar winds \citep{Whittet:1992mz}.

In an effort to find other sources for high-redshift dust, recent models have looked at the ejecta of core-collapse supernovae (SNe) as possible formation sites, which can occur on timescales shorter than the evolution of low-mass stars \citep{Dwek:1980ly}.  Since the inner ejecta of core-collapse SNe are generally cold, dense, and metal-enriched, they are conducive to producing significant amounts of new dust grains.  By utilizing theories of nucleation and grain growth made by \cite{Kozasa:1987gf}, two groups \citep{Kozasa:1989ve, Kozasa:1991ul, Todini:2001qf} have attempted to estimate the dust mass formed in expanding SNe ejecta of varying metallicities, with results ranging from $\sim$0.1--0.3 $M\subscript{\odot}$.  These initial results were extended by \cite{Nozawa:2003pd} and \cite{Schneider:2004lq} to the case of zero-metallicity SNe with significant production of O, Si, and other heavy elements, as would be created by Population III stars.

Although SNe are appealing as potential ``dust factories", current observations of nearby SNe appear to fall short of the required dust masses \citep[$\sim$0.2  $M\subscript{\odot}$, or possibly as high as 1 $M\subscript{\odot}$ per event;][]{Dwek:2007lr} by factors of 10-100.  \cite{Sugerman:2006sx} produced a review of Type II SN 2003gd in which they used radiative transfer models to conclude that up to 0.02 $M\subscript{\odot}$ of dust is present in the ejecta, but this is contradicted by \cite{Meikle:2007dq} who found only $4 \times 10^{-5} M\subscript{\odot}$ of newly formed dust using mid-IR observations. Similar values were found in a young supernova remnant (SNR) located in the Small Magellanic Cloud (SMC) studied by \cite{Stanimirovic:2005rr}, who claimed to observe 10$^{-3} M\subscript{\odot}$ of hot dust.  \cite{Kotak:2009lr} assert a direct detection of ejecta dust in Type II-plateau SN 2004et, but again only estimate a few times 10$^{-4}$ $M\subscript{\odot}$.  One promising observation for SNe as dust factories carried out by \cite{Dunne:2003zr} argued for the existence of $\sim$3 $M\subscript{\odot}$ of dust in the direction of Cassiopeia~A based on a submillimeter detection, but it was quickly contradicted by \cite{Krause:2004cr} who attributed this emission to interstellar dust contained in an adjacent molecular cloud. In the Large Magellanic Cloud (LMC), \cite{Williams:2006ly} observed four SNRs using the 24 and 70 $\mu$m bands on {\it Spitzer Space Telescope} and inferred dust masses of 0.01-0.1 M$\subscript{\odot}$. More recently, \cite{Rho:2008qf, Rho:2009ul} used {\it Spitzer} to study Cassiopeia~A and young supernova remnant 1E0102-7219 to find 0.020-0.054 M$\subscript{\odot}$ and 0.014 M$\subscript{\odot}$ of dust respectively.  To continue the hunt for dust in Cassiopeia~A, data from AKARI and the Balloon-borne Large Aperture Submillimeter Telescope (BLAST) were used by \cite{Sibthorpe:2009sf} to find 0.06 M$\subscript{\odot}$ of new ``cold" (T $\sim$ 35 K) dust, which when combined with previous ``warm" (T~$\sim$~100K) dust estimates, could make the dust-factory argument more plausible.

Observational tests that seek to confirm the predictions made by models of SNe are difficult.  It is necessary to disentangle far-IR backgrounds from the thermal dust emission and separate the newly formed ejecta dust from grains contained in circumstellar and interstellar material. One must also determine what fraction of the dust can be destroyed by the interaction between SNR reverse shocks and the ejecta. As a result of these difficulties, some dust may be located in undetected, unshocked ejecta.  SN 1987A presents a particularly good case of the challenge of producing accurate dust mass estimates, since it has been inferred that dust formation is almost certainly occurring, but only 10$^{-3} M\subscript{\odot}$ was detected \citep{Dwek:2006pw, Dwek:2007cf}.  Observed changes in the optical and bolometric fluxes and asymmetric red and blue emission-line wings indicate that more dust is most likely present \citep{McCray:1993ek, McCray:2007fn}.  Furthermore, since SN 1987A is a young remnant, the reverse shock has not yet impacted the inner ejecta material --- a location that is very likely to harbor additional dust.

The question arises as to whether or not the fast-moving dust ($V\subscript{ej} \ge 1000$ km s$^{-1}$) can survive its impact with the reverse shock, at which point it will be subject to thermal sputtering and grain-grain collisions.  Since the relative speed between the shock and the ejecta clumps can be large, the temperature of the ejecta can be raised to a level that might completely destroy these newly formed grains.  Recent work by \cite{Nozawa:2007rt} investigates the interaction between reverse shocks and SN ejecta through one-dimensional hydrodynamic simulations.  They conclude that 20-100\% of the dust mass is destroyed, depending on the ambient gas density and the initial energy of the SN explosion.  \cite{Bianchi:2007fp} also used a combination of numerical and semi-analytic methods to conclude that only a small fraction of the dust surives --- as low as 7\% in some of their models.  Further efforts were made by \cite{Nath:2008vk} who performed one-dimensional analytic calculations of the reverse shock impacting ejecta material that followed a power-law density profile.  They found that only 1-20\% of the dust mass in the form of silicates and graphites was sputtered, depending on the steepness of the profile and the grain size distribution.

While the above studies have treated the ejecta material primarily as a continuous medium, observations of SNR ejecta indicate that in reality the material is knotted and clumpy \citep{Fesen:2006qy, Hammell:2008sf}.  This lends itself well to the classic ``cloud-crushing" problem \citep{Woodward:1976pd} in which an over-dense clump/cloud of material is impacted by a shock wave.  Such a problem has been studied using numerical hydrodynamic simulations under in the context of an ISM cloud getting hit by the forward shock of a supernova explosion \citep{Bedogni:1990li, Stone:1992wh, Klein:1994uq, Mac-Low:1994fj, Orlando:2005rm, Patnaude:2005zl, Nakamura:2006qy}.

In this paper, we investigate the effects of reverse shocks on SN ejecta by performing three-dimensional cloud-crushing simulations and focusing on the case of an individual ejecta clump encountering a reverse shock. In Section \ref{methods}, we describe the code that we have used to carry out these simulations and our particular simulation set up.  In Section \ref{sims}, we discuss the simulations examined in this paper, which include comparisons to previous work and runs that serve as our means of computing reverse-shock dust destruction.  In Section \ref{dataprocess}, we highlight our methods for calculating dust destruction.  In Section \ref{results}, we present the results of calculations, and we conclude in Section \ref{future} with a summary and discussion of future work.

\section{Methodology}\label{methods}

\subsection{Enzo}
For this study, we use the cosmological, Eulerian adaptive mesh refinement (AMR), hydrodynamics + N-body code, Enzo \citep{Bryan:1997qy, Norman:1999uq, Oshea:2004book}. Owing to the idealized nature of our problem, we do not use any of the cosmological or gravity-solving components of the code.  To handle the fluid dynamics in our work,  we use Enzo's piecewise parabolic method \citep[PPM;][]{Colella:1984qy} for solving the hydrodynamic equations.  For more details on the  inner-workings of Enzo, refer to the cited articles.

The AMR capabilities of Enzo are particularly useful to resolve a range of physical scales or to ensure that only the regions of particular concern remain well resolved, so that we do not waste computation time on unimportant or uninteresting portions of the computational domain.  The refinement algorithm works by taking any cells that meet a user-specified refinement criterion and spawning a new grid of cells with twice the resolution. For example, a three-dimensional cell that satisfies any of the desired refinement criteria will be split into eight cells with cell edges that are now half as long as the parent cell from which they were spawned.  Once this cell is refined to a higher level, all of the new child cells are then checked to see if any still meet the refinement parameters. If so, they will be refined to a higher level in the same way, becoming parents to new children.  Higher and higher levels of refinement can be created to theoretically infinite resolution.  However, since large amounts of refinement can quickly become computationally expensive, a maximum level is often specified when running to the code to halt refinement at the highest resolution that can be afforded by the simulator.  Enzo comes preloaded with the ability to refine based on a variety of parameters: matter overdensity; gradients in pressure, energy, or density; Jeans length; cooling time; and shocks.  It is also possible to define new refinement criteria that are simulation specific.  We have created such a criterion for our cloud-crushing simulations that refines on gas that is initially contained within the cloud.  The implementation of that refinement method is discussed in Section \ref{revshocks}.

In addition to the standard hydrodynamic capabilities of Enzo, we utilize ``tracer particles" in all of our simulations.  The particles are Lagrangian in nature and simply advect with the flow present in the grids.  In the current format, the tracer particles do not alter the state of their surrounding medium and serve only as a means of tracking the conditions (i.e., density and temperature) of the fluid they are embedded in.  However, the machinery for allowing the tracer particles to produce feedback, such as metal injection or dust cooling, exists and will be utilized in our future papers.  Currently, the tracer particles are particularly useful in representing the presence of dust in our simulations, and their use as such will be addressed in Section \ref{tracerparticles}.

For a subset of our simulations, we also include the effects of radiative cooling.  Currently, we use the analytic formula defined by \cite{Sarazin:1987kx}, which approximates the cooling rate as a function of temperature for $Z = 0.5 Z\subscript{\odot}$ gas in ionization equilibrium.  While we expect the ejecta material to be metal enriched, with relative abundances that likely deviate from solar values, for the purposes of this paper we sought only to investigate the effect of the presence of metals (cooling on) versus a complete lack of metals (cooling off).  Future work will include more variable abundances to allow for the enrichment of the ejecta by metals such as C, O, Si, Mg, Mn, and Fe which have been shown to have significant impacts on the cooling rates \citep{Santoro:2006yq} and are key ingredients in most dust grains.  When implementing these adjustable abundances and altered cooling rates, we plan to use a Cloudy \citep{Ferland:1998sf} cooling method similar to \cite{Smith:2008vn}.

\subsection{Simulation Setup}\label{simsetup}
To address the proposed problem of a supernova remnant reverse shock impacting a clump of newly formed ejecta, we added to the suite of pre-packaged Enzo simulations.  In this particular problem type, we seek to simulate the previously studied ``cloud-crushing" problem in which a shock is driven into a cloud of over-dense material.  Although previous work focused on the forward shock of a supernova propagating into the ISM, the process is analogous to that of a reverse shock penetrating the clumpy supernova ejecta.

\begin{figure*}[htp]
\centering
  \includegraphics[width=0.75\textwidth]{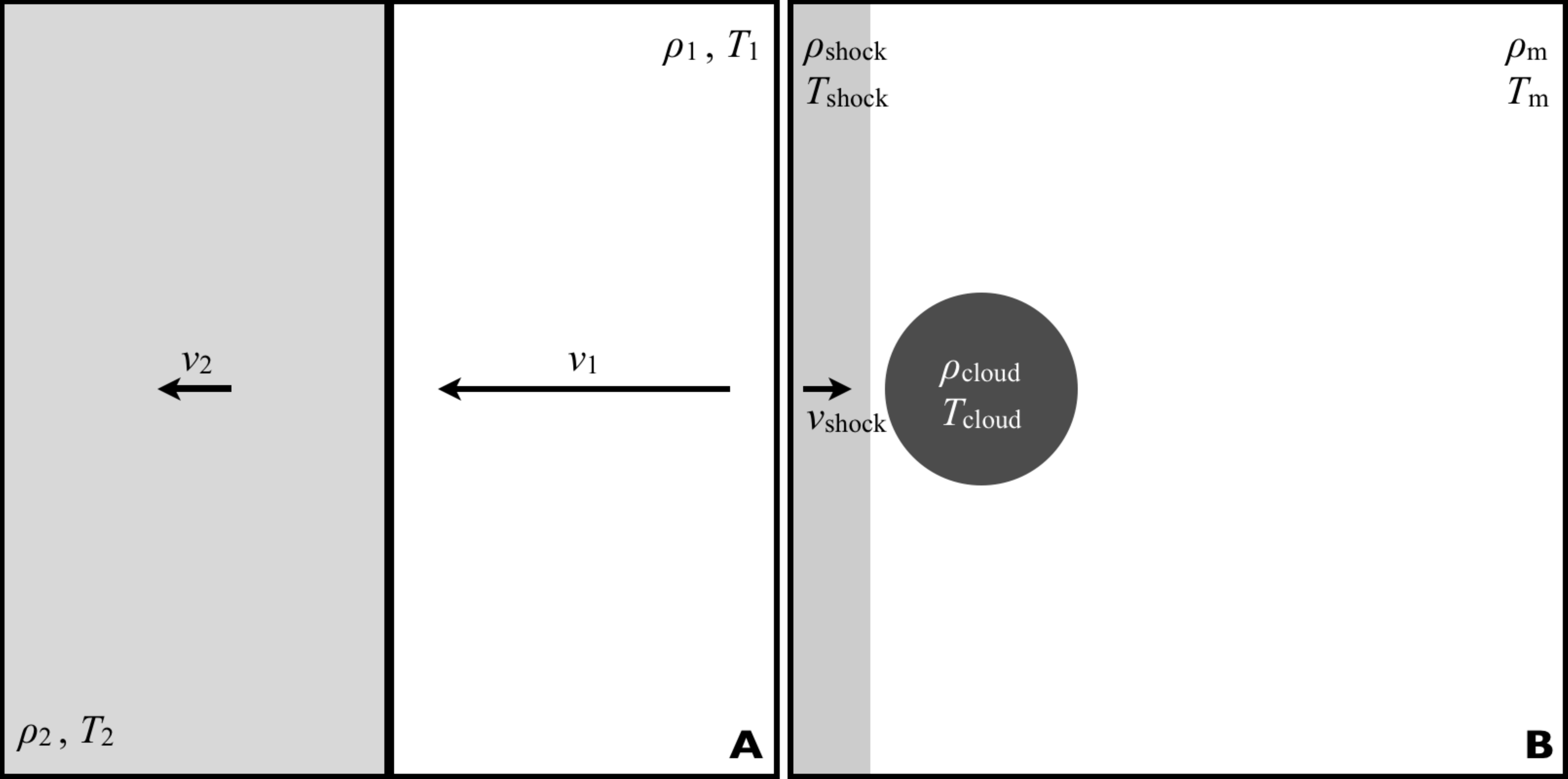}
  \caption{(a) Jump condition quantities in the frame of the shock front, represented by the vertical black line. (b) Depiction of cloud-crushing setup in the frame of the ambient medium.  The medium and the cloud are both at rest. }\label{jumpcrush}
\end{figure*}

The basic nature of our new problem type is to initialize a spherical cloud with constant overdensity, $\chi$, embedded in an ambient medium of density $\rho\subscript{m}$ such that $\rho\subscript{cloud} = \chi\rho\subscript{m}$.  We then send a shock wave with Mach number, $\mathcal{M} = v$/$c\subscript{s}$ where $c\subscript{s}$ is the adiabatic sound speed with $\gamma = 5/3$, in the cartesian $x$-direction toward the cloud.  Additional user-defined parameters required to initialize the simulations include the radius of the cloud, $r\subscript{cloud}$, and the temperature of the cloud, $T\subscript{cloud}$.  All of the remaining quantities needed to start the simulation are derived from the above mentioned user-specified parameters.  These derived values include temperature of the ambient medium, $T\subscript{m}$, velocity of the shock, $v\subscript{shock}$, post-shock density, $\rho\subscript{shock}$, and post-shock temperature, $T\subscript{shock}$.  We set $T\subscript{m}$ so that the cloud remains in pressure equilibrium, while the shock-related values are calculated using the Rankine-Hugoniot jump conditions, which when defined in the frame of the shock front, are given by:
\begin{eqnarray}
\frac{\rho\subscript{2}}{\rho\subscript{1}} &=& \left[\frac{\left(\gamma + 1\right)\mathcal{M}^2}{\left(\gamma - 1\right)\mathcal{M}^2 + 2}\right],
\end{eqnarray}

\begin{eqnarray}
\frac{v\subscript{1}}{v\subscript{2}} &=& \left[\frac{\left(\gamma + 1\right)\mathcal{M}^2}{\left(\gamma - 1\right)\mathcal{M}^2 + 2}\right],
\end{eqnarray}

\begin{eqnarray}
\frac{T\subscript{2}}{T\subscript{1}} &=& \left[\frac{\rho\subscript{1}}{\rho\subscript{2}}\right]\left[\frac{2\gamma\mathcal{M}^2 - \left(\gamma - 1\right)}{\left(\gamma + 1\right)}\right].
\end{eqnarray}
Here, the subscripts 1 and 2 refer to the upstream and downstream quantities of the flow with respect to the shock front and $\mathcal{M} = v\subscript{1}$/$c\subscript{s}$.  See Figure \ref{jumpcrush}a for a schematic of the flow.  From these jump conditions, we shift from the rest frame of the shock front to the rest frame of the ambient medium to define the quantities used in our simulation.  Under this transformation, the upstream and downstream values for density and temperature become the medium and shock quantities, respectively.  However, the velocity of the inflowing material is defined as the difference between upstream and downstream values, $v\subscript{shock} \equiv v\subscript{1} - v\subscript{2}$.  This results in the following set of equations to define the initial conditions for the simulation:
\begin{eqnarray}
\rho\subscript{shock} &=& \rho\subscript{m}\left[\frac{\left(\gamma + 1\right)\mathcal{M}^2}{\left(\gamma - 1\right)\mathcal{M}^2 + 2}\right],
\end{eqnarray} 

\begin{eqnarray}
T\subscript{shock} &=& T\subscript{m}\left[\frac{\rho\subscript{m}}{\rho\subscript{shock}}\right]\left[\frac{2\gamma\mathcal{M}^2 - \left(\gamma - 1\right)}{\left(\gamma + 1\right)}\right],
\end{eqnarray}

\begin{eqnarray}
v\subscript{shock} &=& c\subscript{s} \mathcal{M} \left( 1 - \frac{ \rho\subscript{m}}{ \rho\subscript{shock}} \right),
\end{eqnarray}
where the meanings of $\mathcal{M}$ and $c\subscript{s}$ remain unchanged and the flow is now represented by Figure \ref{jumpcrush}b.

While this formulation of the cloud-crushing problem is sufficient to study the general properties of the shock-cloud interaction and compare to previous works, we made provisions for additional complexities worthy of investigation.  In one avenue of investigation, we allowed the cloud to take on a variety of density profiles.  While the simplest implementation is to define the cloud as a sphere of uniform density, which we explore in some of our simulations, we expect that the boundary between the ejecta clump and the ambient medium should be smoothly varying, analogous to the ISM clouds of \cite{Patnaude:2005zl} and \cite{Nakamura:2006qy}.  In addition, a ``soft" cloud boundary reduces undesirable numerical effects commonly produced by a hard edge.  Therefore, we define the cloud to consist of a smoothly varying envelope surrounding a uniform core for a fraction of our simulations.  Specifically, we allow for two types of cloud envelopes --- one given by a $1/r$ power-law fall-off,
\begin{eqnarray}
\rho(r) &=& \left\{
	\begin{array}{lr}
		\rho\subscript{core} \ \ \ \ \ \ \ \ \ \ \ \ \ \ \ \ \ \ \ ;  0 \le r \le r\subscript{core} \\
			\\
		\rho\subscript{core} + \left[\rho\subscript{m} - \rho\subscript{core}\right]\left[\frac{1/r - 1/r\subscript{core}}{1/r\subscript{cloud} - 1/r\subscript{core}}\right] &\\
			\\
			\ \ \ \ \ \ \ \ \ \ \ \ \ \ \ \ \ \ \ \ \ \ \ \ \ ; r\subscript{core} < r \le r\subscript{cloud}
	\end{array}
\right. ,
\end{eqnarray}
and another by a gaussian fall-off,
\begin{eqnarray}
\rho(r) &=& \left\{
	\begin{array}{lr}
		\rho\subscript{core} \ \ \ \ \ \ \ \ \ \ \ \ \ \ \ \ \ \ \ ; 0 \le r \le r\subscript{core} \\
		\\
		\rho\subscript{core}\left[\frac{\rho\subscript{m}}{\rho\subscript{core}}\right]^{\left(\frac{r - r\subscript{core}}{r\subscript{cloud} - r\subscript{core}}\right)^2} &\\
		\\
		\ \ \ \ \ \ \ \ \ \ \ \ \ \ \ \ \ \ \ \ \ \ \ \ \ ; r\subscript{core} < r \le r\subscript{cloud}
	\end{array}
\right. ,
\end{eqnarray}
where $r\subscript{core}$ is the inner radius at which the cloud is assumed to have a uniform density and $\rho\subscript{core} = \chi\rho\subscript{m}$ is the density of that core.  In the case of a uniform cloud $\rho\subscript{core}$ and $\rho\subscript{cloud}$ are interchangeable as indicated in Figure \ref{jumpcrush}b.

For all of the simulations presented in this paper, we have treated the shock as a constant inflow of material.  This is representative of the scenario in which the width of the shock is much large than the radius of the cloud (effectively infinite).  For the situation in which the shock is of finite width and comparable in size to that of the cloud, it would be necessary to send a time-limited pulse of post-shock material.  Such methods are being investigated for future work.

\subsection{Cloud Tracking}
Finally, we implement two additional features that allow us to track the evolution of the ejecta clump.  First, we populate all of our clouds with a uniform density of tracer particles to represent dust contained in the ejecta material.  This allows us to track the movement of the dust while the cloud is crushed, as well as the nature of the medium in which the dust is embedded.  We can then post-process the history of each tracer particle to calculate dust destruction by assigning each particle some distribution of dust-grain properties.  Second, we add a simple scalar field that is advected with the flow in the same manner as density, but does not alter the state of the system.  We use this field to represent ``cloud material", $\xi$, by initializing the values to be non-negligible only in cells that fall within the radius of the cloud.  This field is motivated purely by computational necessity, and all cloud material values are dimensionless and devoid of physical meaning. To avoid potential round-off errors in the numerics, we set cells that are outside the cloud radius to be small but non-zero (10$^{-20}$).  The non-negligible values are set to be dependent on the density of the cloud such that a cell with initial density $\rho\subscript{core}$ is assigned a cloud material value of $\xi=1$.  This means that a uniform density cloud would be filled with 1's while a cloud with a power-law or gaussian envelope would have 1's in the core and then a fall-off in cloud material that scales directly with the density profile.  This field proves to be particularly useful as a refinement criterion for the AMR component of Enzo to ensure that the hydrodynamic evolution of the cloud is tracked with sufficient resolution.  Section \ref{revshocks} discusses its uses as a refinement criterion.

\section{The Simulations}\label{sims}

\subsection{Dimensionality and Comparison to Previous Works}\label{comparework}
Before addressing the ejecta-reverse shock interaction, we present simulations in which we demonstrate Enzo's ability to successfully handle the cloud-crushing problem.  We also compare it to previous studies that investigated the interactions between the forward shock and the ISM.  In this simulation, we construct a computational volume that consists of $512^3$ cells and a physical size of $2.47 \times 10^{18}$ cm (0.8~pc) on each side.  We define the cloud of over-dense material to have a radius $r\subscript{cloud} = 3.09 \times 10^{17}$ cm (0.1~pc) and an over-density factor $\chi =$ 10 in a higher resolution nested grid such that there are 128 cells per cloud radius, a physical resolution of $2.41 \times 10^{15}$ cm ($\sim$161~AU) per cell edge.  This choice is motivated by \cite{Klein:1994uq} and \cite{Orlando:2005rm} who found that this provides sufficient resolution to capture the physical properties of the problem.  We then use inflow boundary conditions to send material with the appropriate shock values to produce a $\mathcal{M} = 10$ shock wave in the direction of the cloud.  We allow the simulation to run for a time $t = 3t\subscript{cc}$, where $t\subscript{cc} = \chi^{1/2}r\subscript{cloud}/v\subscript{shock}$ is the cloud-crushing time as defined by \cite{Klein:1994uq}, at which point the cloud flows out of the domain.  We find that this set of input parameters compares closely to simulations carried out by \cite{Bedogni:1990li}, \cite{Stone:1992wh}, \cite{Klein:1994uq}, and \cite{Orlando:2005rm} and therefore creates a simulation that can be compared to those previous works.  We maintain the $2.41 \times 10^{15}$ cm per cell edge resolution over the computational domain that encases the cloud for the full duration of the simulation. 

\begin{figure*}[htp]
\centering
  \includegraphics[width=0.9\textwidth, angle=0]{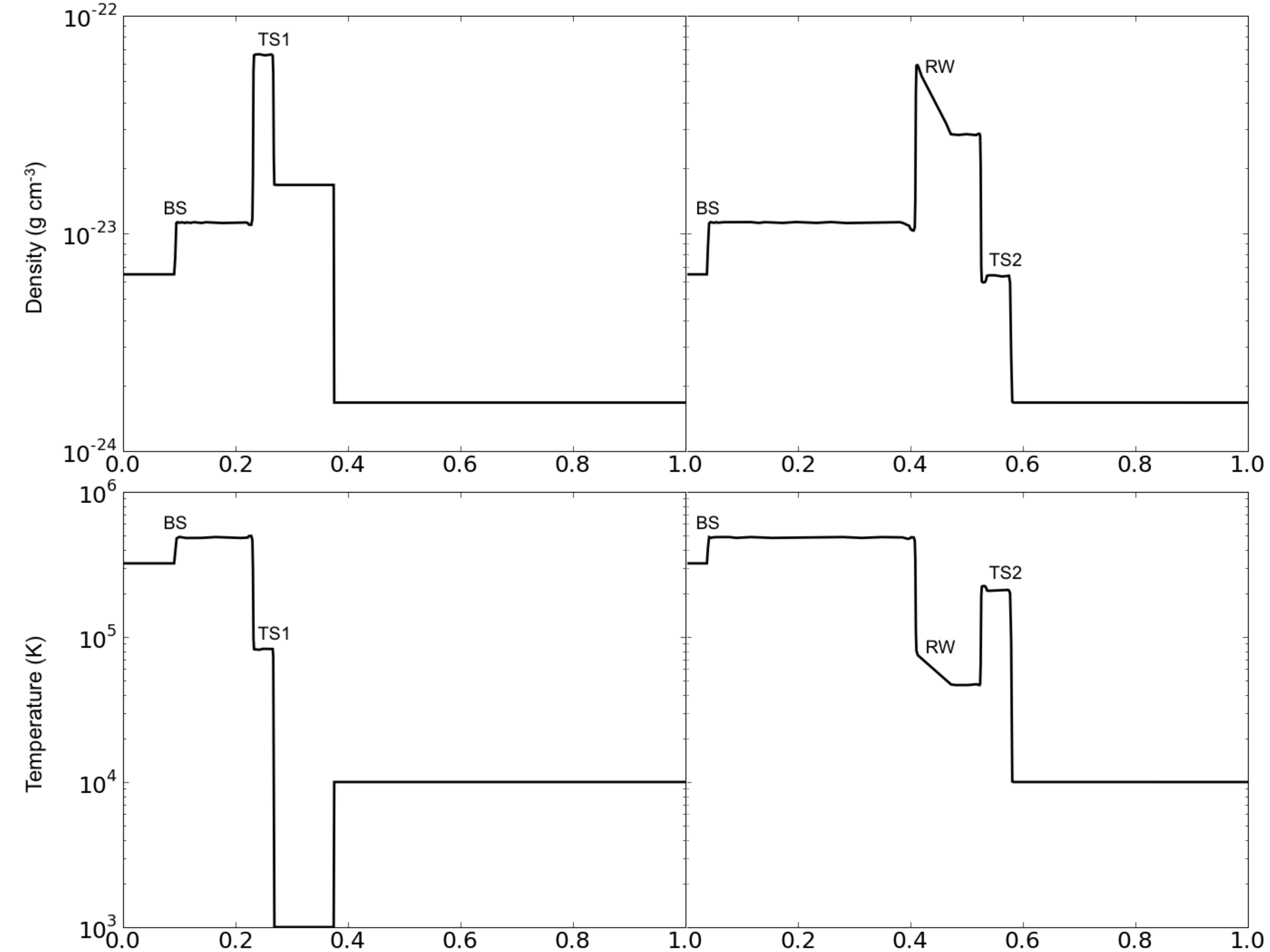}
  \caption{Density (top) and temperature (bottom) profiles for the 1D comparison simulation.  The left column corresponds to $t = 0.52t\subscript{cc}$, and the right column corresponds to $t = 1.40t\subscript{cc}$, where $t = 0$ is when the shock first impinges upon the cloud edge. The labels BS, TS1, TS2, and RW mark the locations of the bow shock, first transmitted shock, second transmitted shock, and rarefaction wave respectively.}\label{1Dcrush}
\end{figure*}

In order to explore the effects of dimensionality on this problem, we investigate the evolution of this particular set up by starting first with a one-dimensional (1D) simulation and then progressing to higher dimensions.  The 1D case corresponds to a scenario in which the inflowing shock impacts an effective ``wall" of cloud material.  This situation lends itself well to tracking the various transmitted and reflected shocks present in the problem as well as for identifying these features in higher dimension simulations.  The first example of such shock propagation occurs when the inflowing shock impacts the cloud wall and transmits a slower shock through the cloud material and a bow shock is reflected back towards the inflow.  When the first transmitted shock crosses the back end of the cloud wall, another shock is transmitted into the ambient medium, while a rarefaction wave travels in the direction opposite the inflow through the newly-shocked cloud material.  At this point, additional ringing occurs as waves bounce back and forth inside the cloud wall.  See Figure \ref{1Dcrush} for density and temperature profiles of the 1D simulation at various points in time which display the above mentioned features.

\begin{figure*}[htp]
\centering
  \includegraphics[width=0.975\textwidth, angle=0]{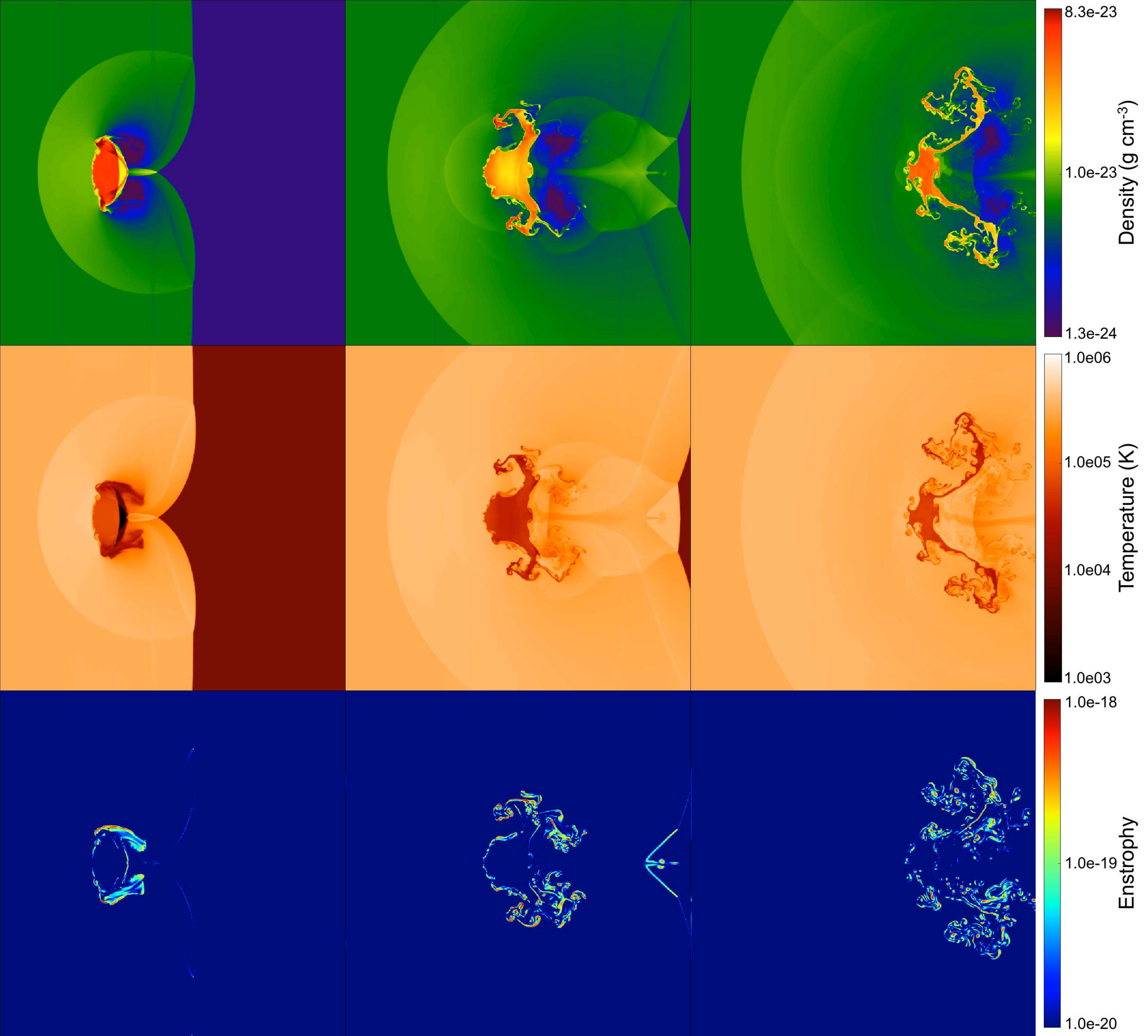}
  \caption{Density (top), temperature (middle), and enstrophy (bottom) for the 2D comparison simulation.  The left column shows the state of the system at $t = 0.81t\subscript{cc}$, the middle at $t = 1.69t\subscript{cc}$, and the right at $t = 2.56t\subscript{cc}$}\label{2Dcrush}
\end{figure*}

For the two-dimensional (2D) case, we find many of the same features as in the 1D case, with the exception that the inflowing shock is capable of wrapping around and re-converging behind the cloud before the shock transmitted into the cloud traverses the full extent of the cloud.  This is because the cloud no longer has the wall-like presence it had in the 1D case, and is now a circular object surrounded by ambient medium.  As a result, an additional reverse shock is driven through the cloud opposite the flow as the inflowing shock envelops the cloud and exerts added pressure on all sides.  The second dimension also allows for vortex creation, as material is stripped off the outer edges of the cloud and pushed into the evacuated region directly behind the cloud by the high pressure inflow material that surrounds it.  In our 2D realization of this problem, we find qualitatively similar results to the work of \cite{Bedogni:1990li} --- further confirmation that Enzo provides adequate machinery to study this problem.  Figure \ref{2Dcrush} shows the results of this 2D run in density, temperature, and enstrophy, the square of the magnitude of the vorticity vector, $|\mathbf{w}|^2 = |\mathbf{\nabla}\times\mathbf{v}|^2$.

\begin{figure*}[htp]
\centering
  \includegraphics[width=1.0\textwidth]{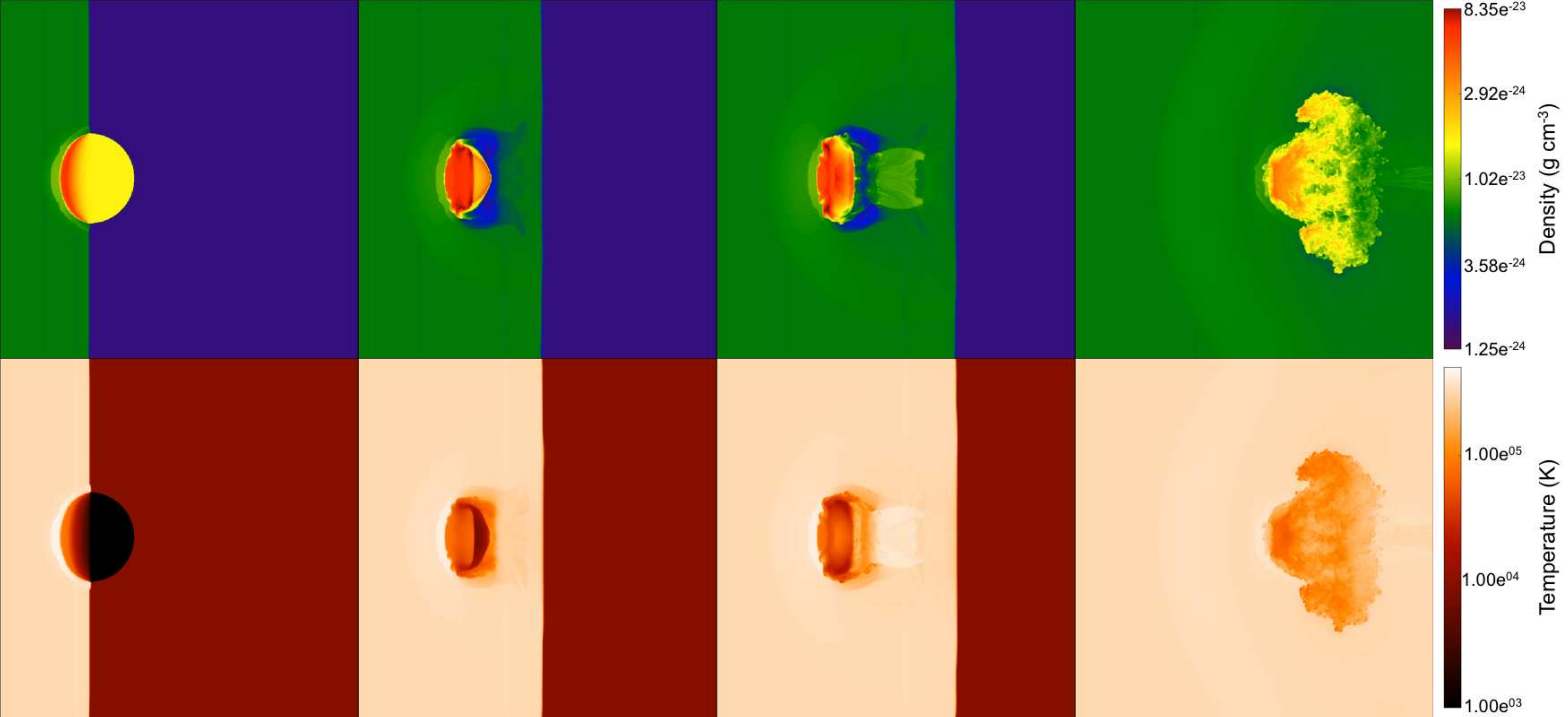}
  \caption{Density (top) and temperature (bottom) projected onto the $x$-$y$ plane weighted by cloud material, $\xi$.  Refer to Section \ref{simsetup} for a description of the cloud material.  From left to right, the snapshots correspond to times $t = 0.23t\subscript{cc}$, $t = 0.72t\subscript{cc}$, $t = 1.02t\subscript{cc}$, and $t = 2.47t\subscript{cc}$, where $t = 0$ is when the shock first impinges upon the cloud edge. }\label{comparesim}
\end{figure*}

In Figure \ref{comparesim}, we see that the time evolution of the projected density for the three-dimensional (3D) simulation also compares favorably with previous works, particularly when we track the development of various features as a function of time.  As in the 1D and 2D cases, when the inflowing shock first impacts the cloud, we see a slower shock transmitted into the cloud and a reflected bow shock that propagates into the post-shock medium while the cloud becomes laterally flattened.  By time $t = 0.75t\subscript{cc}$, the shock has flowed completely past the cloud and converged directly behind the cloud on the $x$-axis.  At this point, the cloud is bathed in the high-pressure, post-shock medium and is compressed in all directions; an additional shock is driven back into the cloud opposite the direction of the inflow.  Eventually, around $t = 1.2t\subscript{cc}$, the cloud begins to re-expand into the post shock medium owing to the compression caused when the shock first passed over it.  Finally, for $ t \gtrsim 2.0t\subscript{cc}$, the cloud is dominated by hydrodynamic instabilities and takes on a highly non-uniform, fragmented, and filamentary structure as it becomes comoving with the inflowing material.  The evolution of vorticity in the 3D simulation becomes even more complex than the 2D case, as the third dimension allows for the stretching of vortex tubes and leads ultimately to a greater level of symmetry-breaking turbulence.  All of these features were seen by \cite{Stone:1992wh}, \cite{Klein:1994uq}, and \cite{Orlando:2005rm} at comparable moments in time.  Figure \ref{comparesim} also shows the same evolution in projected temperature.

\begin{figure*}[htp]
\centering
	\includegraphics[width=1.0\textwidth]{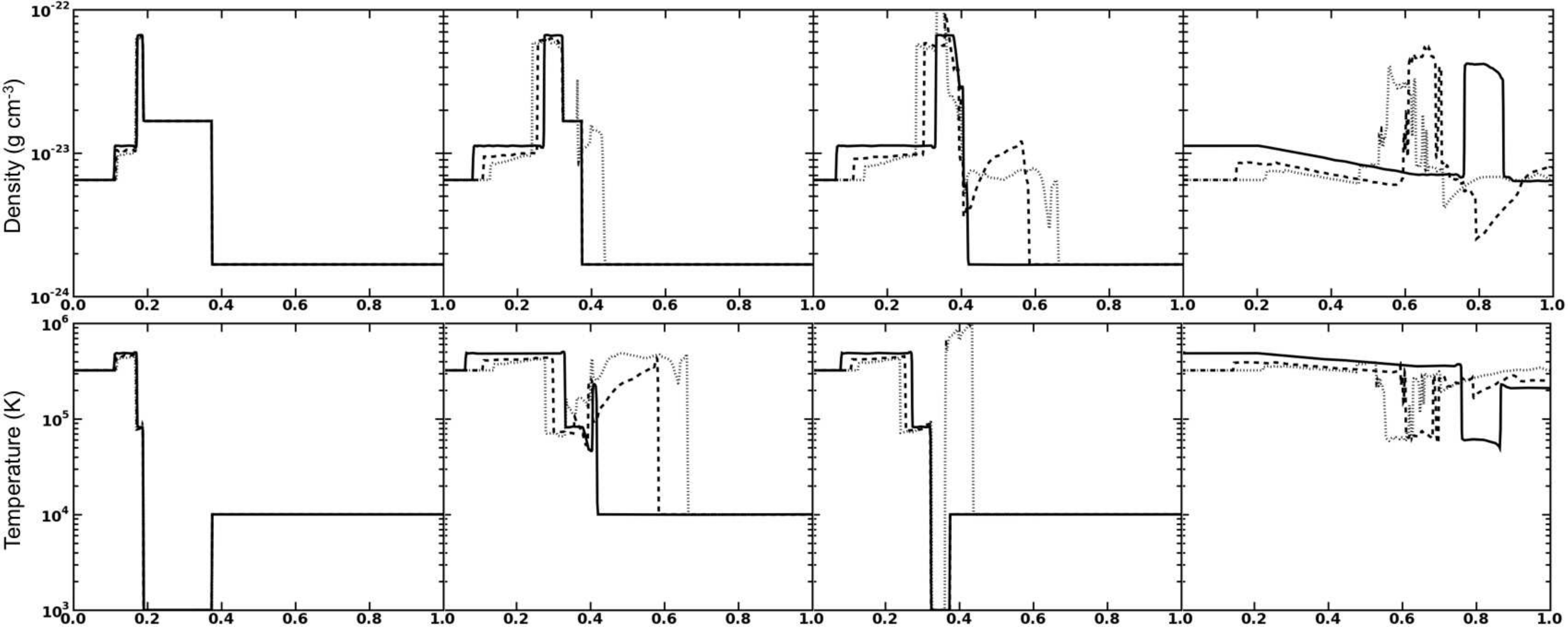}
  \caption{Density (top) and temperature (bottom) profiles for 1D (solid line), 2D (dashed line), and 3D (dotted line) comparison simulations.  The x-axis is presented in code units. From left to right, the panels correspond to times $t = 0.23t\subscript{cc}$, $t = 0.72t\subscript{cc}$, $t = 1.02t\subscript{cc}$, and $t = 2.47t\subscript{cc}$.}\label{allDoverlay_dens}
\end{figure*}

To produce a more direct comparison of the three cases of varying dimensionality, we found it useful to look at 1D profiles of density and temperature as a function of time.  For the 1D data, these profiles are plots of the data as in Figure \ref{1Dcrush}, while for the 2D and 3D simulations, they are found by tracing a path through the center of the computational domain that runs parallel to the $x$-axis.  Figure \ref{allDoverlay_dens} presents these profiles for select time steps.  Upon inspection, the 2D and 3D cases are considerably more complex than the 1D case, but they still show similar evolution, especially in the time period before the inflowing shock wraps around the cloud in the 2D and 3D cases.   We also point out that the cloud material moves across the computational domain more rapidly for lower dimension simulations.  This can be explained by the fact that the momentum imparted by the planar shock has fewer degrees of freedom to transport into.  For the 1D case, all of the material is forced to move in the $x$-direction by the inflow, but in the 2D and 3D cases it can be diffused into the $y$- and $z$-directions.
 
\subsection{Reverse Shocks and Ejecta Clumps}\label{revshocks}
While the simulations we use to study the interaction between a clump of ejecta and a supernova remnant reverse shock are qualitatively similar to the case of a forward shock penetrating the ISM, the values of the input parameters are considerably different.  In the case of the reverse shock impacting newly formed ejecta, the relative velocity between the cloud and the shock wave is generally an order of magnitude higher ($\sim$10$^3$ km s$^{-1}$) than the shock-ISM case ($\sim$10$^2$ km s$^{-1}$), and the overdensity of the ejecta clump is on the order of $\chi \sim$ 1000 versus the $\chi \sim 10$ values assumed for ISM clouds.

In addition to the modifications of the input parameters, we also made use of the variable cloud structure to explore what role, if any, the density profile of the cloud played in determining the amount of dust that would be sputtered as the ejecta are shocked.  Using the full array of configurable options in our cloud-crushing problem type (shock width, cloud structure, radiative cooling, cloud overdensity, shock velocity), we ran a grid of simulations to determine the parameters that strongly influence the dust destruction induced by the shock-cloud interaction.  Specifically, we investigate various combinations of power-law and gaussian envelopes; cloud over-densities of $\chi =$ 100 and 1000; shock velocities of 1000, 3000, and 5000 km s$^{-1}$; and no cooling and the analytic cooling of \cite{Sarazin:1987kx}. For a list of the simulations and their corresponding parameters refer to Table~\ref{gridtable}. 

\begin{deluxetable*}{lrrrrrr}
\tabletypesize{\scriptsize}
\tablecaption{Simulations Parameters \label{gridtable}}
\tablewidth{0pt}
\tablehead{\colhead{Simulation} &
	\colhead{Cloud Envelope \tablenotemark{a}} &
	\colhead{$\chi$} &
	\colhead{$v\subscript{shock}$} &
	\colhead{$\mathcal{M}$} &
	\colhead{$t\subscript{cc}$} &
	\colhead{Cooling \tablenotemark{b}} \\
	\colhead{} &
	\colhead{} &
	\colhead{} &
	\colhead{(km s$^{-1}$)} &
	\colhead{} &
	\colhead{(yrs)} &
	\colhead{}}
\startdata
1	&	Gaussian		&	100	&	1000		&	27.83	&	31.69	&	None	\\
2	&	Gaussian		&	100	&	1000		&	27.83	&	31.69	&	SW87	\\
3	&	Gaussian		&	100	&	3000		&	83.49	&	10.56	&	None	\\
4	&	Gaussian		&	100	&	3000		&	83.49	&	10.56	&	SW87	\\
5	&	Gaussian		&	100	&	5000		&	139.15	&	6.34		&	None	\\
6	&	Gaussian		&	100	&	5000		&	139.15	&	6.34		&	SW87	\\
7	&	Gaussian		&	1000	&	1000		&	8.80		&	100.21	&	None	\\
8	&	Gaussian		&	1000	&	1000		&	8.80		&	100.21	&	SW87	\\ 
9	&	Gaussian		&	1000	&	3000		&	24.60	&	33.40	&	None	\\ 
10	&	Gaussian		&	1000	&	3000		&	24.60	&	33.40	&	SW87	\\
11	&	Gaussian		&	1000	&	5000		&	44.00	&	20.04	&	None	\\ 
12	&	Gaussian		&	1000	&	5000		&	44.00	&	20.04	&	SW87	\\
13	&	Power-law	&	1000	&	1000		&	8.80		&	100.21	&	None
\enddata
\tablenotetext{a}{For the Gaussian and Power-law envelopes, $r\subscript{core} = 0.7r\subscript{cloud}$}
\tablenotetext{b}{For the cooling, ``SW87" represents analytic formula by \cite{Sarazin:1987kx}}
\end{deluxetable*}

All of the simulations have root-grid dimensions of 512$\times$256$\times$256, which provides a physical resolution of $1.25 \times 10^{15}$ cm ($\sim$84~AU) per cell edge.  At this level, the radius of the cloud spans 8 cells for a physical radius of 10$^{16}$ cm.  In addition to the base-level resolution, we include a problem-specific refinement criterion that allows for two additional levels of resolution.  Cell refinement is controlled using the previously mentioned cloud material field and flags cells for refinement based on a dimensionless cloud material ``mass", $m\subscript{\xi} = \xi V\subscript{cell}$, where $V\subscript{cell}$ is the volume of the cell.  To determine whether or not a cell should be flagged it must have a cloud material mass greater than $m\subscript{flag} = m\subscript{min}r^{l\alpha}$, where $m\subscript{min}$ is the minimum mass for refinement, $r$ is the refinement factor, $l$ is the current refinement level of the cell, and $\alpha$ is a tunable exponent used to control whether the refinement acts in a Lagrangian ($\alpha=0$), super-Lagrangian ($\alpha < 0)$, or sub-Lagrangian ($\alpha > 0$) manner.  For all simulations presented in this paper, we use $m\subscript{min}=10^{-8}$, $r=2$, and $\alpha=0$, where $m\subscript{min}$ is a dimensionless number.  By allowing the grid to refine to two levels beyond the root grid, we achieve a physical resolution of $3.125 \times 10^{14}$ cm ($\sim$21~AU) on the highest level.  Although these reverse-shock simulations do not have the same number of cells per cloud radius as presented in the comparison case, we find that this particular combination of initial root-grid resolution, cloud material refinement criterion, and two-level refinement is sufficient to track and study the properties of the interactions between the reverse shock and the ejecta cloud.  Additionally, the final dust masses do not change significantly with increased resolution.

We set the duration of our reverse-shock simulations based on the longest computationally manageable timescale for the parameters that produce the longest cloud-crushing time.  For the above-mentioned parameter space, this corresponds to simulations with $r\subscript{cloud} = 10^{16}$~cm, $\chi = 1000$, and $v\subscript{shock} = 1000$ km s$^{-1}$, giving $t\subscript{cc} \approx 100$ yrs.  With this constraint, we are able to run these simulations to a physical time of $t \approx 10^3$~yrs $\approx 10t\subscript{cc}$.  Therefore, we run all simulations to a minimum of 10 cloud-crushing times which results in varied physical times.

Owing to the nature of the inflowing shock and relevant time scales, we encountered a possible simulation limitation in which the cloud material was often blown out of the back end of the computational domain by the shock medium.  We solved this limitation and kept the majority of the cloud material in the simulation box for the desired duration by allowing the simulations to run up to a point in which the cloud is nearly comoving with the inflowing shock material and then shift into the inflow frame.  We execute this by pausing the calculation at a suitable moment in time, subtracting off the inflow velocity, $v\subscript{shock}$, from the $x$-velocity of every cell, and resuming the simulation.  After the simulation is restarted, the cloud evolves mostly statically in terms of bulk motion.  Since the temperatures and densities remain unaltered, the cloud's relative motions are identical to what they would have been in the initial rest frame.

In studying the evolution of these ``ejecta-crushing" simulations, we find that at early times the simulations agree with the previously discussed 3D comparison simulation, with the exception that the smoothly varying envelope subdues the interactions between the cloud boundary and the inflowing material.  The lack of a sharp density discontinuity decreases the development of features perpendicular to the direction of the inflow and the ``wings" present in the comparison simulation are not seen.  Instead, the outer envelope material is stripped away, allowing the remaining compressed core to expand adiabatically into the surrounding medium.  Also, since we run these simulations for more cloud-crushing times, we see the continued destruction imparted upon the cloud by the constant inflow of material as it gets progressively more shredded.  After approximately five cloud-crushing times, the relative velocities between the cloud material and shock material drop to $\sim$0.1$v\subscript{shock}$ as the cloud becomes entrained in the flow. By $t\sim 8t\subscript{cc}$, the cloud has become well mixed with the inflowing material, with density variations on the few percent level.

One of the most striking differences between the simulations, with and without cooling, is the formation of cold, dense nodules in the simulations with cooling turned on.  The nodules appear to be amplifications of over-densities created in the cloud.  The regions of high density have sufficiently shortened cooling times and therefore cool and collapse approximately isothermally.  As a result, the dense knots are able to persist through their bombardment with inflowing material for a longer fraction of the simulation duration.

\begin{figure*}[htp]
\centering
	\includegraphics[width=1.0\textwidth]{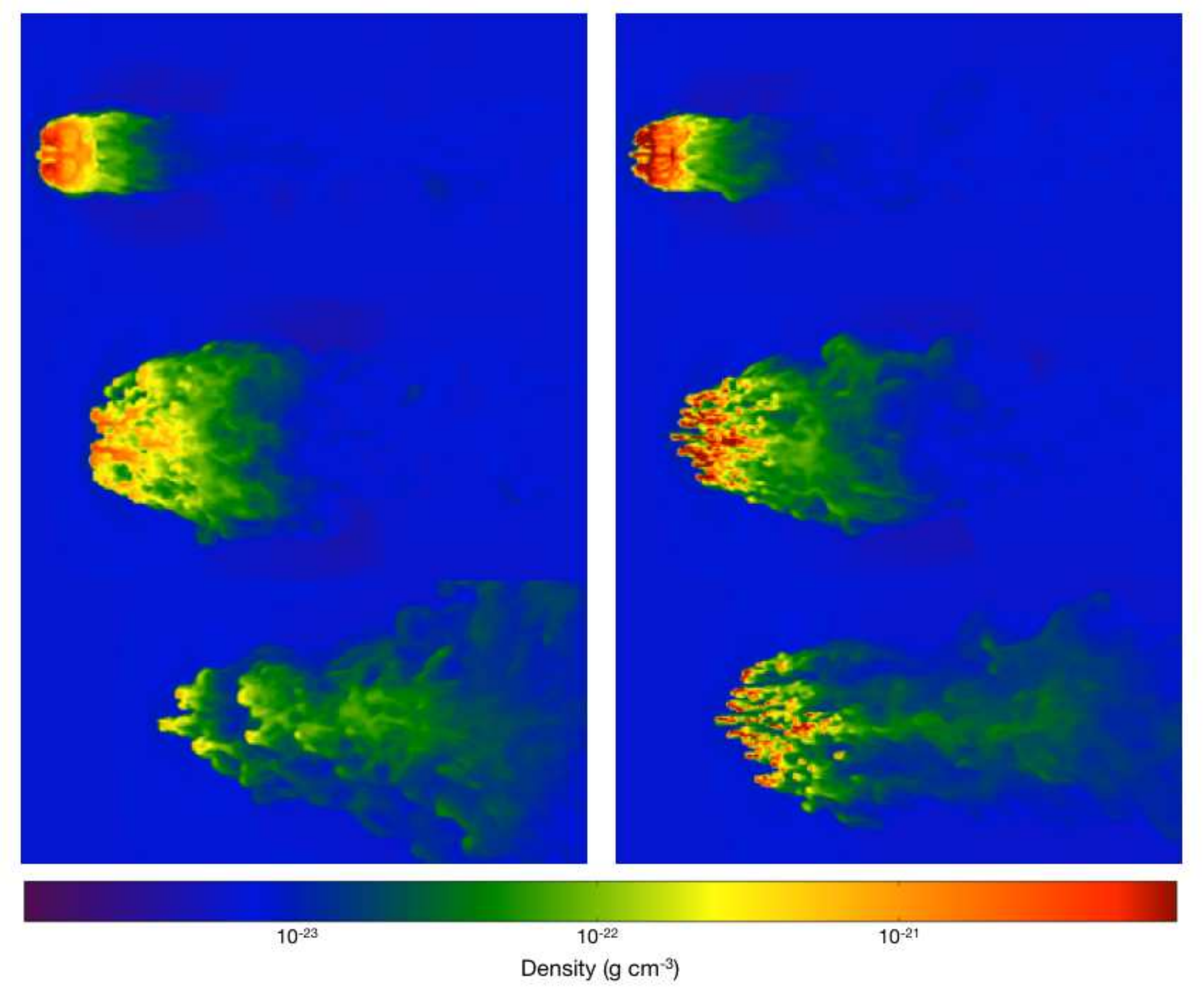}
  \caption{Density projections in the $x$-$z$ plane weighted by cloud material, $\xi$, for the case of $\chi=1000$ and $v\subscript{shock}=1000$ km s$^{-1}$.  The left column shows the evolution for a non-cooling simulation, while the right column shows that for a cooling simulation.  From top to bottom, the time progresses in the following order: $t=1.8t\subscript{cc}$, $t=2.4t\subscript{cc}$, $t=3.0t\subscript{cc}$.}\label{rshocknocool}
\end{figure*}

Figure \ref{rshocknocool} shows the evolution of the ejecta clump to roughly one-third of the full simulation time.  Beyond that point, the cloud continues to expand and mix with the shocked medium.  The formation of high density nodules is apparent in the right column of Figure \ref{rshocknocool}.  We note that \cite{Raga:2007fk} found similar dense protrusions in their 3D simulations of radiative, interstellar bullet flow.  The number of nodules that form tends to result from the balance between shock-heating and cooling.  For higher shock velocities, and hence higher temperatures, the number of persisting over-densities decreases.  

When we run the cooling simulations with an additional level of refinement, we generally find these dense knots to be smaller and higher in number.  This result is expected, as the turbulence created by the shock passing through the cloud will naturally cascade down to the smallest physical scale, which is determined by the resolution of the simulation. Additionally, these nodules are constantly being eroded by inflowing material, and at late times their size will be comparable to the grid-cell size.  However, despite the difference in size and count, these small knots evolve similarly in density and temperature, which results in nearly identical dust destruction.

\section{Data Post-processing}\label{dataprocess}

\subsection{Modeling Grain Distributions Using Tracer Particles}\label{tracerparticles}
\begin{figure}[htp]
\centering
  \includegraphics[width=0.48\textwidth]{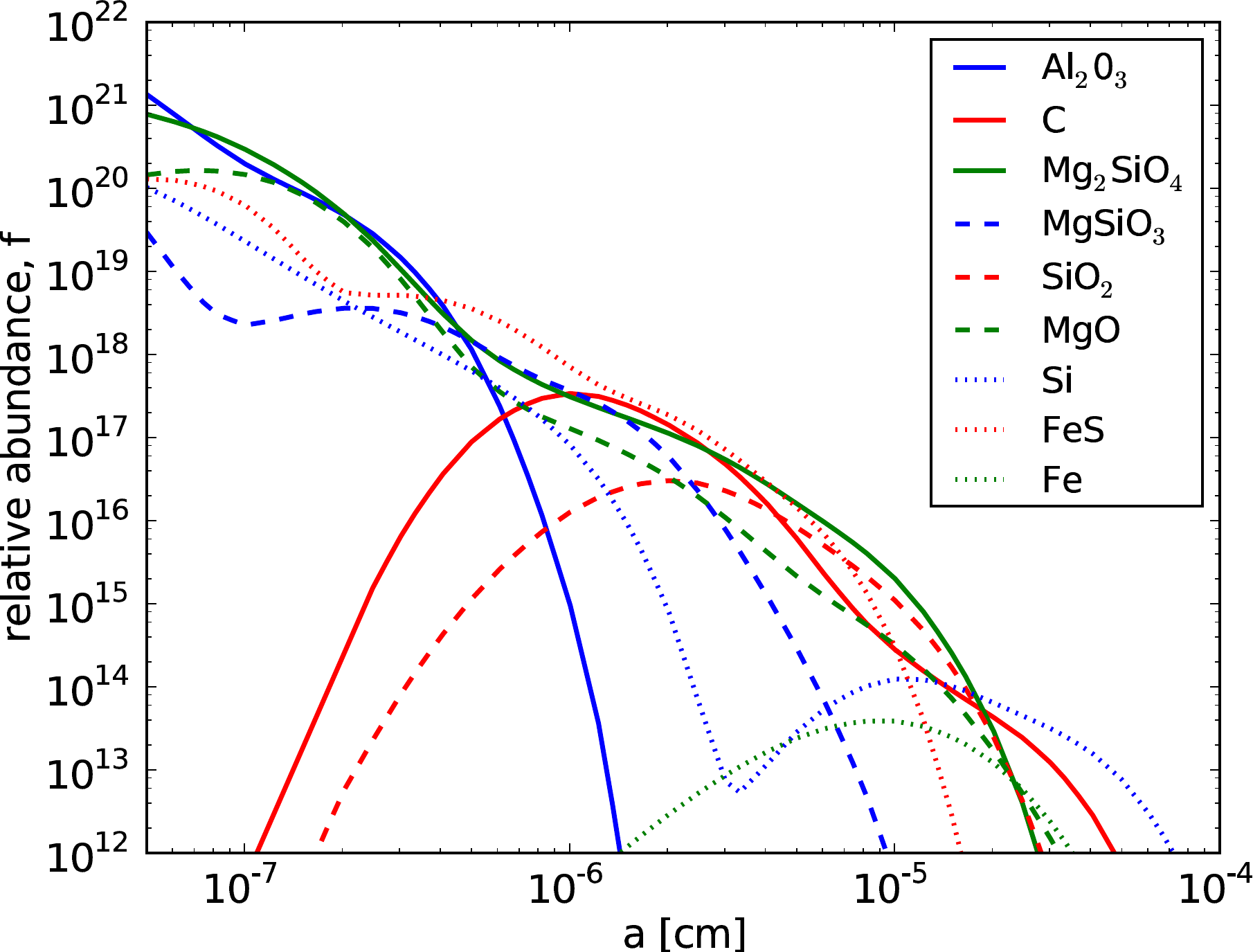}
  \caption{Reproduced grain radius distributions for the species expected in the unmixed ejecta model of a core-collapse supernova with a progenitor mass of 20 M$\subscript{\odot}$ as calculated by \cite{Nozawa:2003pd}.  In situations where data from \cite{Nozawa:2003pd} ended prior to a grain radius of 5$\times$10$^{-8}$ cm, we carried out a linear extrapolation to smaller grain radii to make the data continuous.}\label{nozdistribution}
\end{figure}

As mentioned previously, the clouds in all of our simulations are populated with a uniform distribution of tracer particles.  We use 4096 particles in order to adequately sample the cloud while simultaneously keeping the computational requirements of our post-processing analysis manageable.  Each individual particle is assigned a full distribution of grain radii and therefore represents a distinct population of dust.  The radius distributions used in all of our analysis come from the values calculated by \cite{Nozawa:2003pd} for a core-collapse supernova (CCSN) with a progenitor mass of 20 $M\subscript{\odot}$, which have been reproduced in Figure \ref{nozdistribution}.  Since we assume that the number density of dust grains in the cloud should scale with the surrounding gas density, we scale the relative abundance of grains in the initial distributions by the density of the surrounding medium that the tracer particles are initially embedded in.  Given this assumption, $f(a) = f\subscript{0}(a)(\rho\subscript{tp}/\rho\subscript{max})$, where $a$ is the grain radius, $f\subscript{0}(a)$ is the unscaled abundance, $\rho\subscript{tp}$ is the density of the surrounding medium for a given tracer particle, and $\rho\subscript{max}$ is the value for the most dense medium in which a tracer particle is initialized.  In general, $\rho\subscript{max} = \rho\subscript{core}$. To determine the distribution of grain radii for all the dust contained in the simulation, we sum the radius distributions for all 4096 tracer particles.

We can also use the tracer particles and their respective grain distributions to calculate a proxy for the amount of dust mass in our simulation.  First, we define the number density of grains between $a-\delta a$ and $a$ to be $n(a) = f(a)\delta a$, where $\delta a$ is set by the number of bins used to track the distribution.  Second, we convert this number density to the mass proxy by multiplying by $a^3$.  By summing the amount of mass in each radius bin, we can determine the amount of mass contained in a given tracer particle and then sum tracer particles to get the total dust mass in the simulation.  Rather than work in absolutes and include the densities of particular grain species, we leave this mass proxy as a relative value and track how it evolves from $t=0$ to $t=t\subscript{final}$.

In the same way that it was difficult to keep all of the cloud material inside the finite computational domain, some fraction of the tracer particles also flow off the grid, particularly ones that start out in the cloud envelope.  Since we have no means of tracking the environment that those tracer particles would exist in after the point where they leave the bounds of the simulation, we exclude them from our sputtering calculations for the full duration of the run.  In some simulations the combination of input parameters leads to vigorous cloud expansion that cannot be confined to our computational domain for the desired duration.  This results in situations in which large fractions of the tracer particles ($>$50\%) flow beyond the box boundaries.  To address this fact, we halt the post-processing analysis at the point when the dust mass contained in the lost tracer particles exceeds 5\% of the initial dust mass.  While we would like to carry out our analysis for all simulations on the same timescales, creating simulation domains that are large enough to contain the cloud's expansion for all of parameter space was too computationally expensive to be feasible.  Despite this limitation, we still find our results up to the 5\% cut-off to be robust.

\begin{figure}[htp]
\centering
  \includegraphics[width=0.48\textwidth]{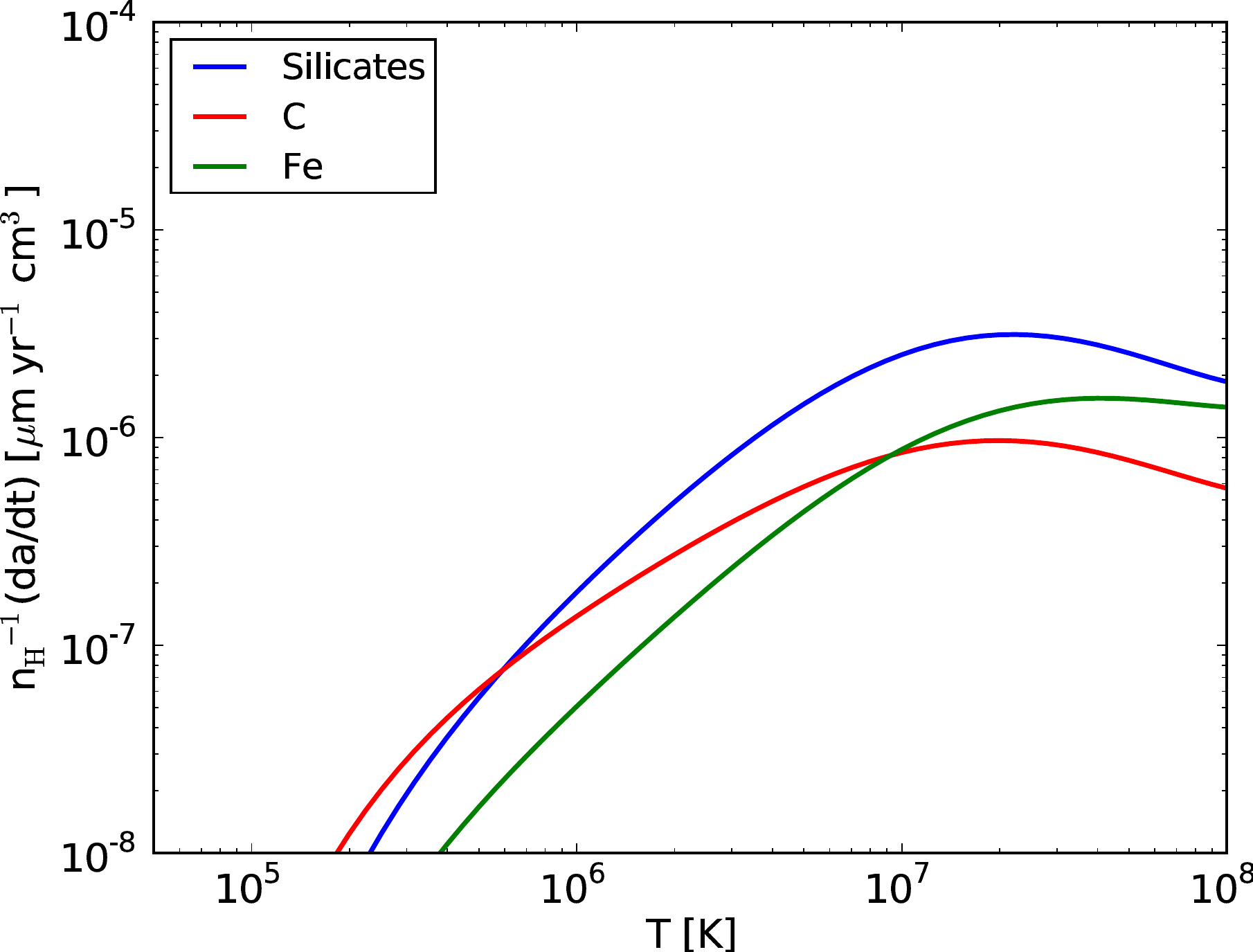}
  \includegraphics[width=0.48\textwidth]{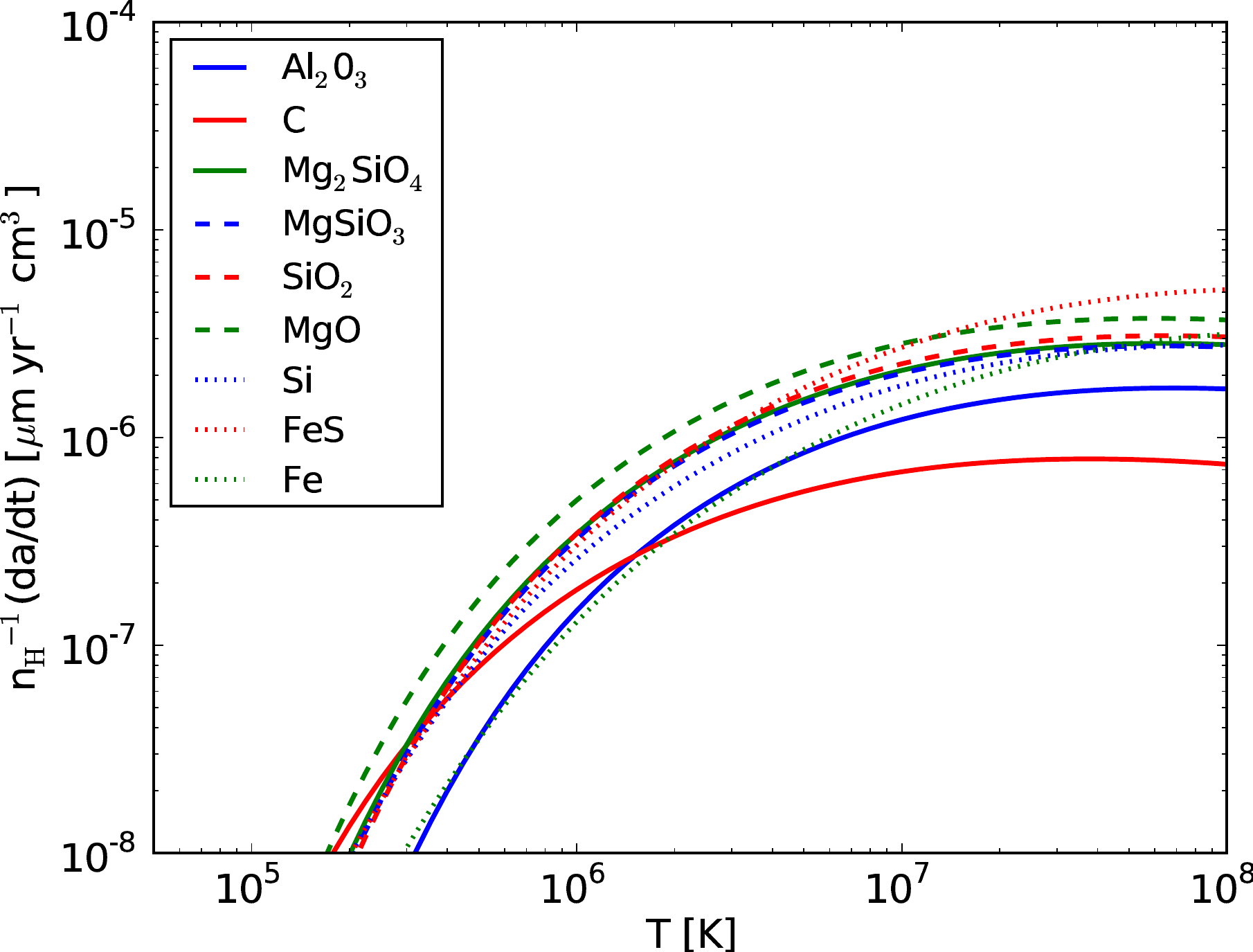}
  \caption{Our adopted thermal sputtering rates based on the polynomial fits of \cite{Tielens:1994kx} (top) and for an elemental composition of $Z = Z\subscript{\odot}$ as calculated by \cite{Nozawa:2006ve} (bottom).}\label{thermalsput}
\end{figure}

\subsection{Dust Sputtering}
Having established our initial grain distributions, we post-process the grains using generalized fifth-order polynomial fits to thermal sputtering from \cite{Tielens:1994kx} and tabulated thermal sputtering rates.  These sputtering rates have been reproduced in Figure \ref{thermalsput}.  We pass the density and temperature of the environment of a given tracer particle to one of the sputtering methods and calculate the erosion rate, $da/dt$, for each time step.  We multiply the erosion rate by the length of the time step and subtract that value from the initial grain radius such that $a\subscript{new} = a\subscript{old} - (da/dt)\Delta t$.  This method allows each population of dust (each tracer particle) to evolve in direct response to its environment for the duration of the simulation.  Using the method mentioned above, we then compute the percentage of dust mass lost as a function of time.  While \cite{Tielens:1994kx} provide sputtering rates for SiC and H$\subscript{2}$O grains, we do not track them our analysis because \cite{Nozawa:2003pd} do not include size distributions for these grains.  In addition, since \cite{Tielens:1994kx} used SiO$\subscript{2}$ as the basis for the sputtering rates of silicates, we use the grain distribution of SiO$\subscript{2}$ as our starting point for silicate sputtering. 

\section{Results}\label{results}
We present the sputtering results for all 13 simulations listed in Table \ref{gridtable}.  Specifically, we study the changes in grain radius distributions from their starting to ending values and the evolution of the dust mass contained in the tracer particles as a function of time.  For a condensed view of the key focus of this paper, dust survival rate, for all combinations of simulations and sputtering methods, we refer the reader to Table \ref{dustsurvival}.  One caveat is that Simulations 8 ($\chi=1000$, $v_{shock}=1000$~km~s$^{-1}$, cooling on), 9 ($\chi=1000$, $v\subscript{shock}=3000$~km~s$^{-1}$, cooling off), and 10 ($\chi=1000$, $v_{shock}=3000$~km~s$^{-1}$, cooling on) do not span the full range of cloud-crushing times because they ended up losing more than 5\% of their dust mass due to tracer particles that left the computational domain.  However, since the dust destruction in Simulations 9 and 10 hits a plateau before the simulation analysis ends, the results of these simulations can still be fairly accurately compared.  Unfortunately, Simulation 8 terminates well before any such plateau, leading to misleading final dust mass values.  In addition, we do not devote any significant discussion to Simulation 13 (power-law density envelope, $\chi=1000$, $v\subscript{shock}=1000$ km s$^{-1}$, cooling off) because the dust mass evolution proved to be nearly identical to Simulation 7 (gaussian density envelope with the same initial conditions), as can be seen in Table \ref{dustsurvival}.

\subsection{Grain Distributions}
In order to get a handle on the nature of the dust grain sputtering, we first inspect the changes in the distributions of grain radius.  We find that, regardless of species, grains with radii less than a few times 10$^{-6}$ cm are sputtered to much lower radii and are often completely destroyed.  However, given the time scale of our simulations, larger grains are capable of surviving the full duration of the sputtering, but are shifted to lower radii.  For simplicity, we defined destroyed grains as those that are sputtered to less than $5\times10^{-8}$ cm.

\begin{figure*}[htp]
\centering
  	\subfigure[C]{
  	\includegraphics[width=0.325\textwidth]{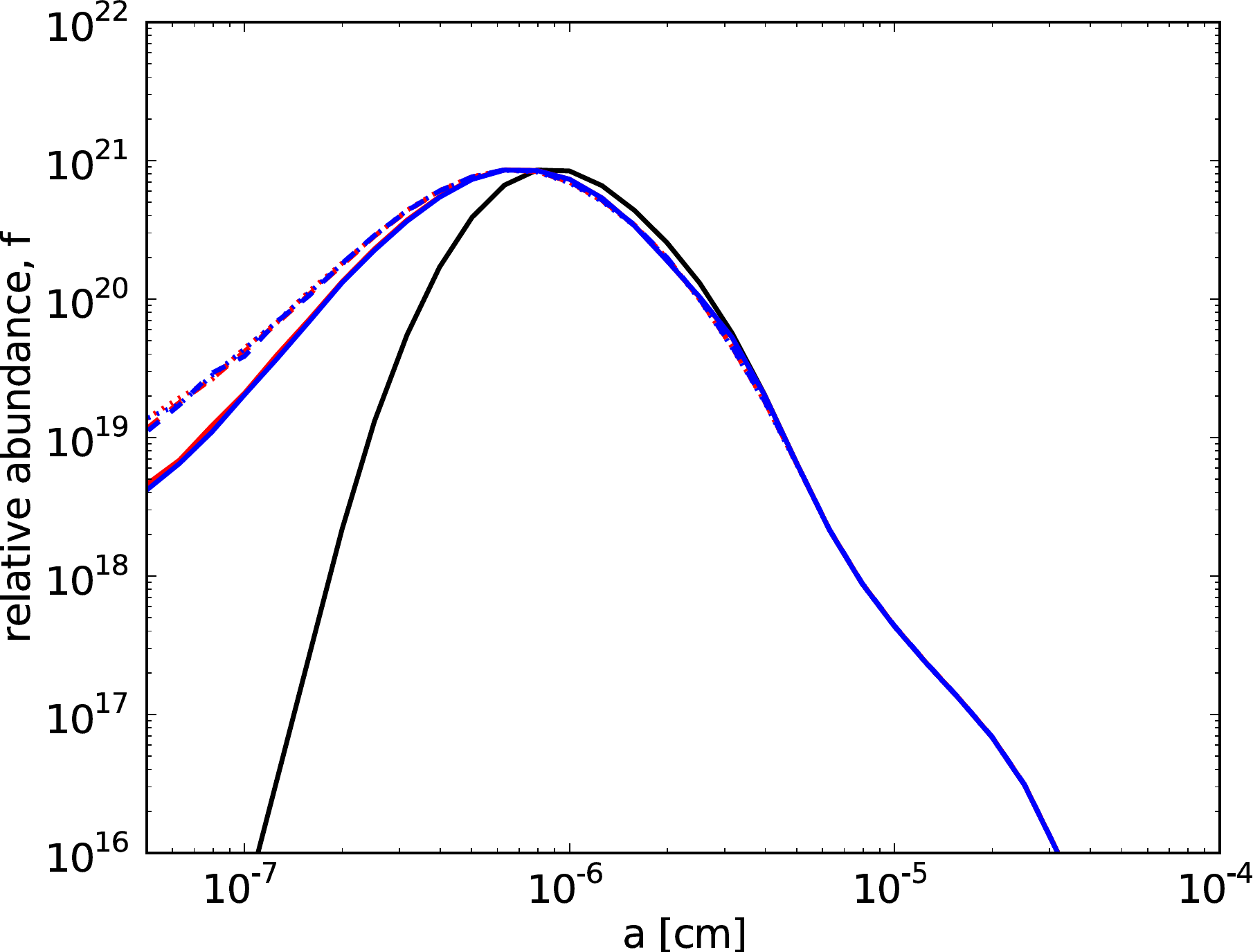}
  	}
	\hspace{-5.00mm}
	\subfigure[Silicates]{
  	\includegraphics[width=0.325\textwidth]{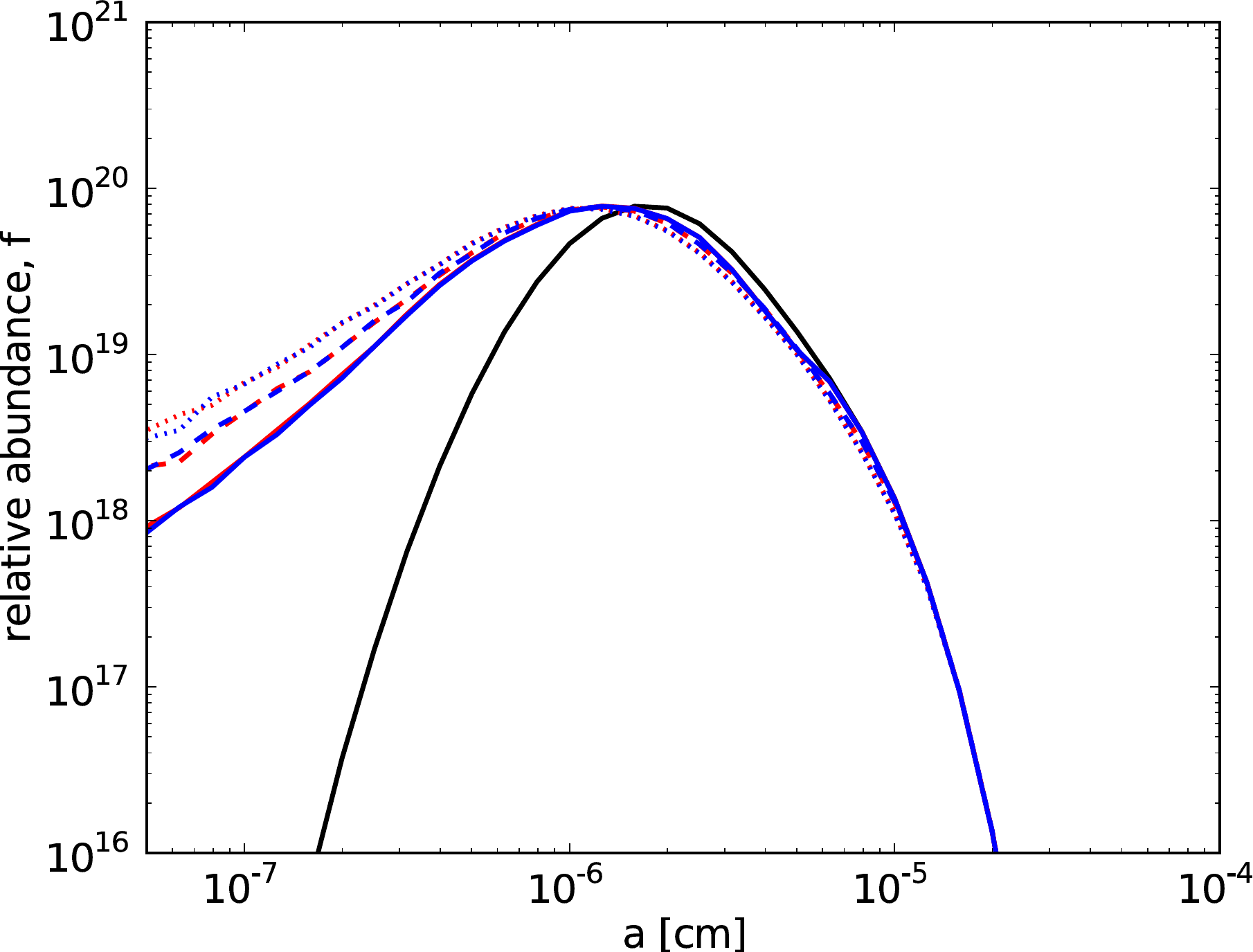}
  	}
	\hspace{-5.00mm}
	\subfigure[Fe]{
  	\includegraphics[width=0.325\textwidth]{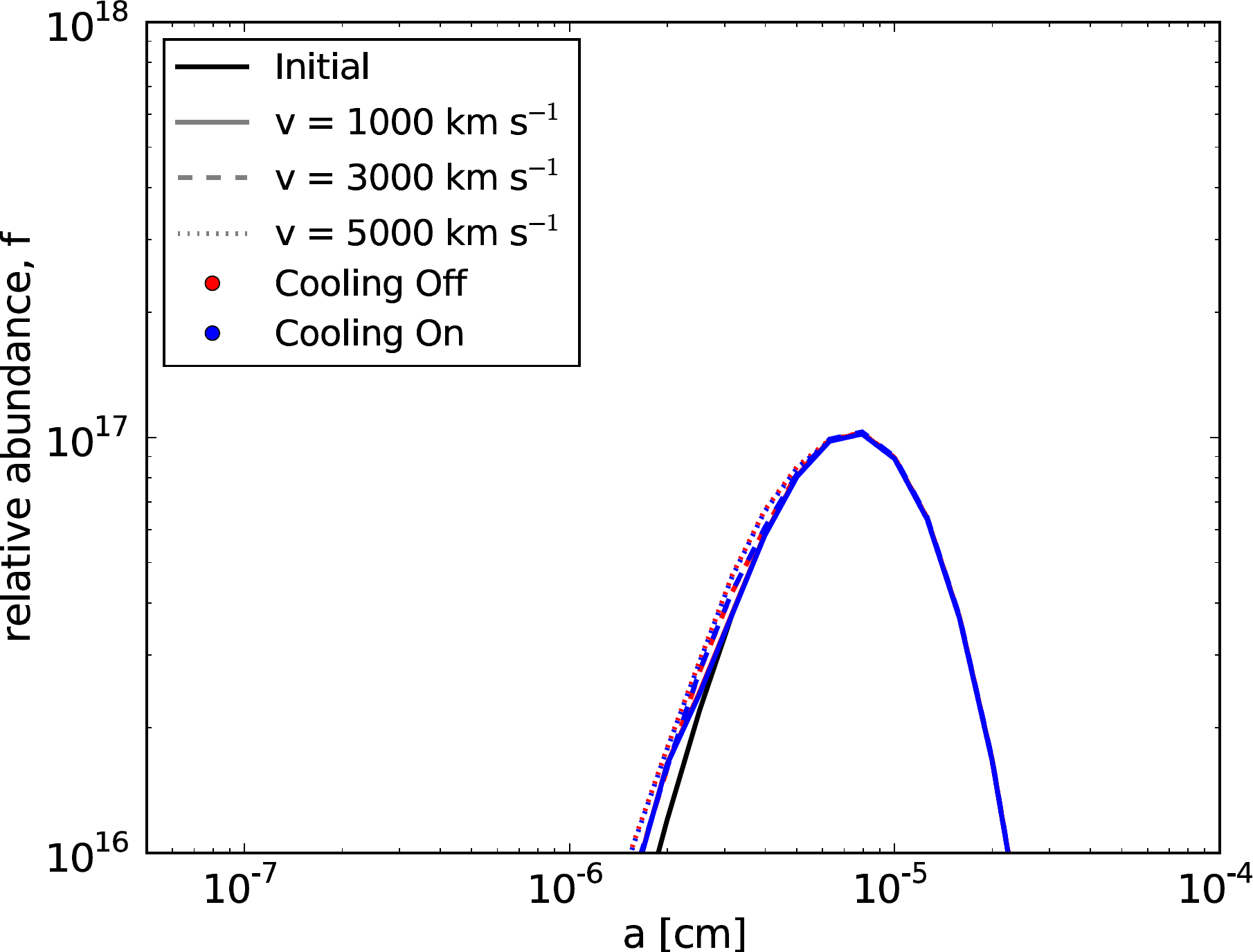}
  	}
  \caption{Grain radius distributions for three of the grain species in \cite{Tielens:1994kx} for simulations with over-density $\chi =100$.  The black line represents the initial distribution, while the colored lines indicate final distributions.  Red signifies simulations with no cooling and blue shows the analytic cooling of \cite{Sarazin:1987kx}.  Solid colored lines indicate simulations with $v\subscript{shock}=1000$ km s$^{-1}$, dashed are $v\subscript{shock}=3000$ km s$^{-1}$, and dotted are $v\subscript{shock}=5000$ km s$^{-1}$.  The grains in this figure were sputtered using the rates from \cite{Tielens:1994kx}.}\label{grainschi100_tiel}
\end{figure*}

In the figures that follow we look at the evolution of grain radius distribution for various locations in parameter space and for two of the three sputtering rate options. In each panel, the solid black line illustrates the initial grain distributions as set by the distributions from \cite{Nozawa:2003pd}.  For the colored lines, solid lines correspond to simulations with the slowest shock velocities, $v\subscript{shock}=1000$ km s$^{-1}$, dashed lines are the simulations with shock velocities of $v\subscript{shock}=3000$ km s$^{-1}$, and dotted lines are the simulations with the highest velocity shocks, $v\subscript{shock}=5000$ km s$^{-1}$.  The colors represent our different options for cooling, where red highlights runs with no cooling and blue highlights runs with the analytic cooling function of \cite{Sarazin:1987kx}.  All figures adhere to the same scheme. 

\begin{figure*}[htp]
\centering
  	\subfigure[C]{
  	\includegraphics[width=0.325\textwidth]{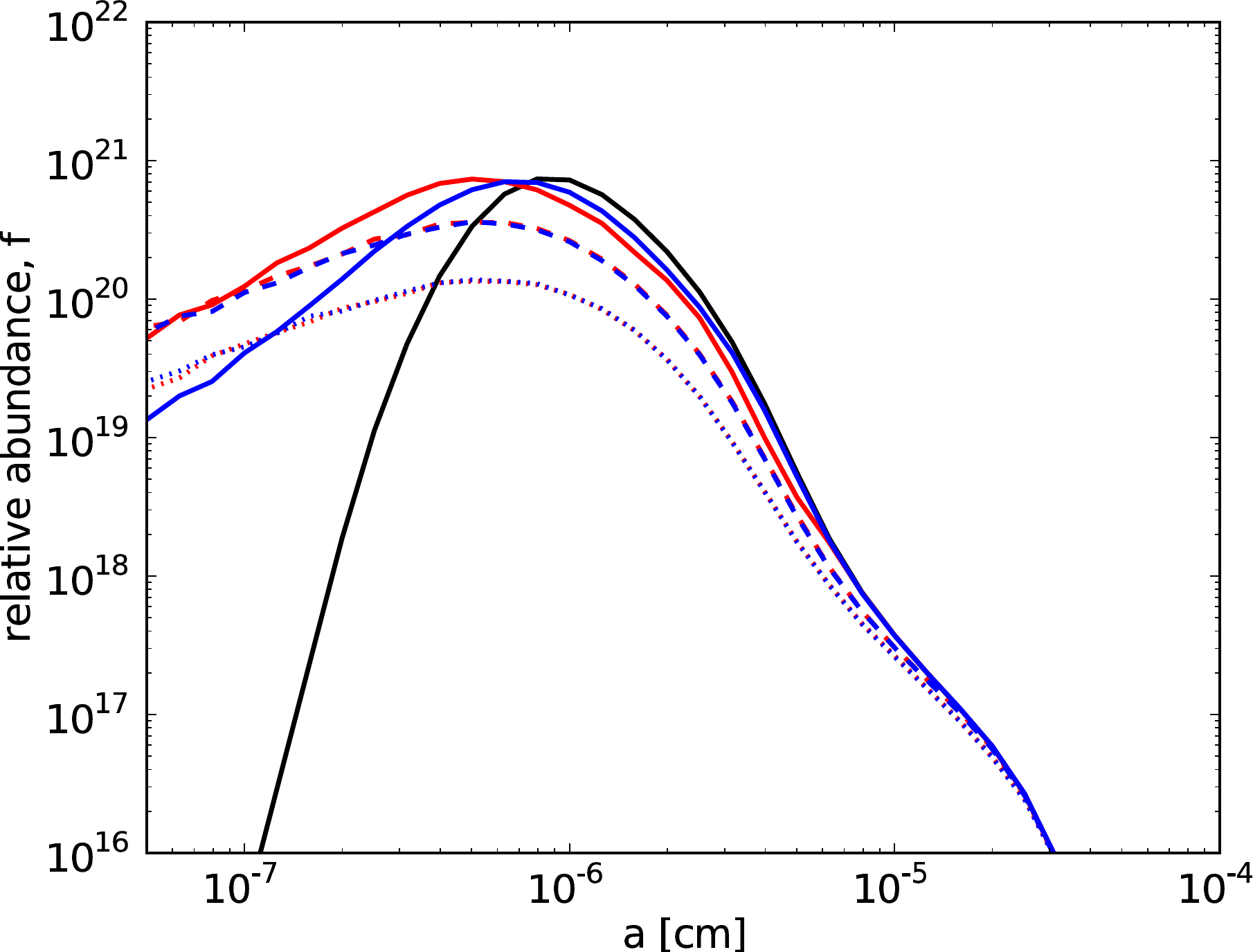}
  	}
	\hspace{-5.00mm}
	\subfigure[Silicates]{
  	\includegraphics[width=0.325\textwidth]{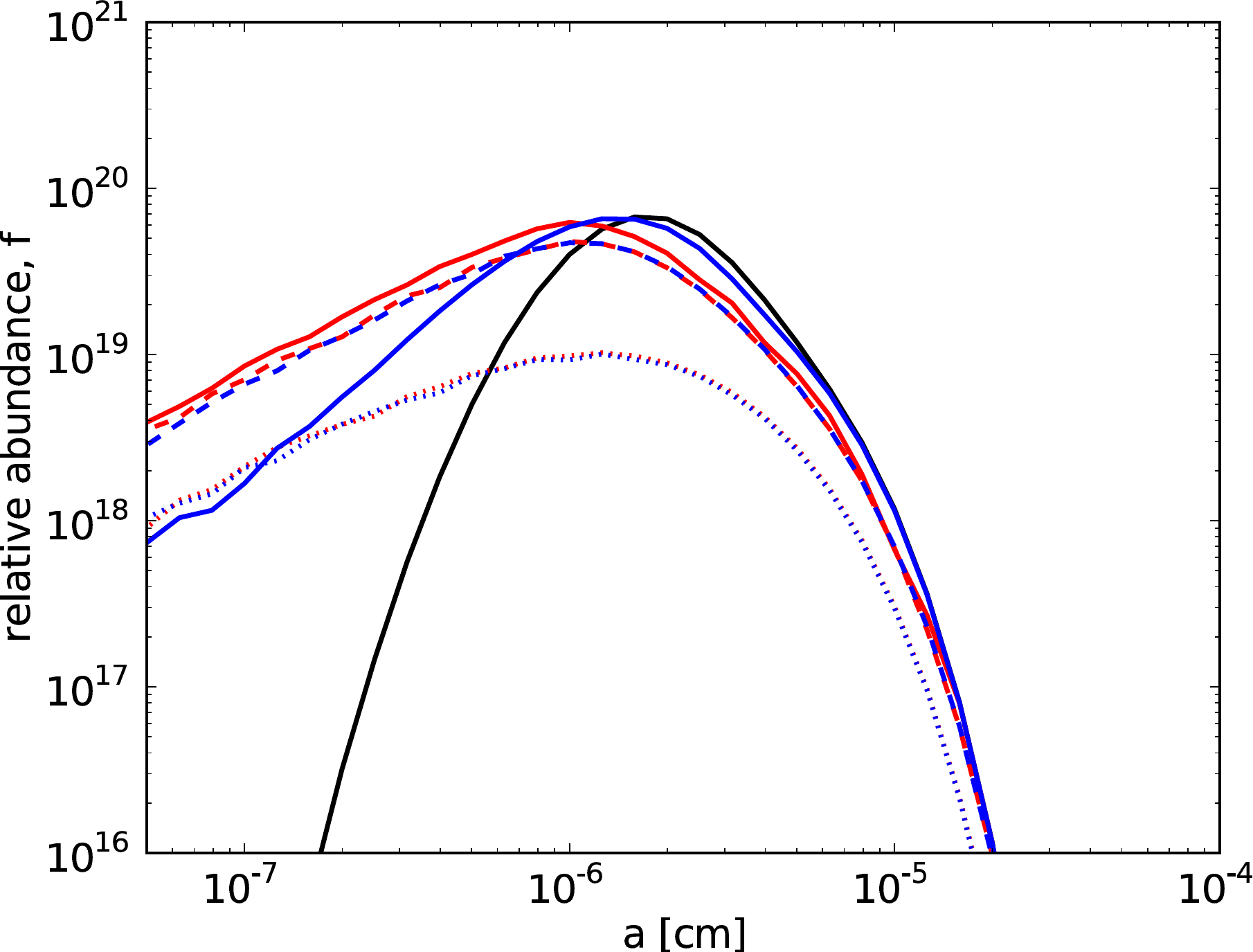}
  	}
	\hspace{-5.00mm}
	\subfigure[Fe]{
  	\includegraphics[width=0.325\textwidth]{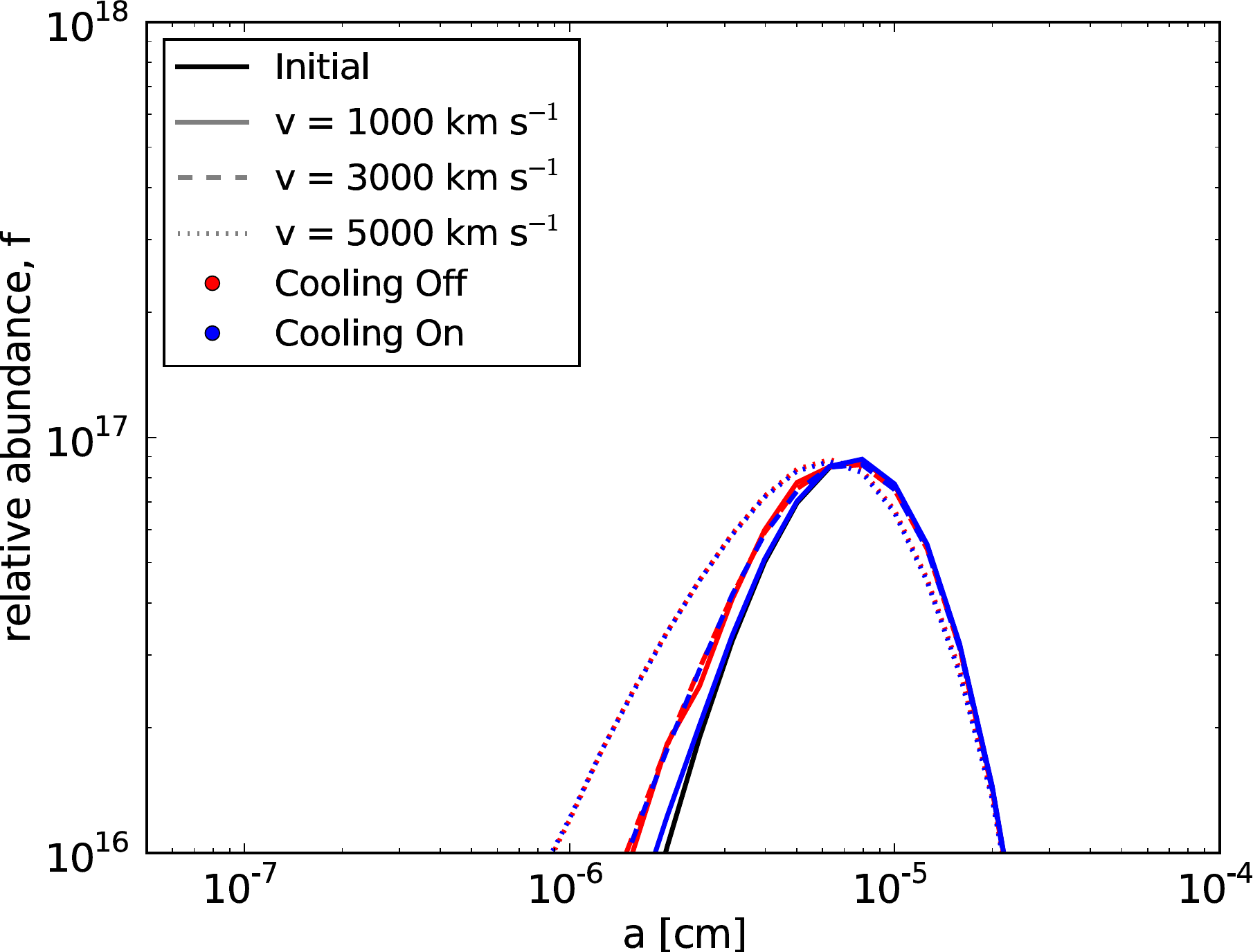}
  	}
  \caption{Same as Figure \ref{grainschi100_tiel}, but for over-density $\chi=1000$.}\label{grainschi1000_tiel}
\end{figure*}

In Figures \ref{grainschi100_tiel} and \ref{grainschi1000_tiel} we see the evolution in three grain species using the sputtering rates from \cite{Tielens:1994kx} for density contrasts $\chi=100$ and $\chi=1000$ respectively.  In both the low-density and high-density clouds, the higher shock velocities are more effective at reducing the size of large grains.  In addition, for $\chi=1000$, the increased density leads to even more dust destruction, with a severe drop off in the small grains as a function of shock velocity.  When we inspect the grain distributions as a function of cooling (red vs. blue), we see little difference between the simulations.  Since the differences are more apparent when looking at dust mass, we discuss these differences in the following section. 

\begin{figure*}[htp]
\centering
	\subfigure[Al$_2$O$_3$]{
  	\includegraphics[width=0.325\textwidth]{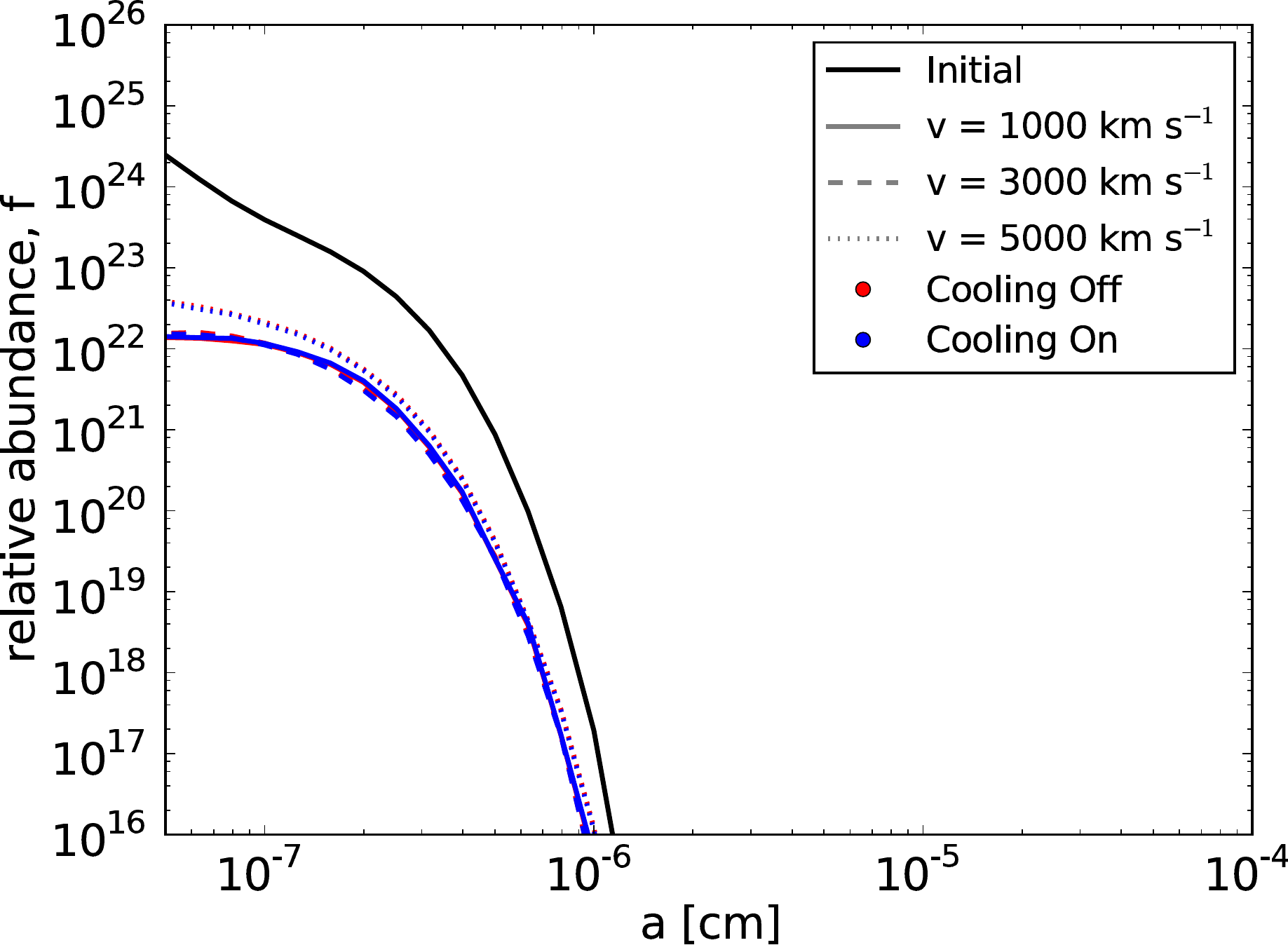}
  	}
	\hspace{-5.00mm}
  	\subfigure[C]{
  	\includegraphics[width=0.325\textwidth]{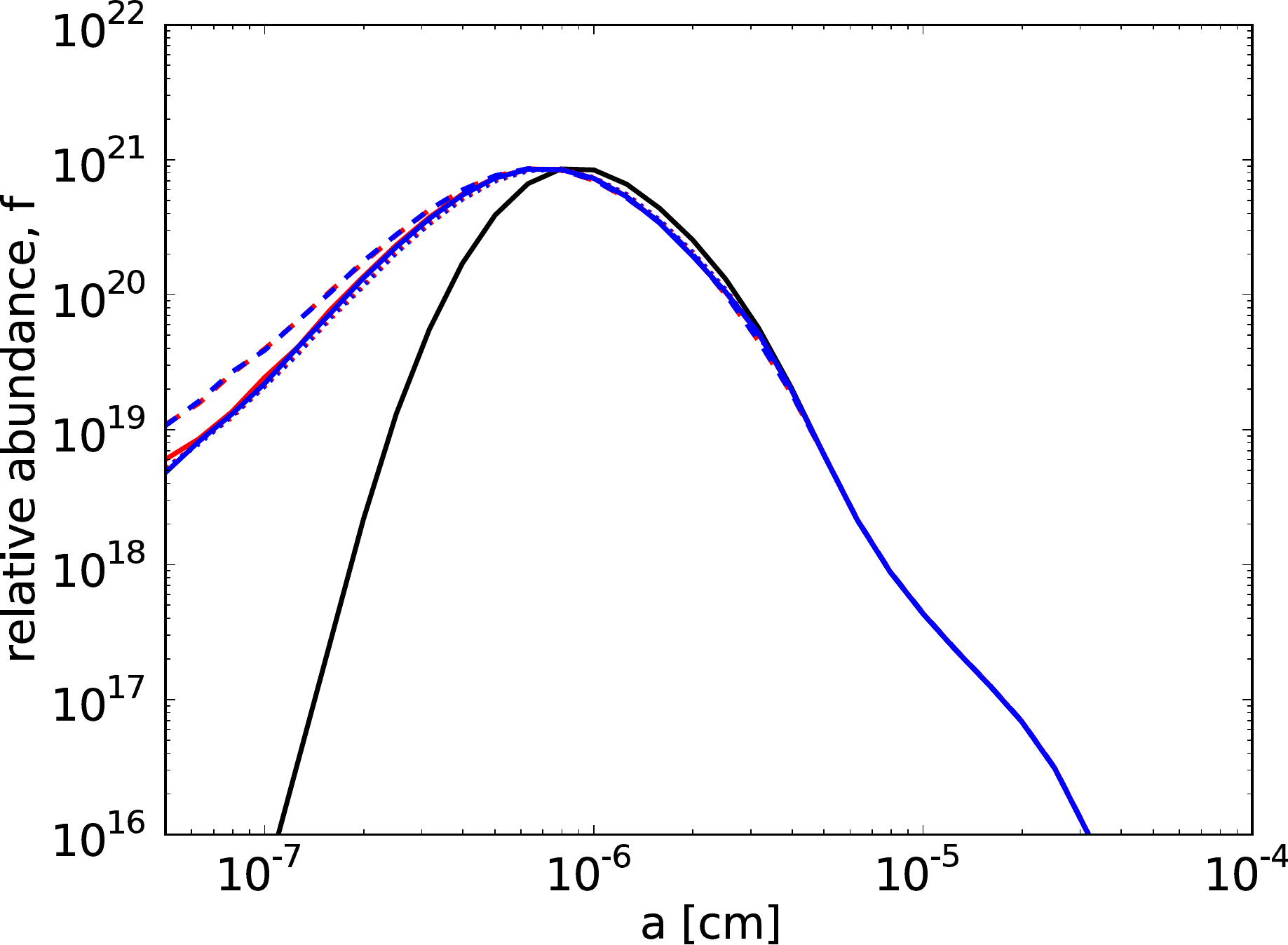}
  	}
	\hspace{-5.00mm}
	\subfigure[Mg$_2$SiO$_4$]{
  	\includegraphics[width=0.325\textwidth]{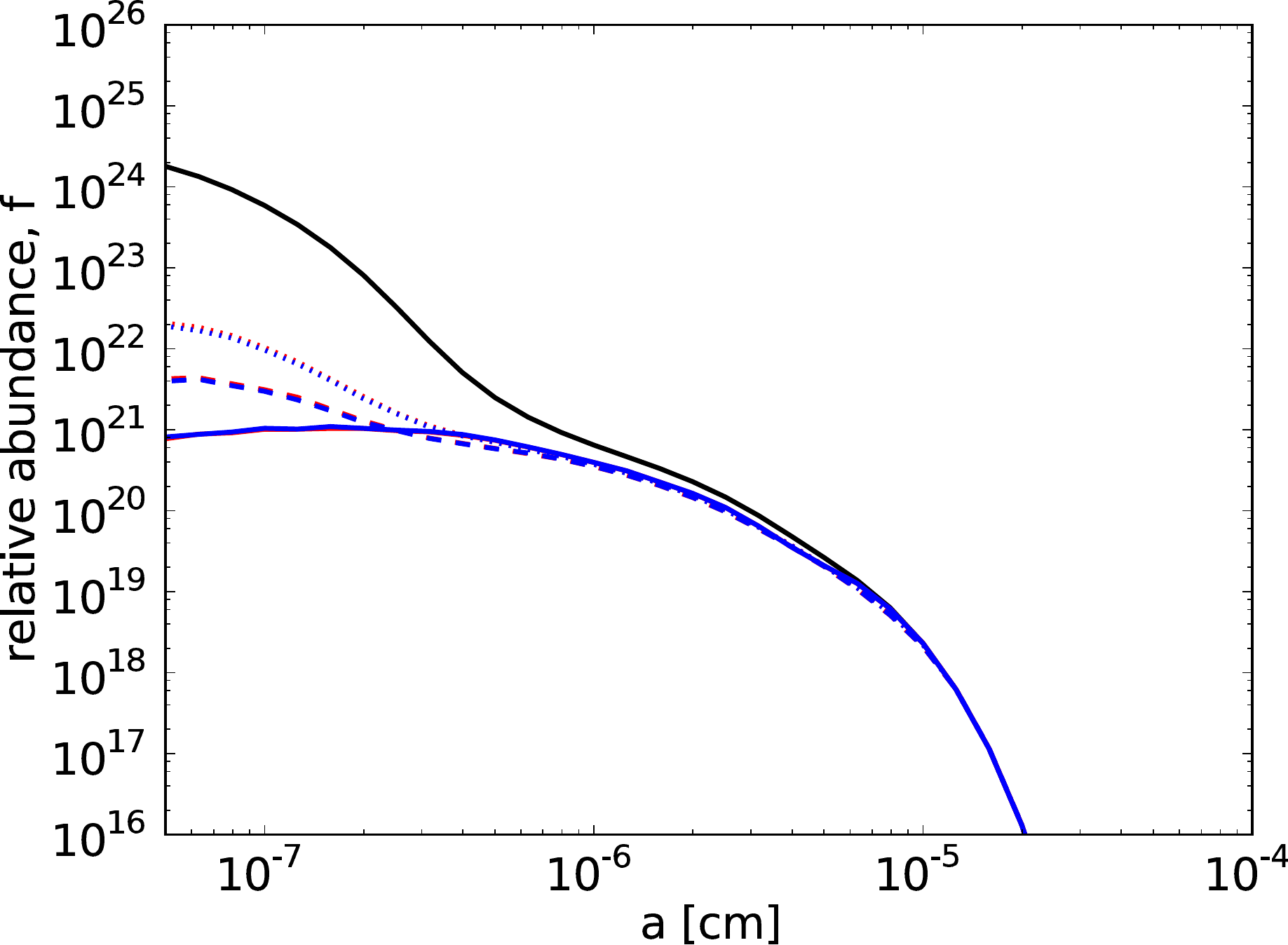}
	}
	\hspace{-5.00mm}
	\subfigure[SiO$_2$]{
  	\includegraphics[width=0.325\textwidth]{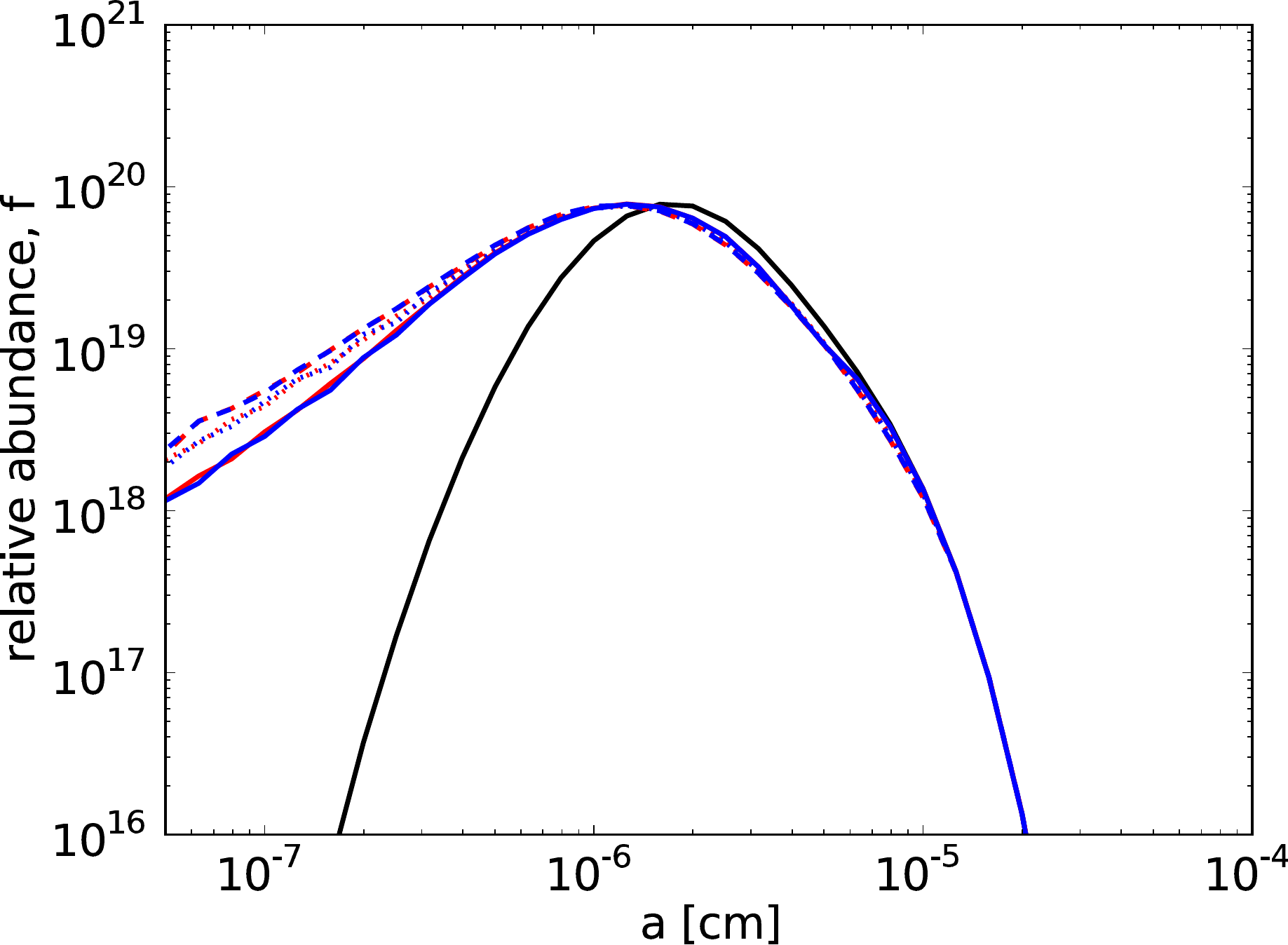}
  	}
	\hspace{-5.00mm}
	\subfigure[FeS]{
  	\includegraphics[width=0.325\textwidth]{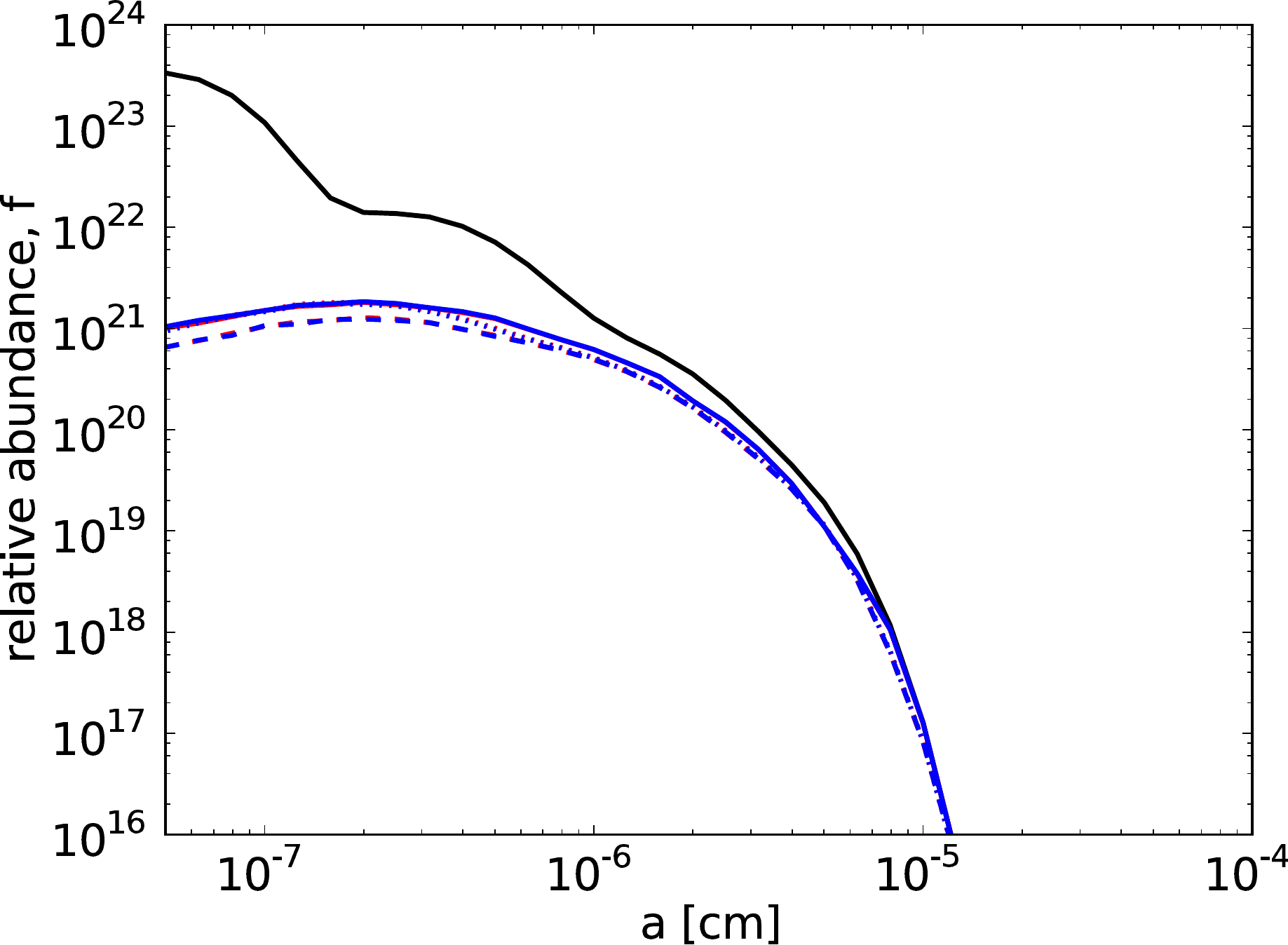}
  	}
	\hspace{-5.00mm}
	\subfigure[Fe]{
  	\includegraphics[width=0.325\textwidth]{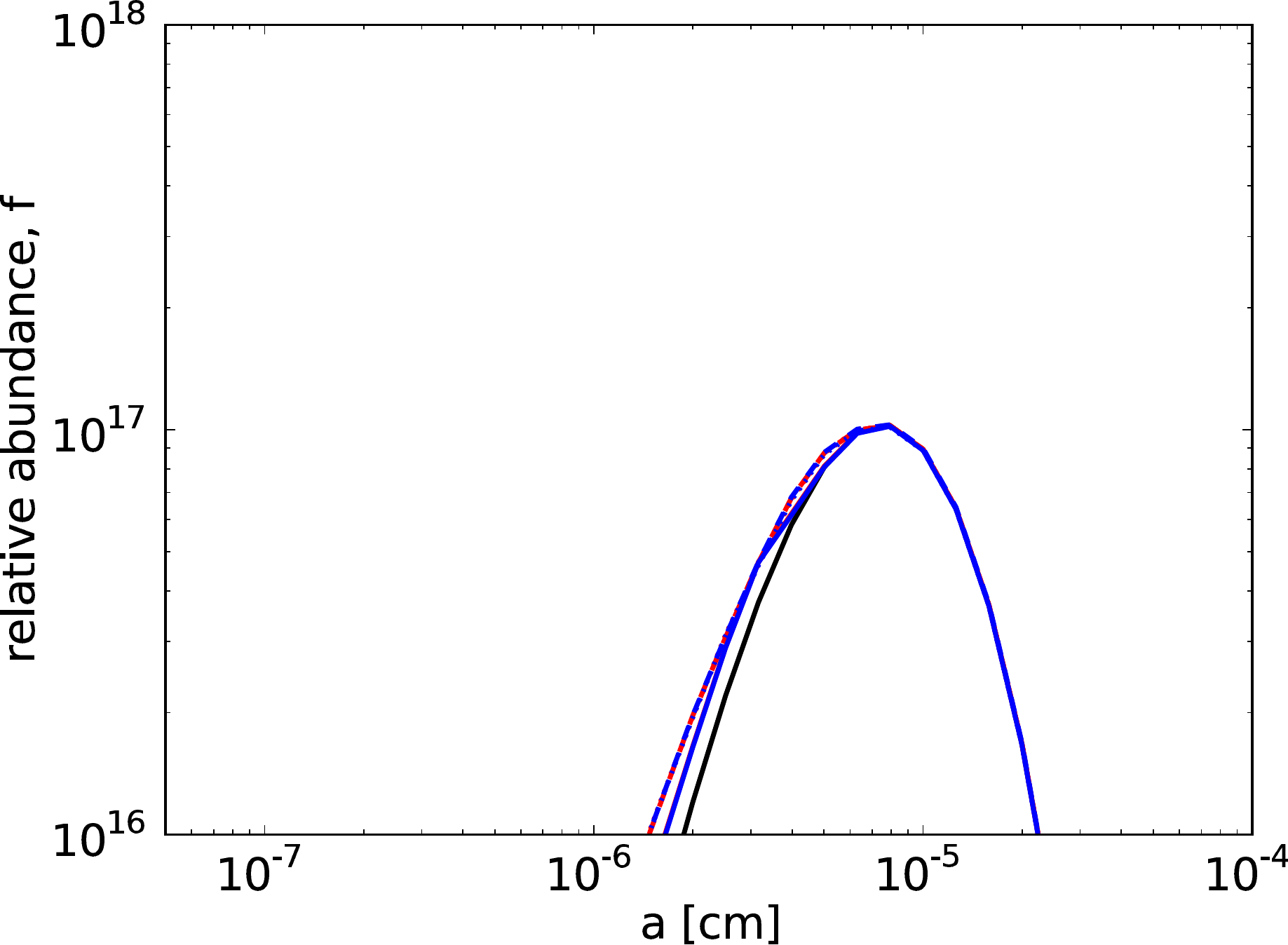}
  	}
  \caption{Grain radius distributions for the six of the grain species in \cite{Nozawa:2003pd} for simulations with $\chi =100$.  The line styles and colors have the same meaning as Figure \ref{grainschi100_tiel}.  The grains in this figure were sputtered using the $Z=Z\subscript{\odot}$ rates from \cite{Nozawa:2006ve}.}\label{grainschi100_noz}
\end{figure*}

\begin{figure*}[htp]
\centering
	\subfigure[Al$_2$O$_3$]{
  	\includegraphics[width=0.325\textwidth]{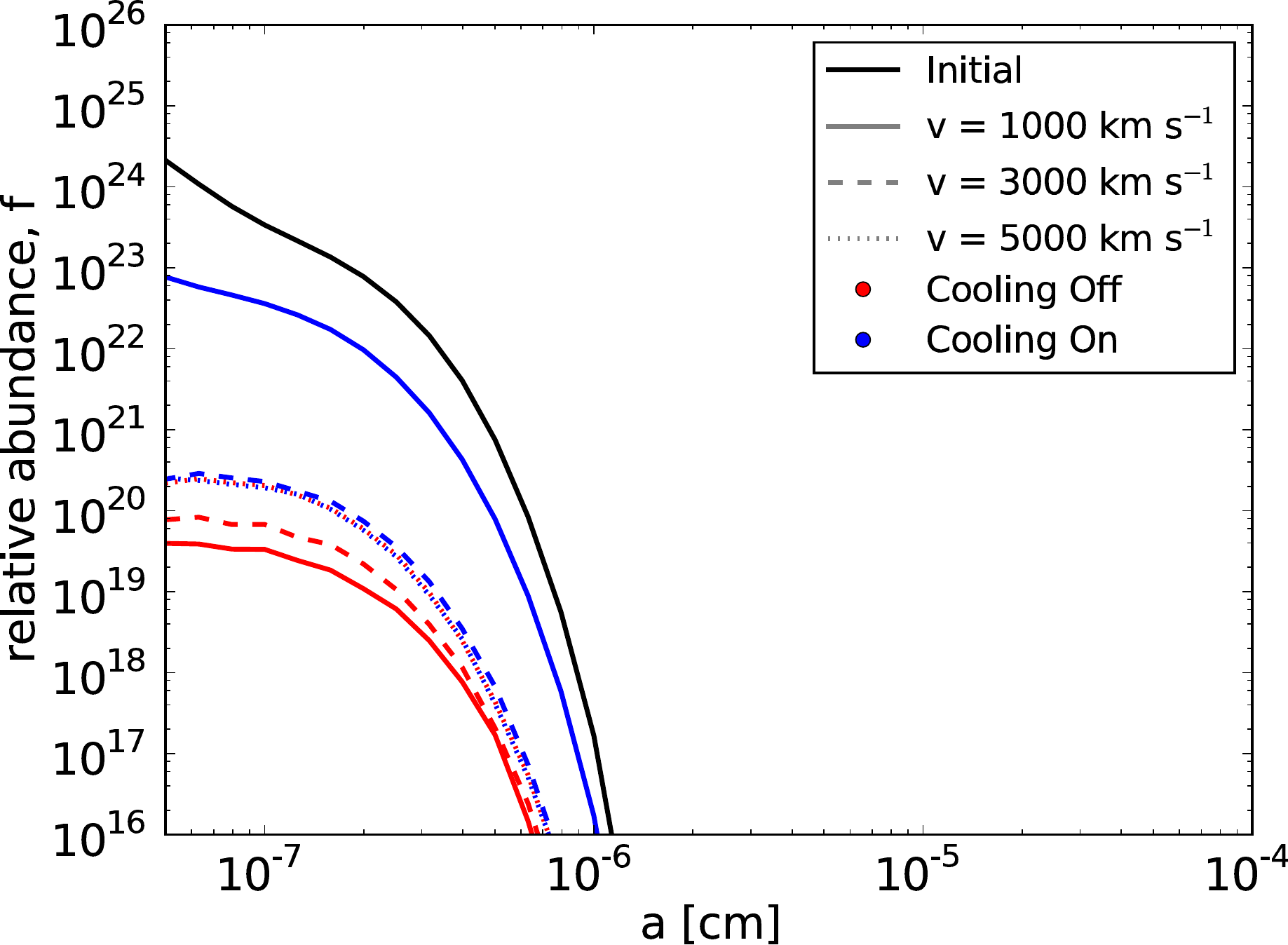}
  	}
	\hspace{-5.00mm}
  	\subfigure[C]{
  	\includegraphics[width=0.325\textwidth]{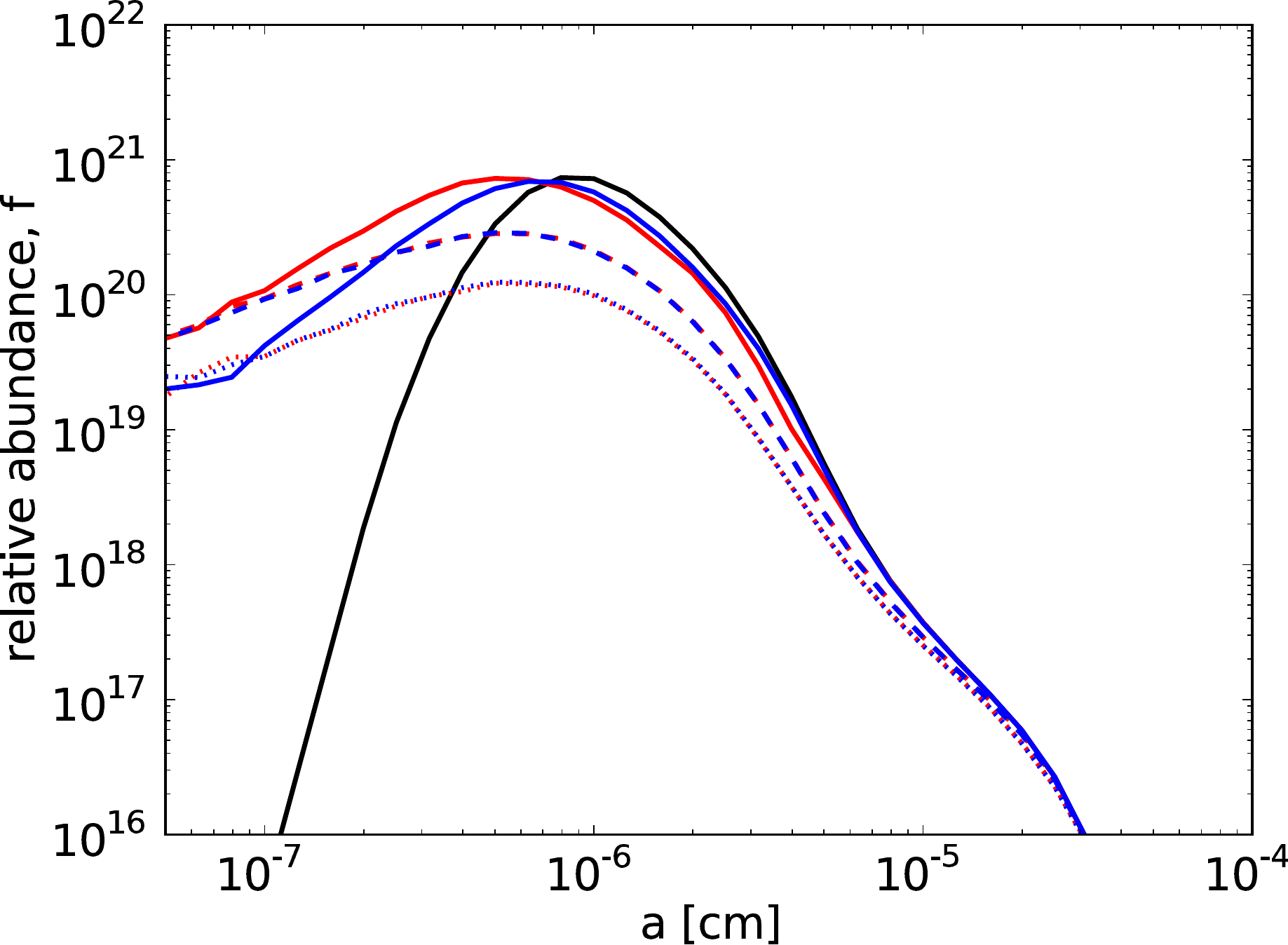}
  	}
	\hspace{-5.00mm}
	\subfigure[Mg$_2$SiO$_4$]{
  	\includegraphics[width=0.325\textwidth]{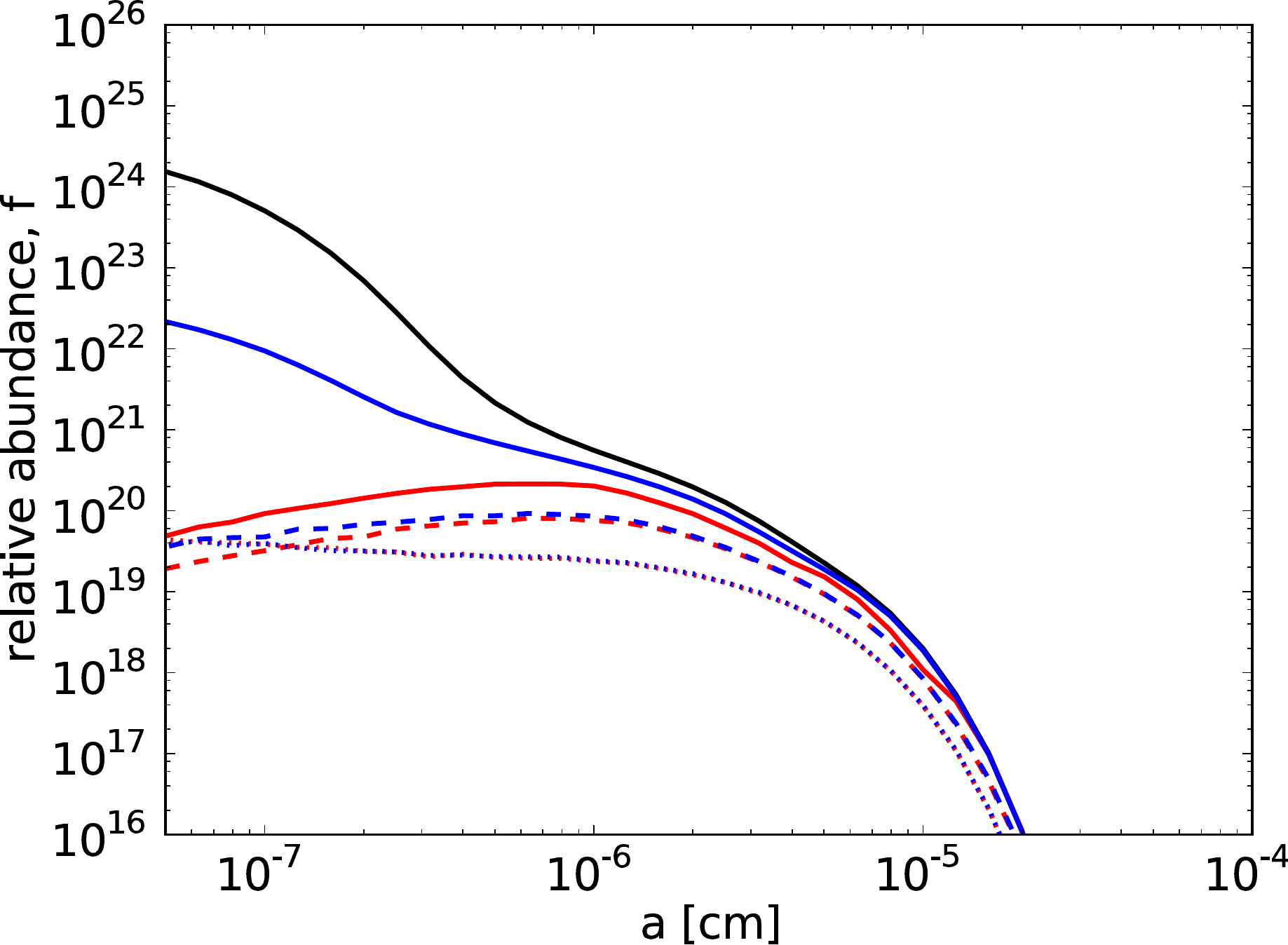}
	}
	\hspace{-5.00mm}
	\subfigure[SiO$_2$]{
  	\includegraphics[width=0.325\textwidth]{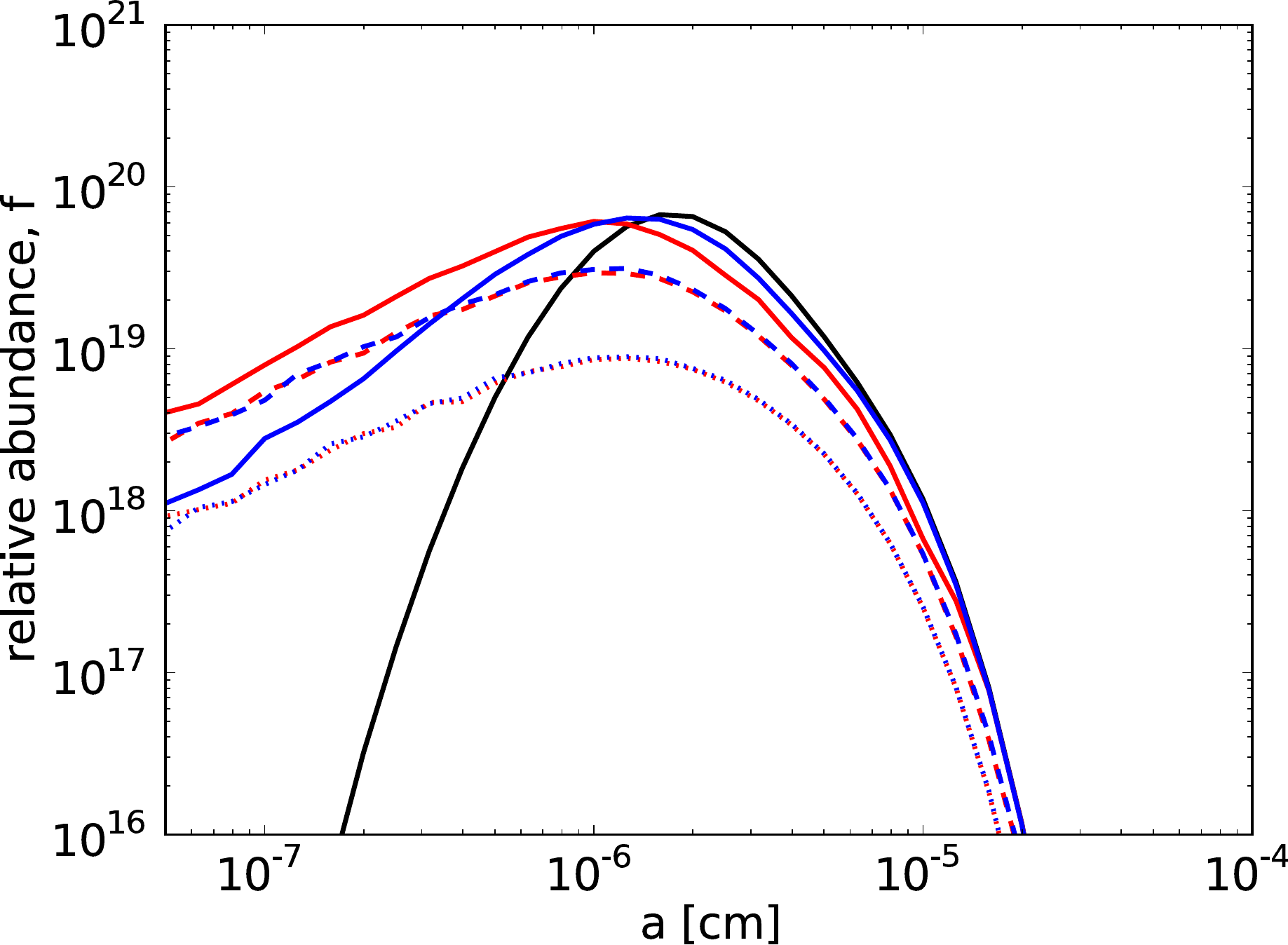}
  	}
	\hspace{-5.00mm}
	\subfigure[FeS]{
  	\includegraphics[width=0.325\textwidth]{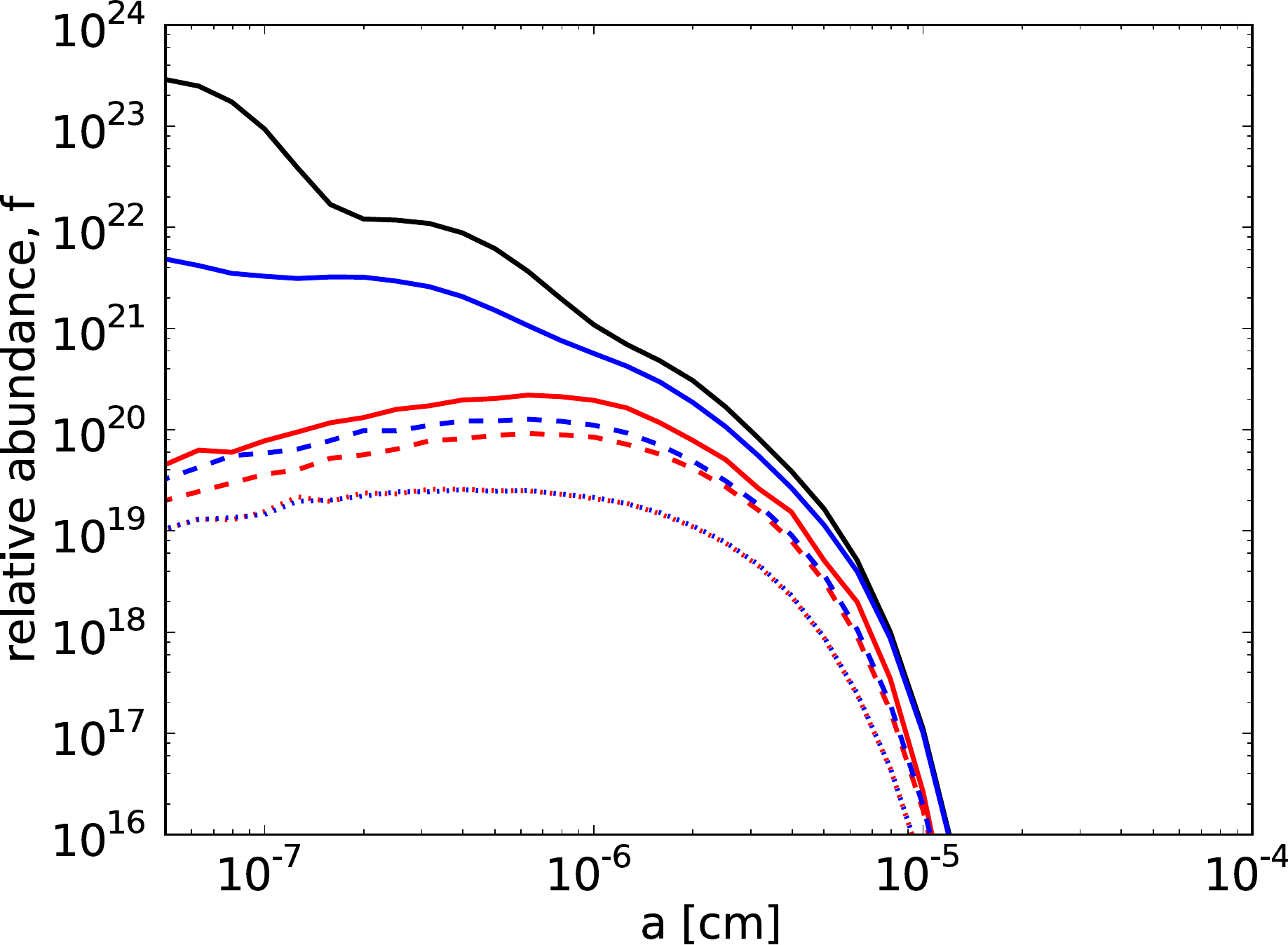}
  	}
	\hspace{-5.00mm}
	\subfigure[Fe]{
  	\includegraphics[width=0.325\textwidth]{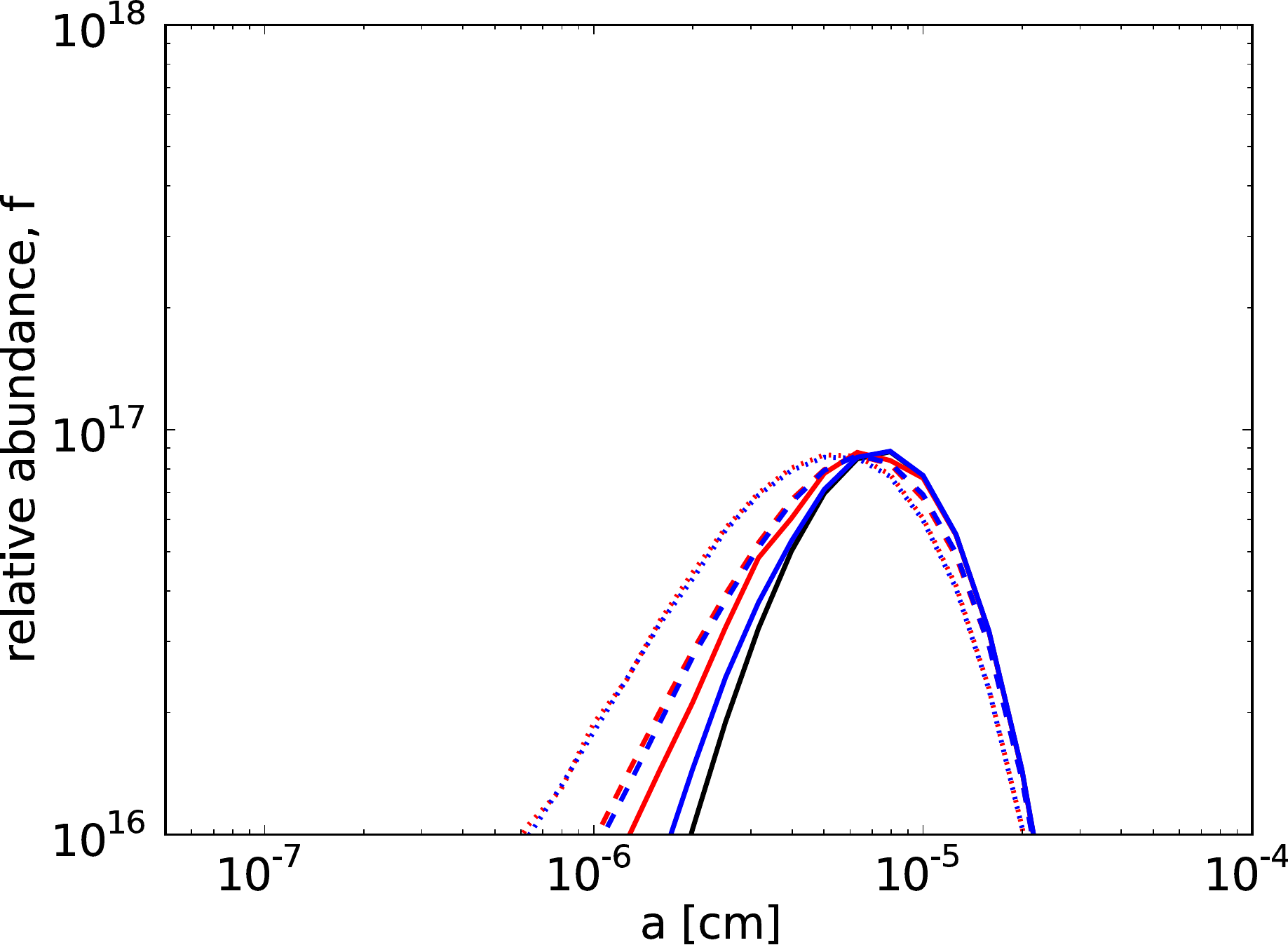}
  	}
  \caption{Same as Figure \ref{grainschi100_noz}, but for over-density $\chi=1000$.}\label{grainschi1000_noz}
\end{figure*}

We show the results of sputtering using the rates from \cite{Nozawa:2006ve} for over-densities of 100 and 100 in Figures \ref{grainschi100_noz} and \ref{grainschi1000_noz} respectively.  In the case of $\chi=100$, the number of small grains sputtered out of existence does not show a clear correlation with shock velocity.  In some cases, a slower shock destroys more grains than a faster shock.  This is likely a result of the cloud being shredded more rapidly in the high-velocity cases, which reduces the density of the cloud material and shuts off sputtering.  There appears to be no significant dependence on whether or not cooling was employed in the simulation. However, we predict that, with higher metal abundances, the differences between cooling and non-cooling simulations should become more pronounced.  For the simulations with $\chi=1000$, grain sputtering depends strongly on shock velocity, as all grains sizes are sputtered more heavily by higher velocity shocks.  Again, we do not see a clear distinction between cooling and non-cooling simulations, even in higher density regimes.  There does seem to be a large discrepancy for the slowest shock velocity, but this stems from the fact that the cooling simulation for that shock velocity did not make it to ten cloud-crushing times.  The differences  between simulations with and without cooling can better be seen in Section \ref{dustevol}, when we look at evolution in dust mass.

\subsection{Dust Mass Evolution}\label{dustevol}
Using the sputtered grain distributions, we convert them into relative dust masses using the method described in Section \ref{tracerparticles}.  In the following figures, we continue with the same color and line style convention, but instead of radius distributions we consider the fractional dust mass of each grain species as a function of cloud-crushing times.  In these plots, any lines that end before 10 cloud-crushing times are terminated because more than 5\% of the original dust mass contained in the particles was lost due to particles that crossed the simulation boundaries.  In addition, for the lower density cloud, effectively zero sputtering of iron grains occurred, and we have omitted those plots.

\begin{figure*}[htp]
\centering
  	\subfigure[C]{
  	\includegraphics[width=0.325\textwidth]{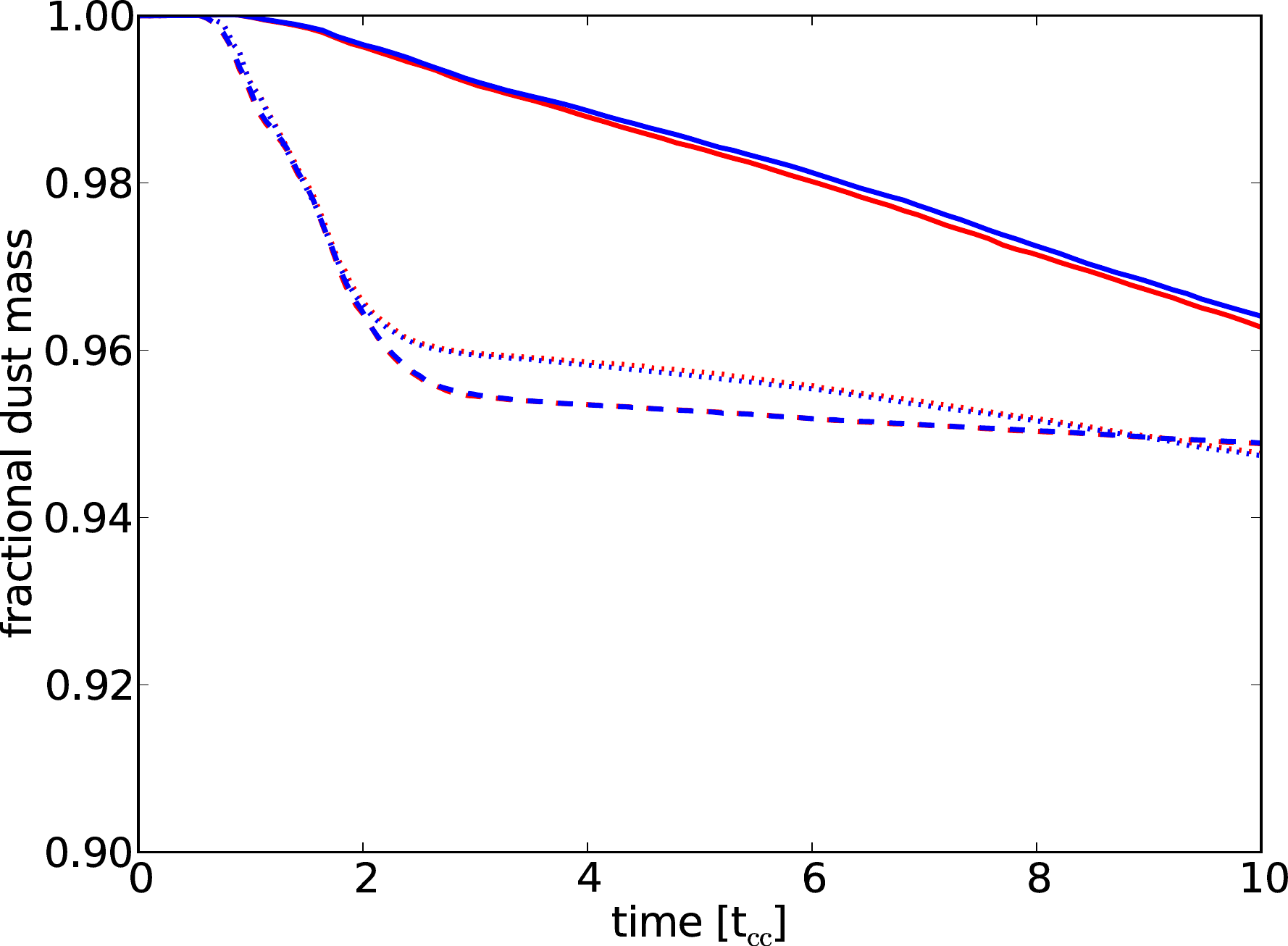}
  	}
	\hspace{-5.00mm}
	\subfigure[Silicates]{
  	\includegraphics[width=0.325\textwidth]{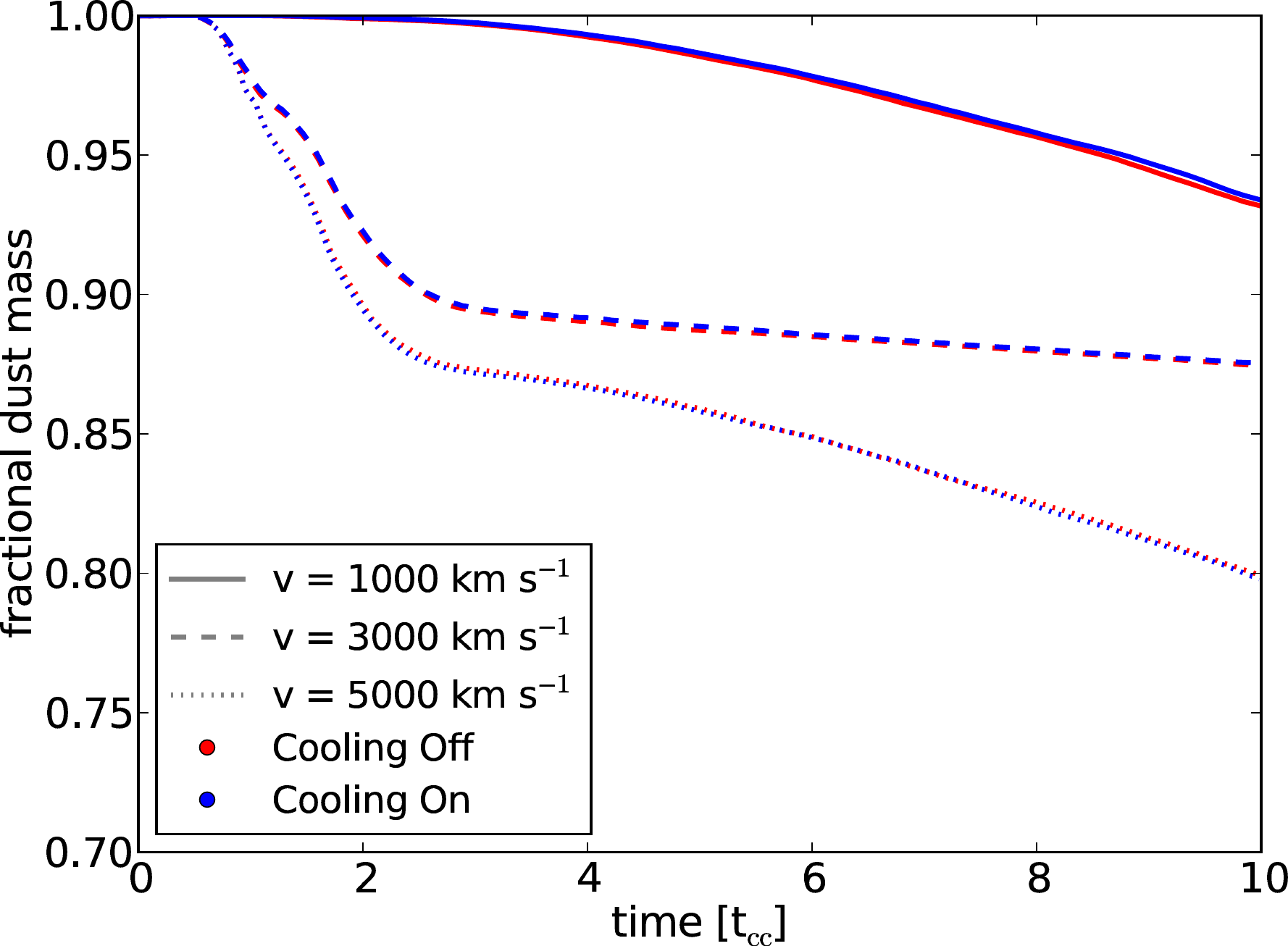}
  	}
  \caption{Dust mass evolution for two of the grain species in \cite{Tielens:1994kx} for simulations with $\chi =100$.  Colors and lines styles are the same as Figure \ref{grainschi100_tiel}. The grains in this figure were sputtered using the rates from \cite{Tielens:1994kx}.} \label{dustmasschi100_tiel} 
\end{figure*}

\begin{figure*}[htp]
\centering
  	\subfigure[C]{
  	\includegraphics[width=0.325\textwidth]{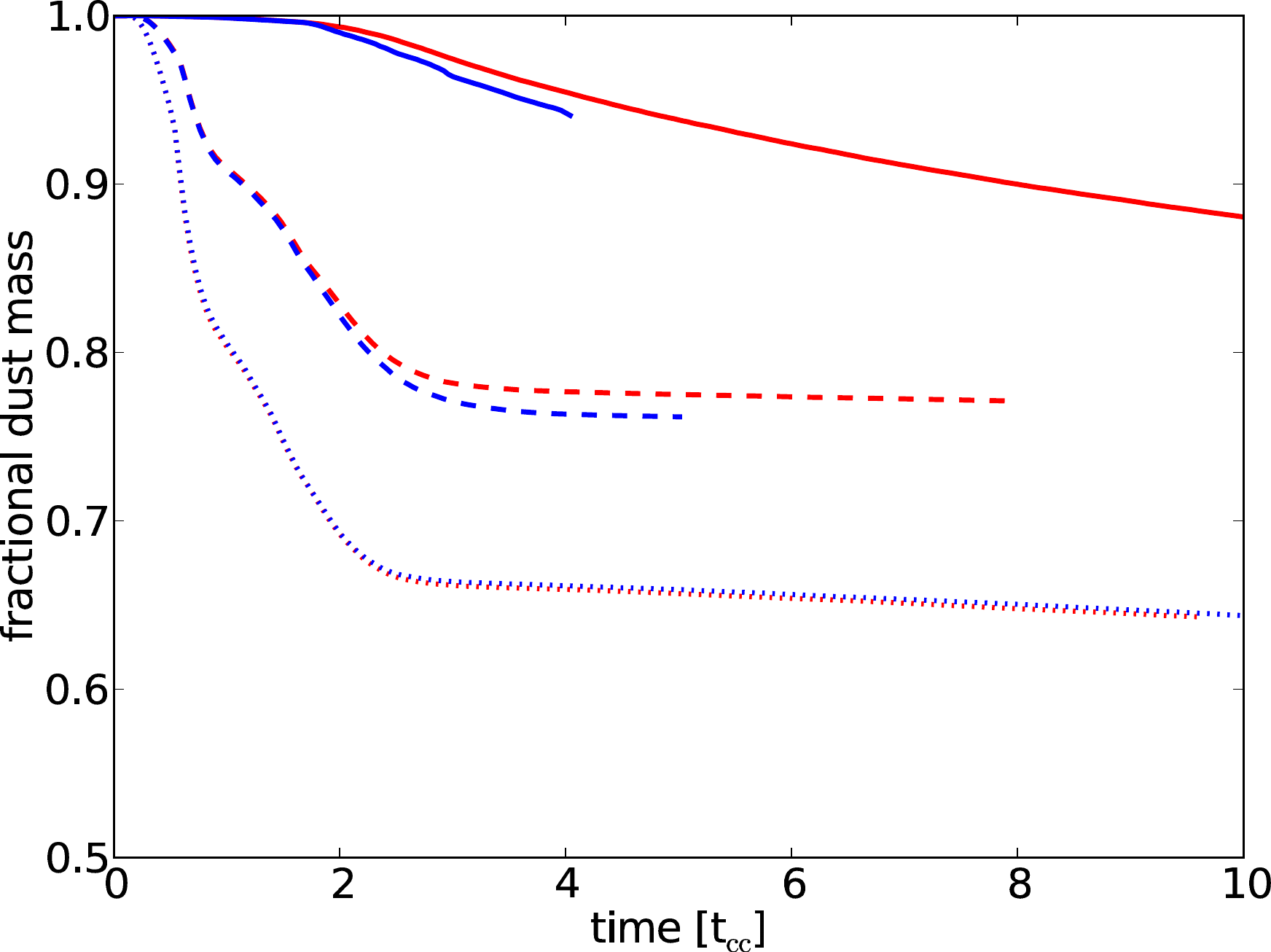}
  	}
	\hspace{-5.00mm}
	\subfigure[Silicates]{
  	\includegraphics[width=0.325\textwidth]{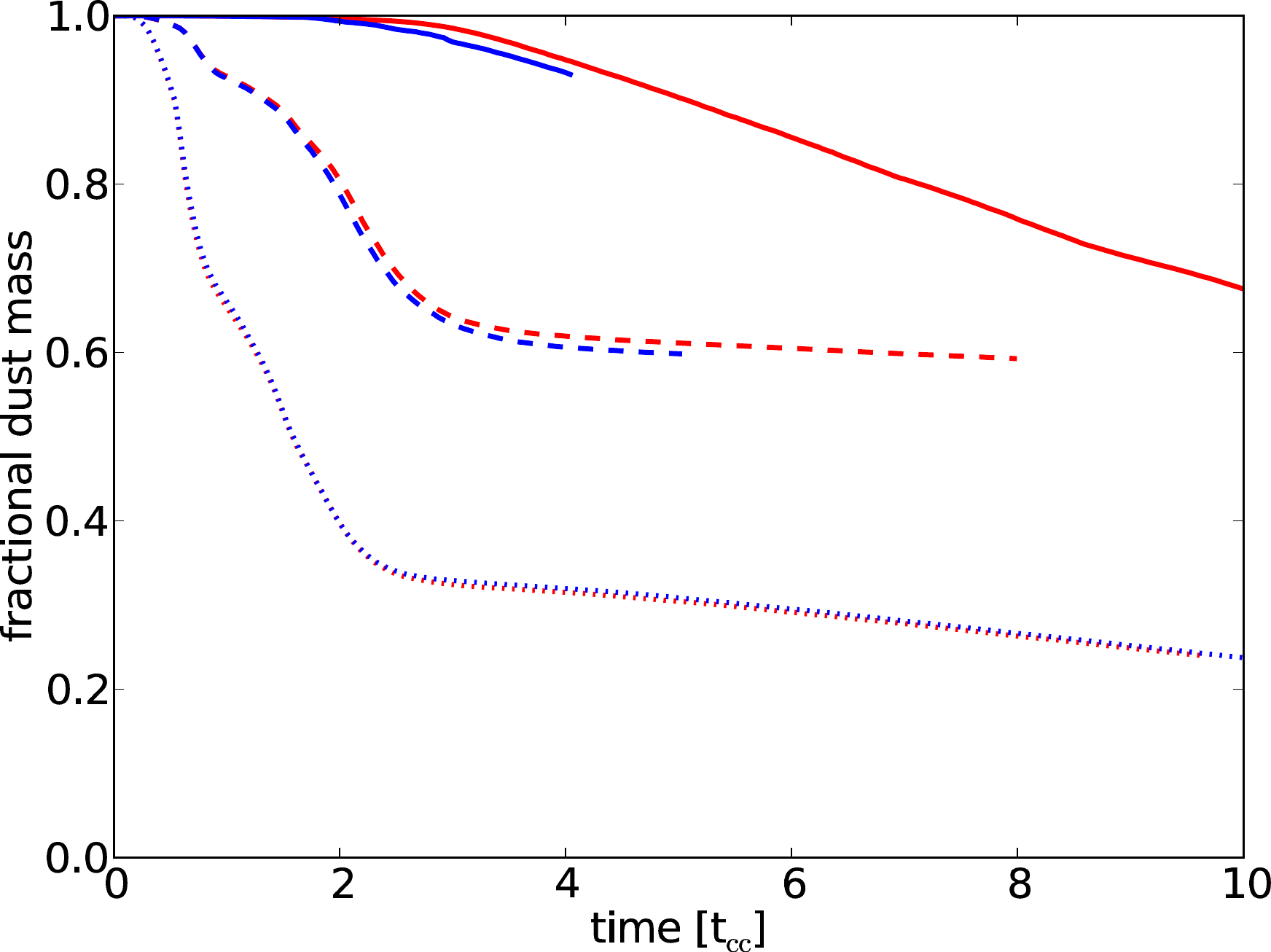}
  	}
	\hspace{-5.00mm}
	\subfigure[Fe]{
  	\includegraphics[width=0.325\textwidth]{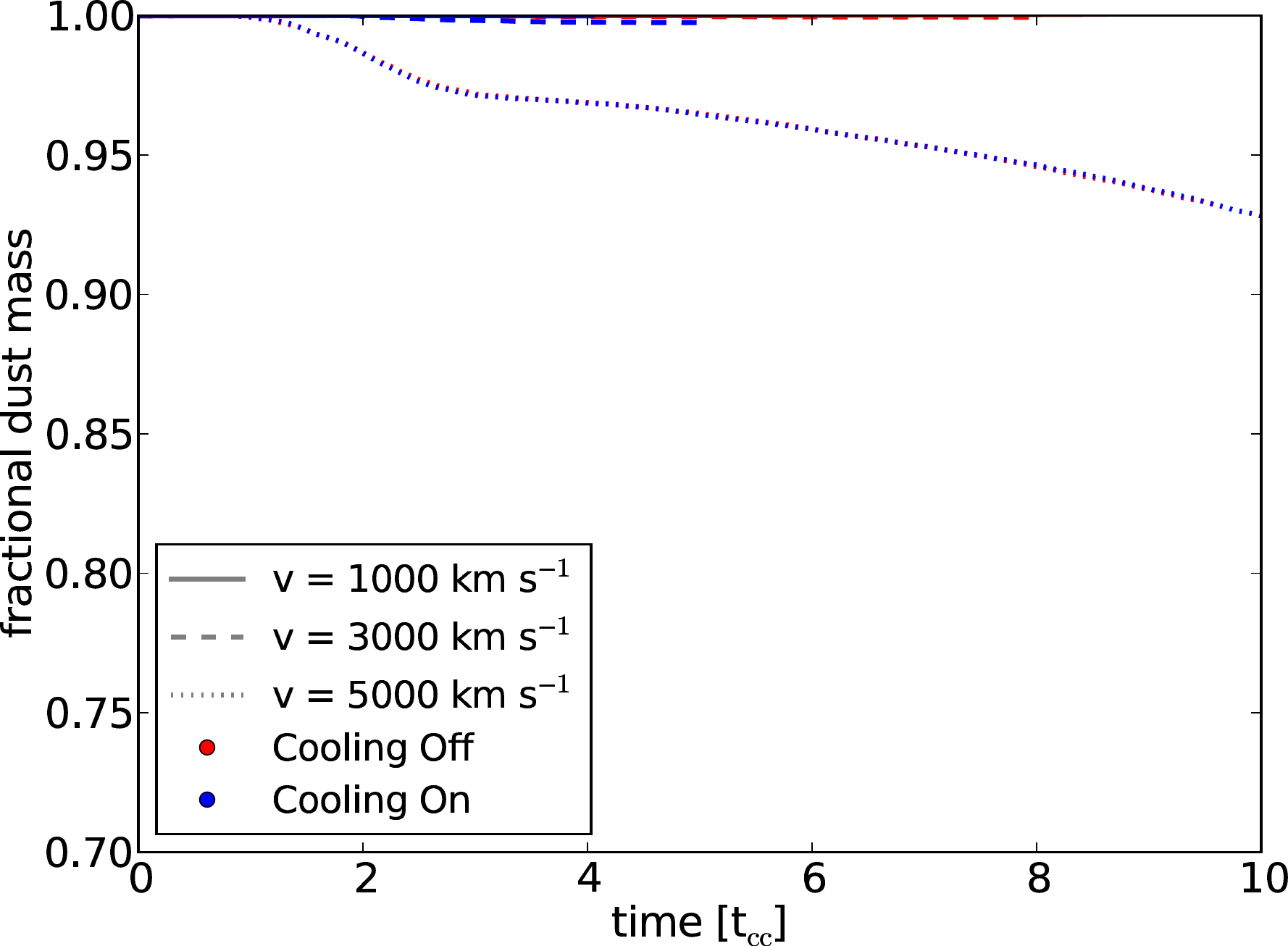}
  	}
  \caption{Same as Figure \ref{dustmasschi100_tiel}, but for over-density $\chi=1000$.  In addition, we include the evolution of Fe grains.}\label{dustmasschi1000_tiel}
\end{figure*}

We again look first at the results when we use the sputtering rates of \cite{Tielens:1994kx}.  Following the previous order, Figure \ref{dustmasschi100_tiel} shows the mass evolution for $\chi=100$, and Figure \ref{dustmasschi1000_tiel} shows the same for $\chi=1000$.  For the low over-density cloud, with the exception of Fe grains, which were negligibly sputtered, we find the scatter in percentage of dust remaining at the end of the simulation is rather small.  However, the path that each simulation takes to get to its final dust mass varies based on the shock velocity.  The high-velocity shock runs show a rapid drop in dust mass and then dust destruction slows considerably, while the dust mass in the low velocity shock runs gradually falls to its final level.  In the higher over-density cloud, the dependence on shock velocity is much more pronounced.  Higher shock velocities lead to greater dust destruction -- as much as 50\% additional destruction for certain grain species.  Comparing the red and blue lines in the cooling versus non-cooling simulations, we only see differences in the higher density cloud, after the initial drop in dust mass.  At later times, we detect a slight splitting in the dust mass evolution tracks in which the simulations with cooling turned on appear to destroy a small fraction of additional dust.  Our explanation for the diverging behavior is that if the medium is capable of cooling, it will slightly drop the pressure in the expanding cloud and keep the medium in which dust is embedded at a higher density.  Since the cloud has been raised to high shock temperatures at late times, where the sputtering rates are relatively constant with temperature, differences in density will play the dominant role in controlling sputtering.

\begin{figure*}[htp]
\centering
	\subfigure[Al$_2$O$_3$]{
  	\includegraphics[width=0.325\textwidth]{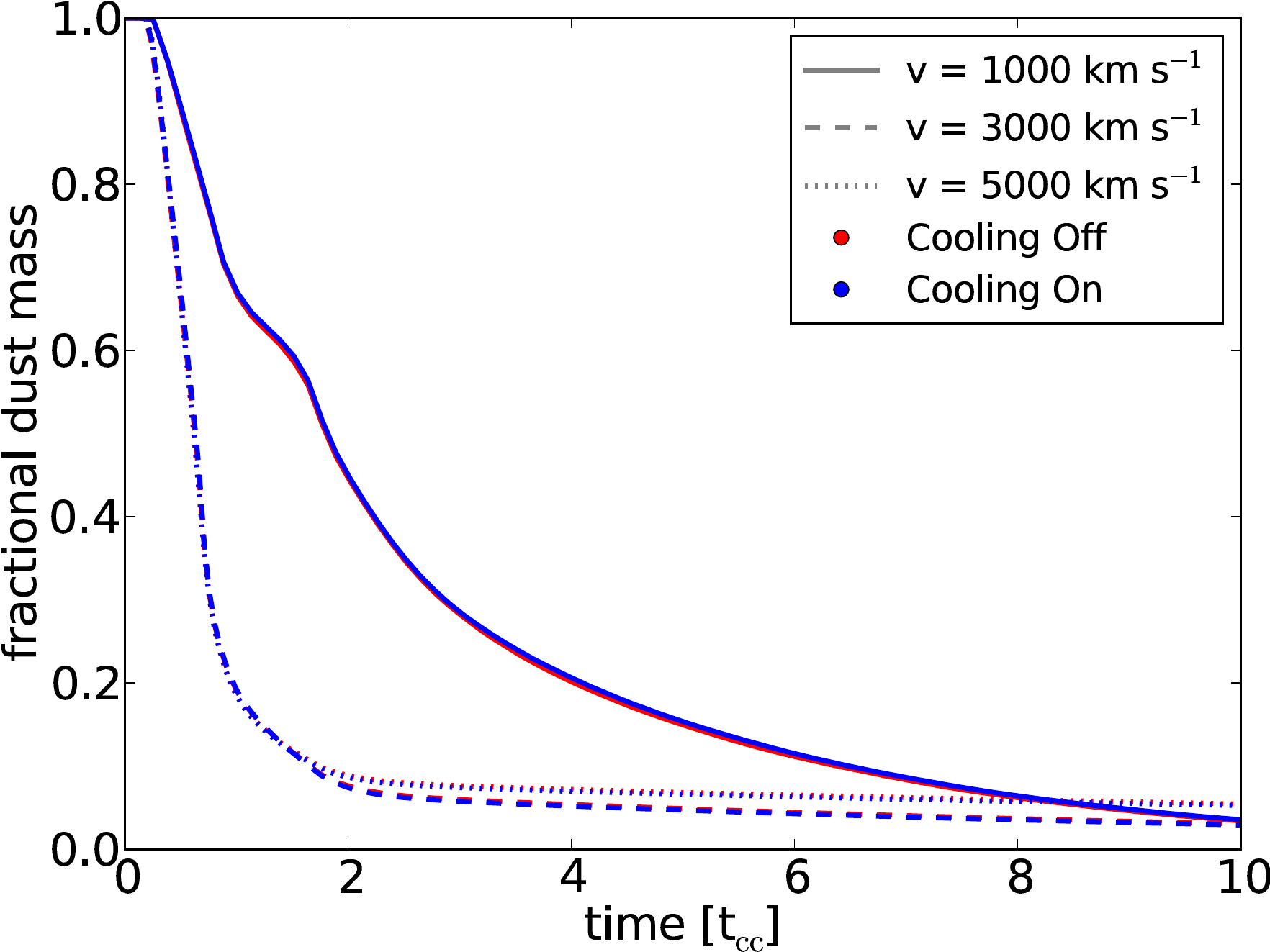}
  	}
	\hspace{-5.00mm}
  	\subfigure[C]{
  	\includegraphics[width=0.325\textwidth]{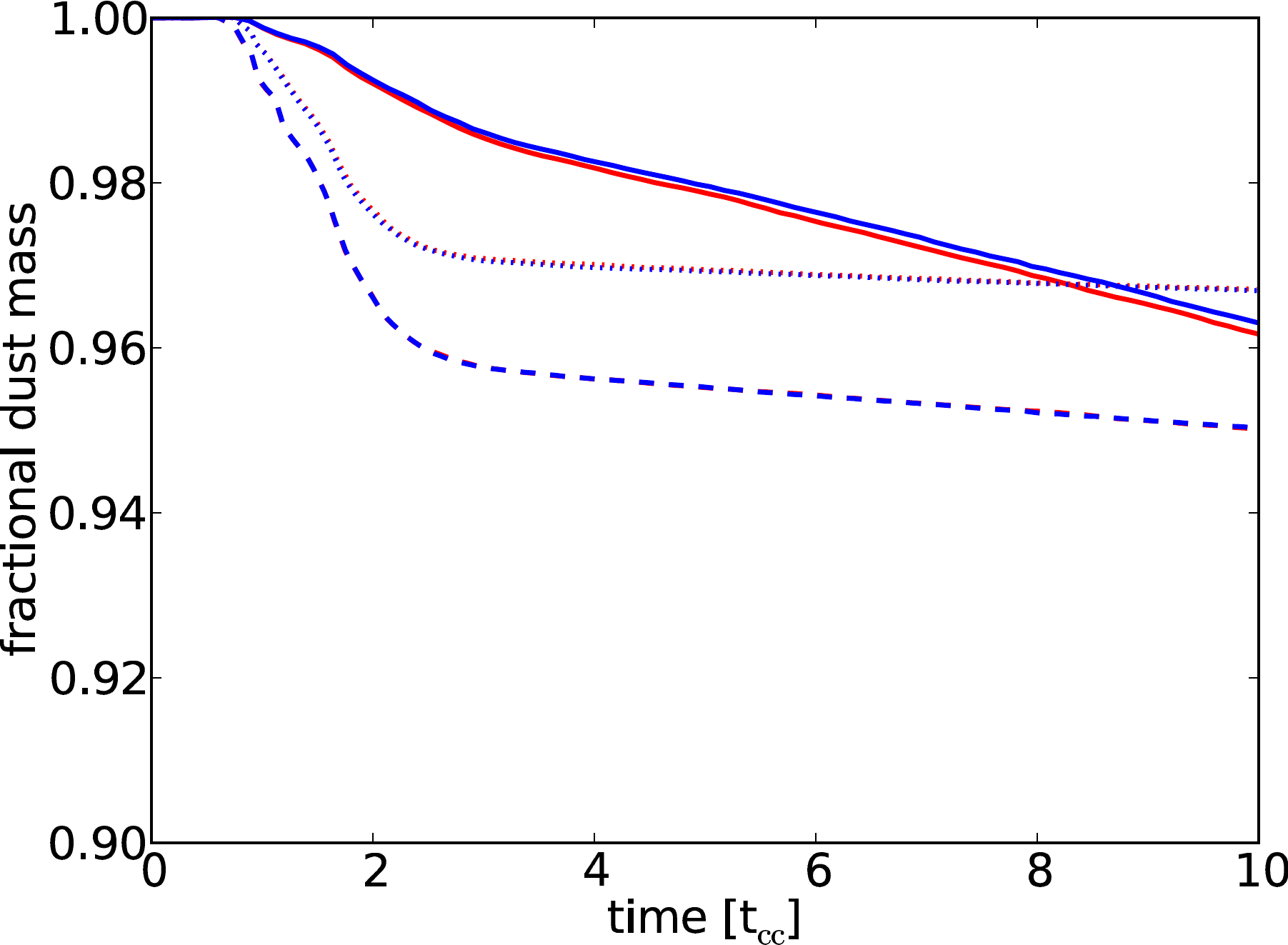}
  	}
	\hspace{-5.00mm}
	\subfigure[Mg$_2$SiO$_4$]{
  	\includegraphics[width=0.325\textwidth]{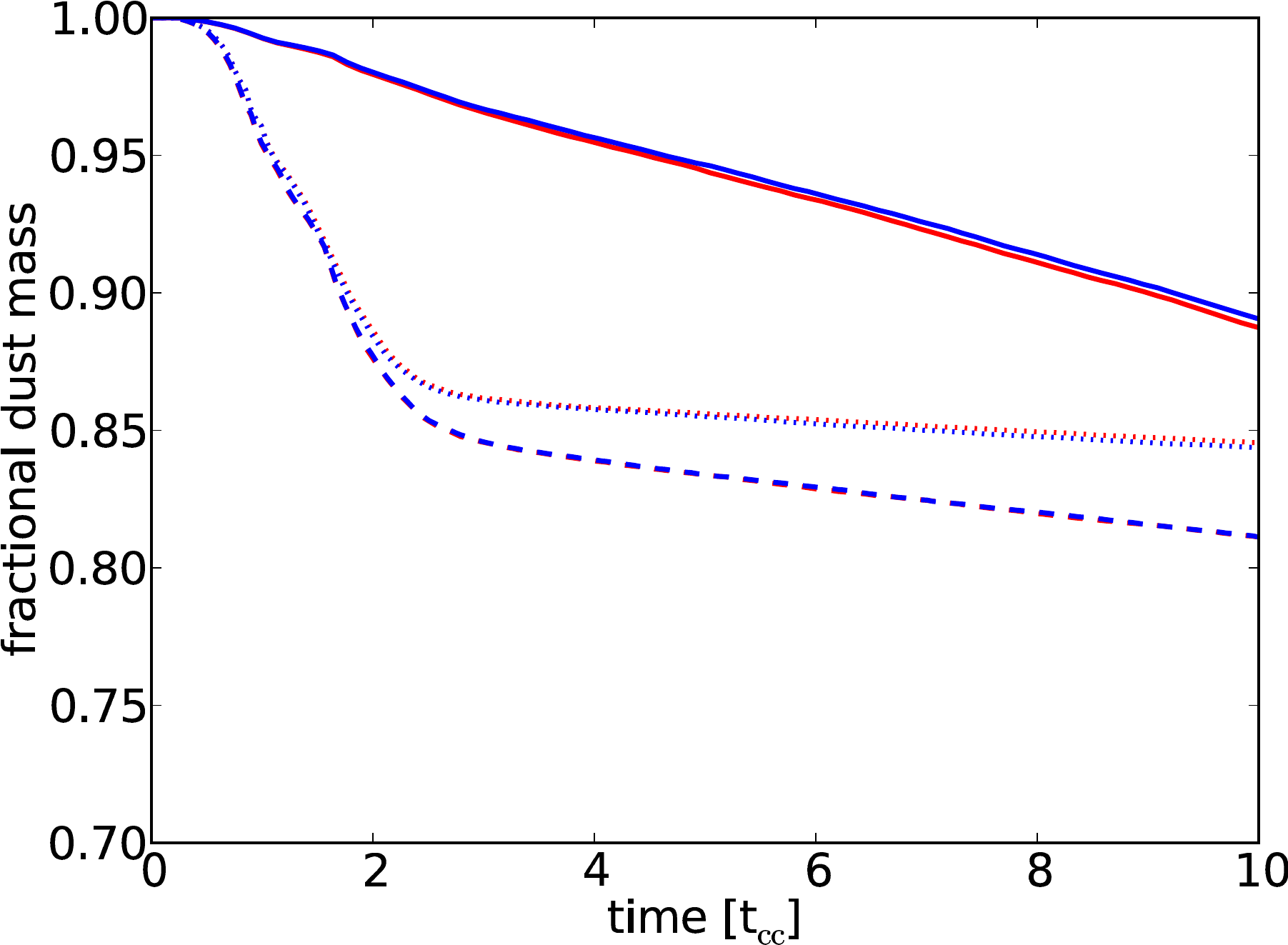}
	}
	\hspace{-5.00mm}
	\subfigure[SiO$_2$]{
  	\includegraphics[width=0.325\textwidth]{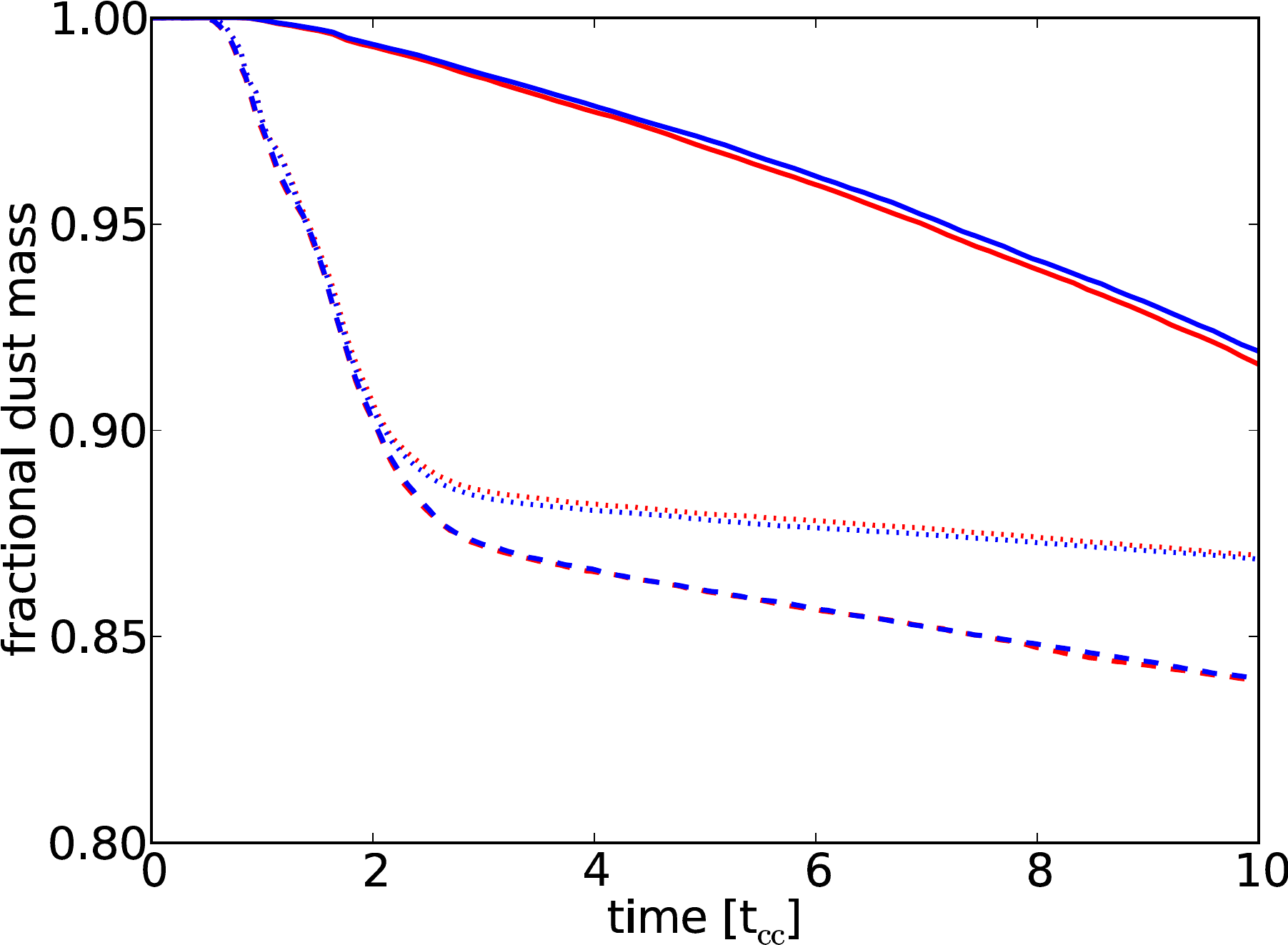}
  	}
	\hspace{-5.00mm}
	\subfigure[FeS]{
  	\includegraphics[width=0.325\textwidth]{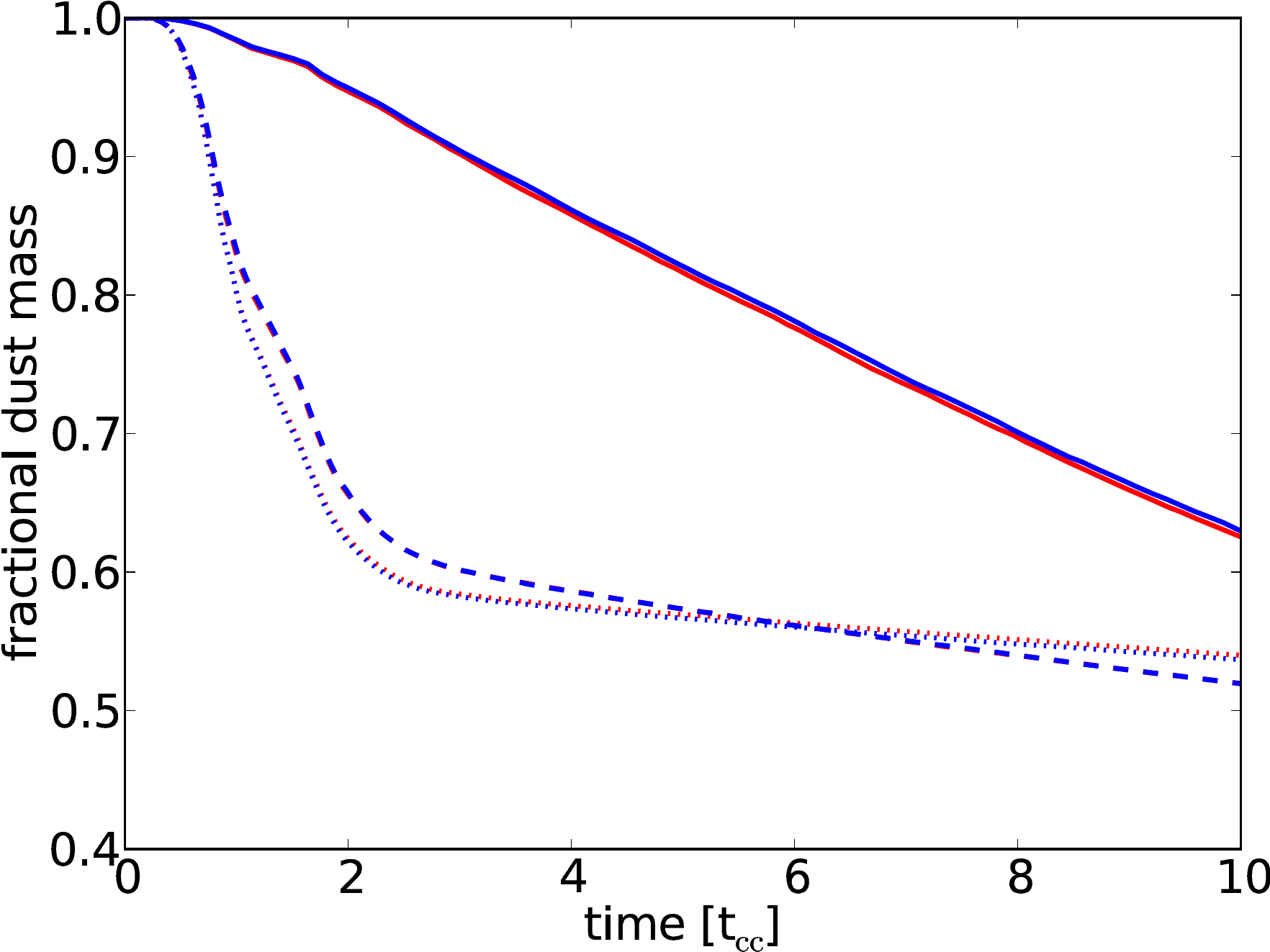}
  	}
  \caption{Dust mass evolution for five of the grain species in \cite{Nozawa:2003pd} for simulations with $\chi =100$.  Colors and line styles are the same as Figure \ref{dustmasschi100_tiel}. The grains in this figure were sputtered using the $Z=Z\subscript{\odot}$ rates from \cite{Nozawa:2006ve}.}\label{dustmasschi100_noz}
\end{figure*}

\begin{figure*}[htp]
\centering
	\subfigure[Al$_2$O$_3$]{
  	\includegraphics[width=0.325\textwidth]{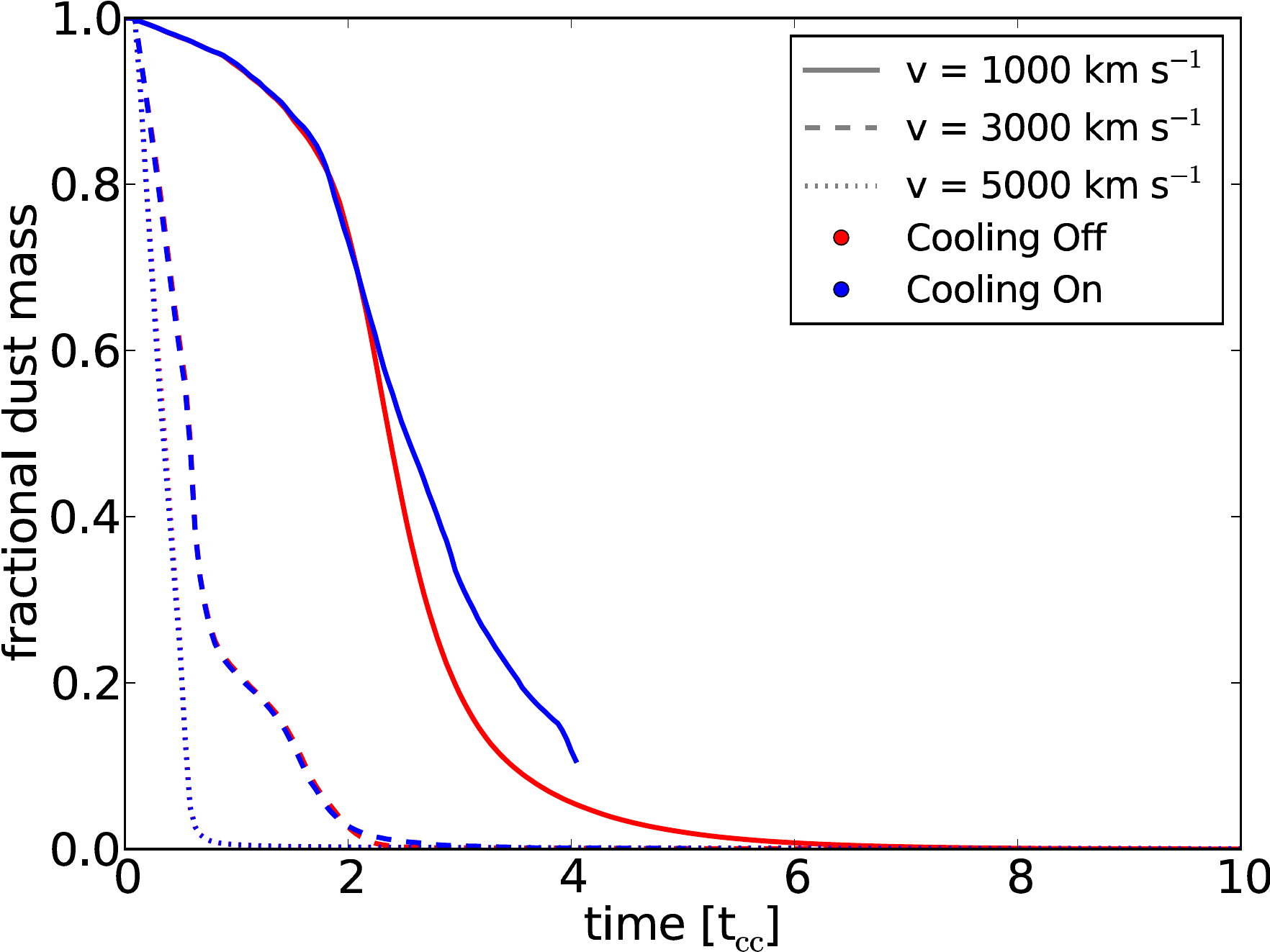}
  	}
	\hspace{-5.00mm}
  	\subfigure[C]{
  	\includegraphics[width=0.325\textwidth]{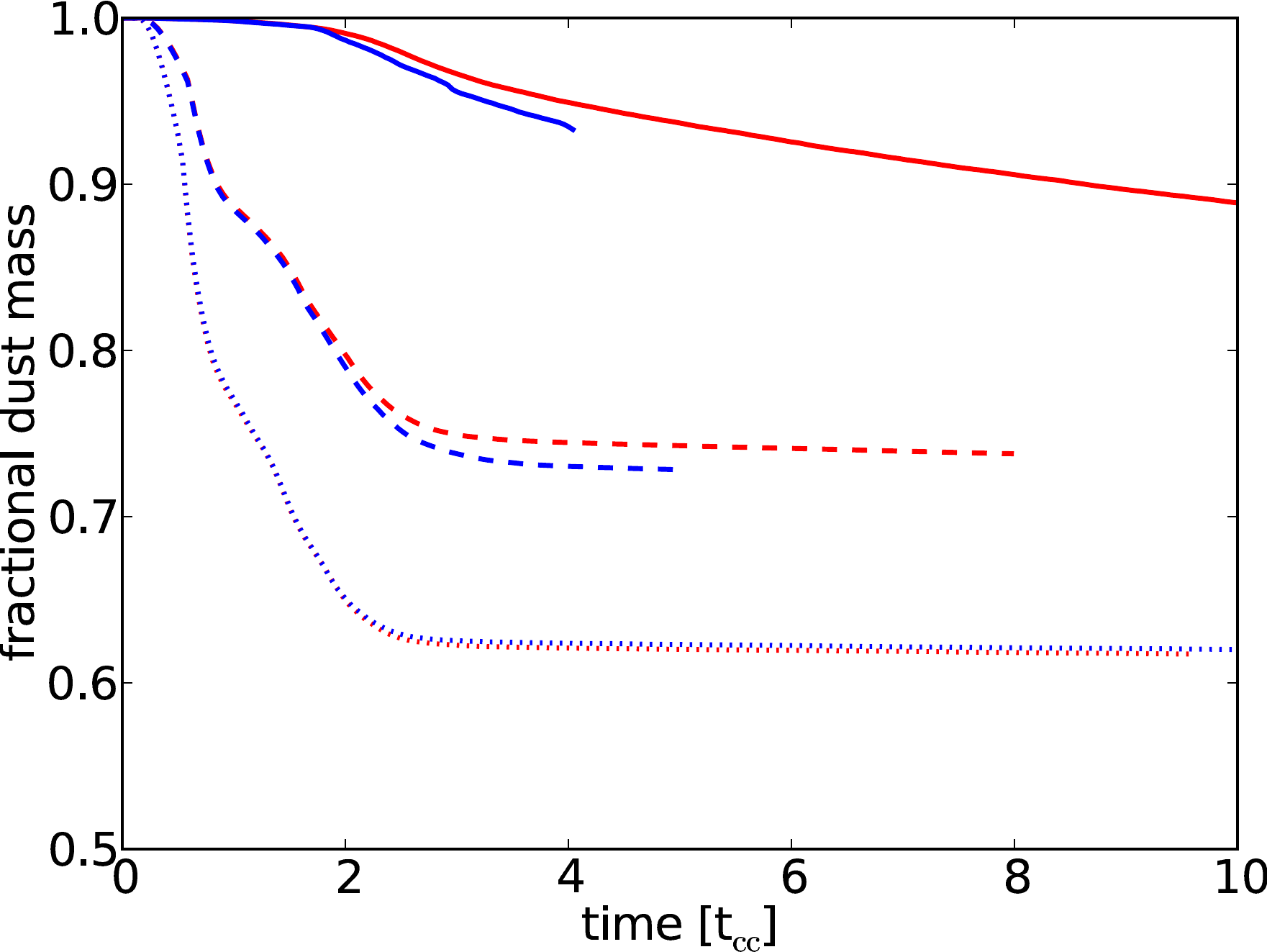}
  	}
	\hspace{-5.00mm}
	\subfigure[Mg$_2$SiO$_4$]{
  	\includegraphics[width=0.325\textwidth]{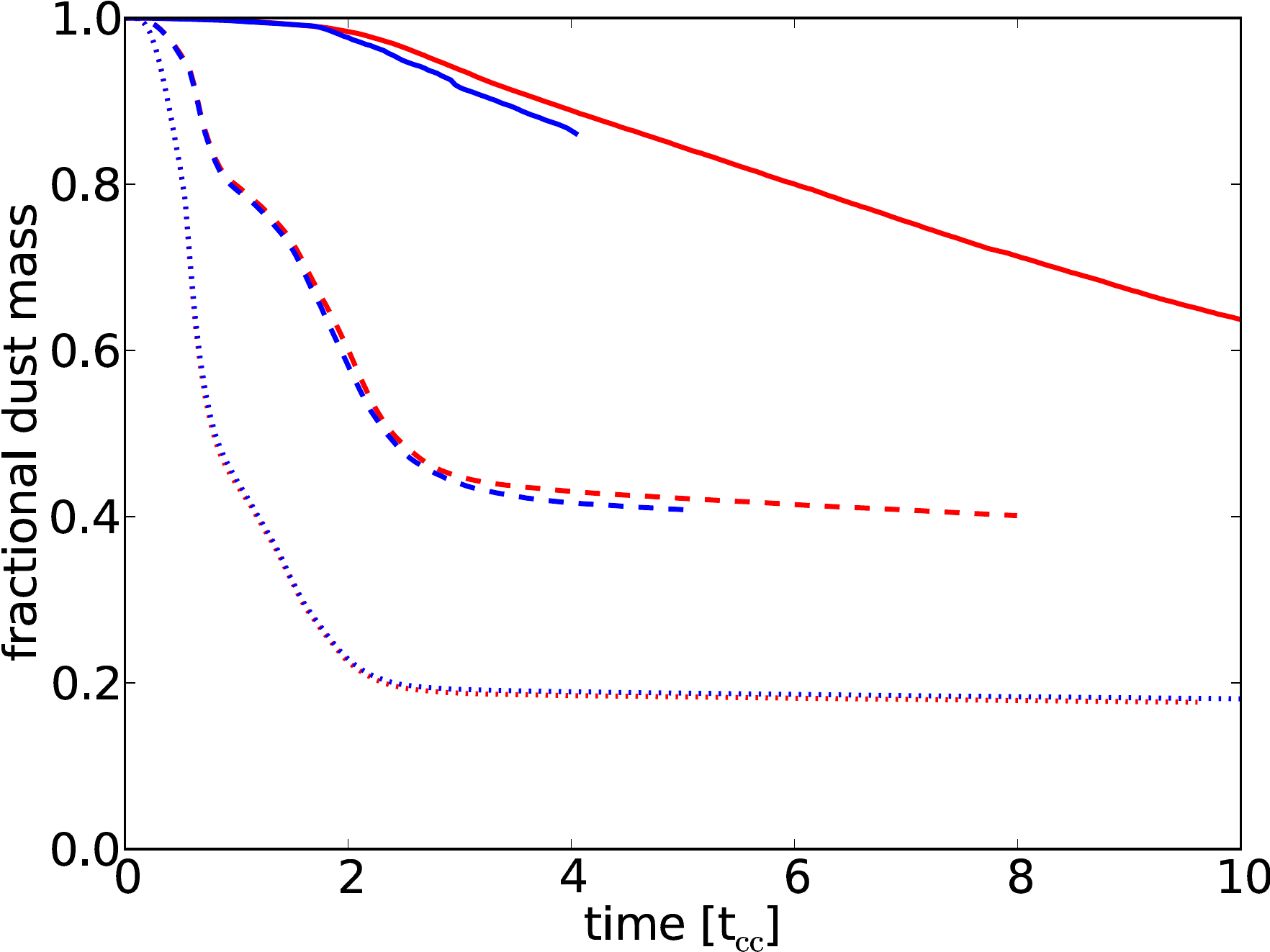}
	}
	\hspace{-5.00mm}
	\subfigure[SiO$_2$]{
  	\includegraphics[width=0.325\textwidth]{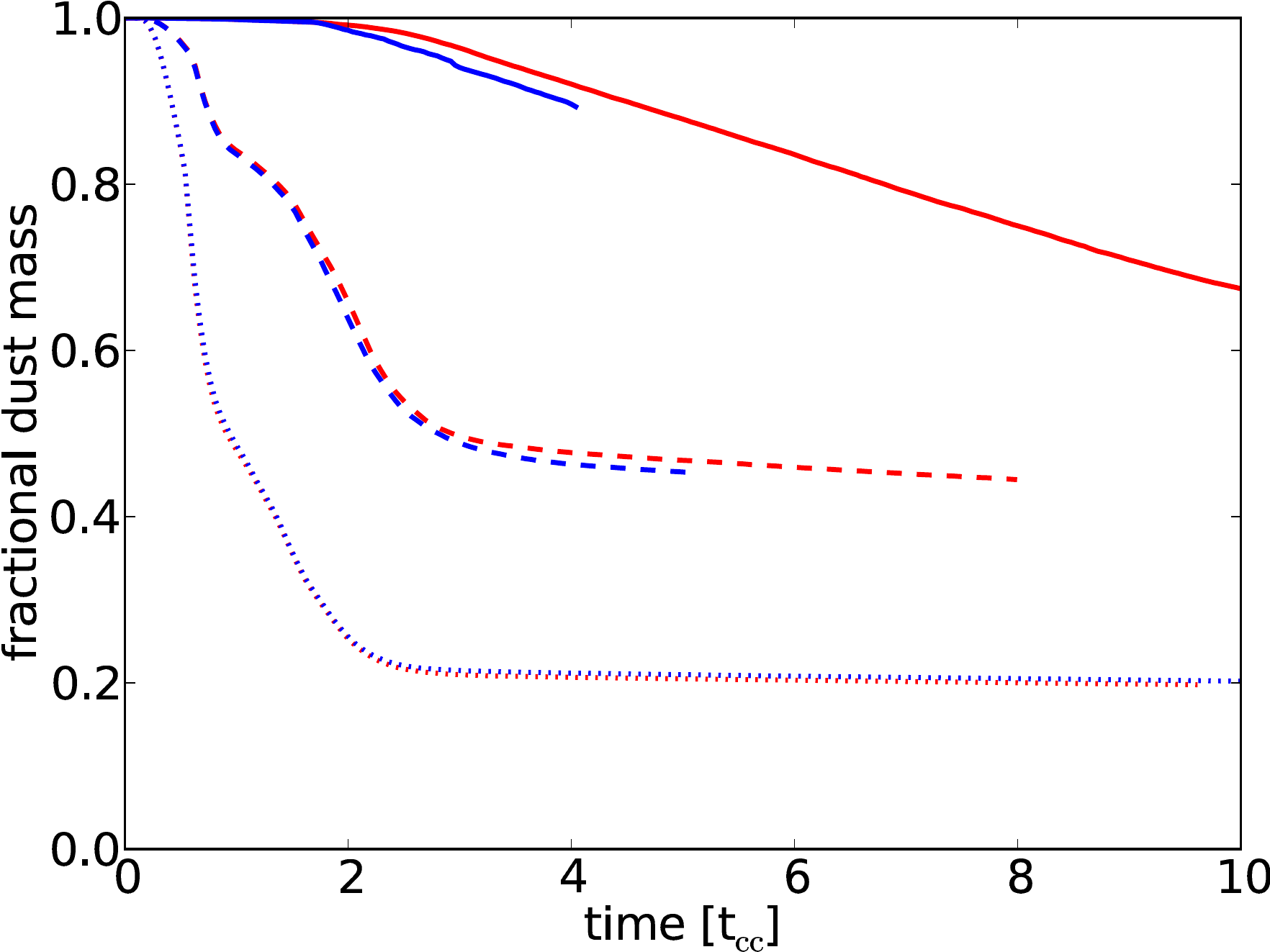}
  	}
	\hspace{-5.00mm}
	\subfigure[FeS]{
  	\includegraphics[width=0.325\textwidth]{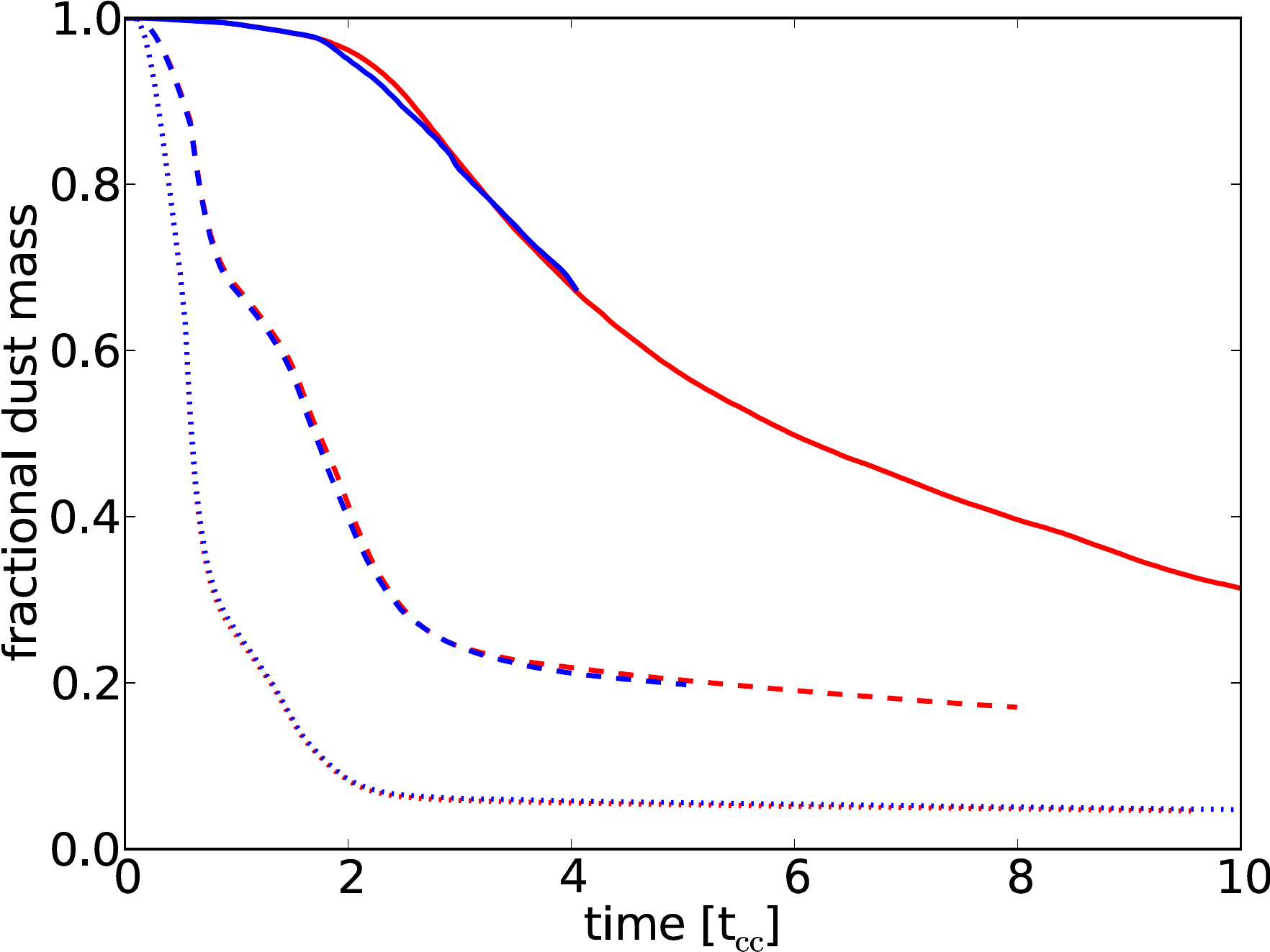}
  	}
	\hspace{-5.00mm}
	\subfigure[Fe]{
  	\includegraphics[width=0.325\textwidth]{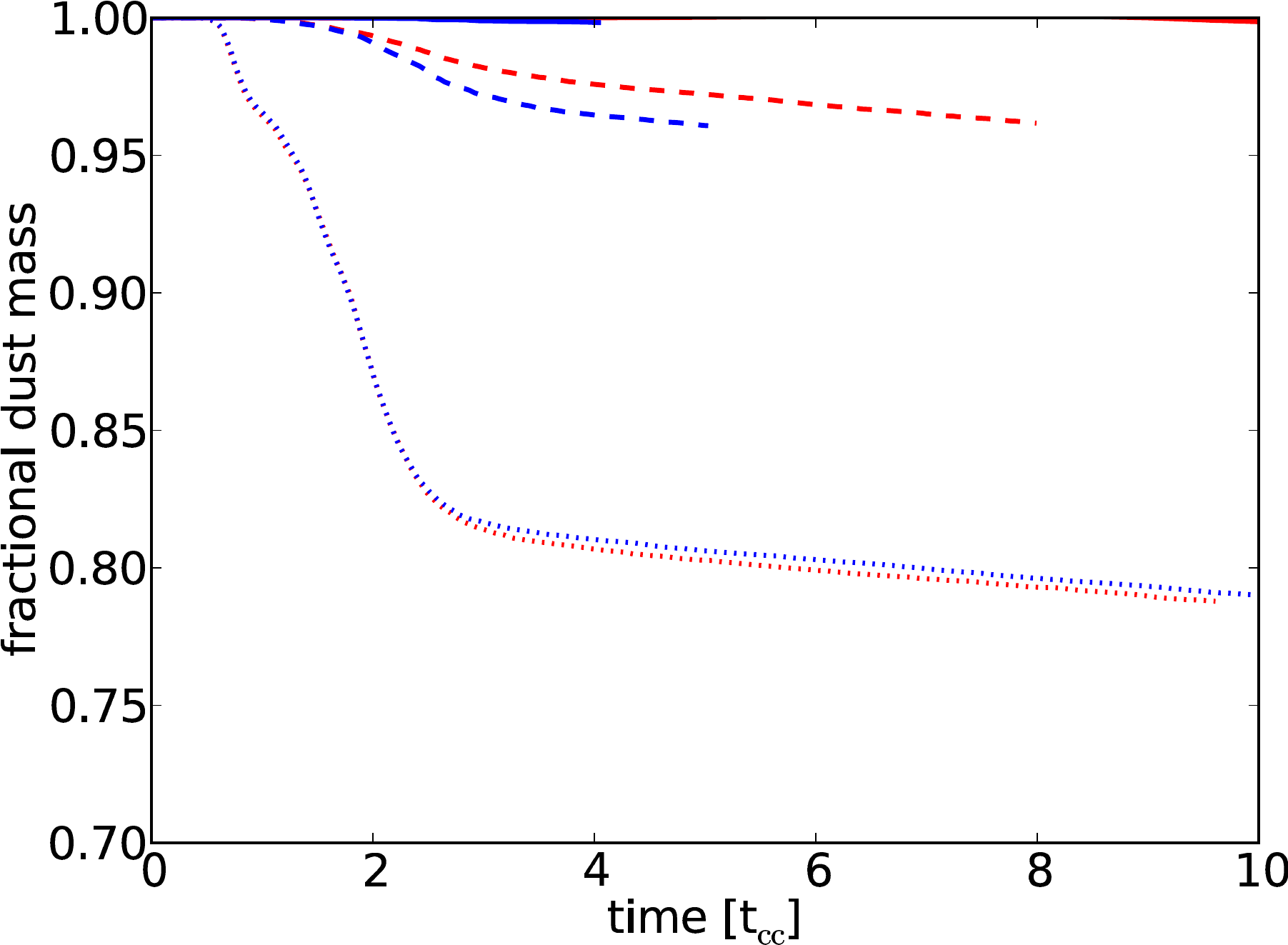}
  	}
  \caption{Same as Figure \ref{dustmasschi100_noz}, but for over-density $\chi=1000$.  In addition, we include the evolution of Fe grains.}\label{dustmasschi1000_noz}
\end{figure*}

We carry out the same inspection for \cite{Nozawa:2006ve} in Figure \ref{dustmasschi100_noz} ($\chi=100$) and Figure \ref{dustmasschi1000_noz} ($\chi=1000$).  We immediately see that, similar to the results from the polynomial sputtering rates, the variation in final dust mass is fairly small for the low-density cloud.  We also see that the trend of faster sputtering at early times for higher velocities shocks holds true in these cases as well.  One particular difference between the results from these sputtering rates and those of \cite{Tielens:1994kx} is that, for the high-velocity shocks, the fractional dust mass seems to plateau at late times, as where in the earlier figures the dust mass evolution still has a negative slope at the end of the simulation.  This may indicate that longer simulation run times are required for the simulations that did not stop evolving.  For the higher density density cloud, we again see that the final dust mass fractions are much more widely varied.  For the sputtering rates of \cite{Nozawa:2006ve}, we see a non-negligible amount of iron sputtering for the medium velocity shock simulations of the high-density cloud which is not present in the polynomial sputtering rate calculations.  We also see the same effects of cooling in the higher density cloud mentioned previously.   In addition, one interesting difference between the non-cooling and cooling simulations shows up most prominently in the evolution of Al$_2$O$_3$ (see Figure \ref{dustmasschi1000_noz}).  The non-cooling simulation appears to have a more rapid fall off in dust mass for the case of $v\subscript{shock} = 1000$ km s$^{-1}$ than the cooling simulation.  However, even though the cooling simulation did not reach ten cloud-crushing times, it seems unlikely that it would not have also arrived at total Al$_2$O$_3$ grain destruction.  Since this particular dust species consists of only small grains, it is possible that the slight drop in plasma temperature caused by the cooling is able to slow the grain destruction, despite the increased density effect mentioned above.

Finally, in an effort to summarize the results of all 13 simulations, Table \ref{dustsurvival} lists the final dust mass fraction for all species for all simulations.  In reviewing these results, we find that grain species that have initial distributions weighted toward large grain radii have higher survival rates over 10 cloud-crushing times.  Furthermore, since the sputtering rates scale linearly with density, simulations with higher density clouds all show substantial sputtering.  The higher density clouds also tend to show a much stronger scaling with shock velocity than do their low-density counterparts.  There appears to be no strong dependence on the two sputtering rate methods, most differences are on the tenths of a percent to few percent level.  While the simulations including cooling did show small deviations from those simulations without cooling in the dust mass evolution plots, in most cases the end results for surviving dust mass fraction are indistinguishable.

\begin{deluxetable}{lccc}
\tabletypesize{\scriptsize}
\tablecolumns{10}
\tablecaption{Final Dust Mass Fraction \label{dustsurvival}}
\tablewidth{0pt}
\tablehead{\colhead{Sim.} &
	\colhead{Al$_2$O$_3$} &
	\colhead{C} &
	\colhead{Mg$_2$SiO$_4$}}
\startdata
1	&	0.034	&	 0.963 / 0.962	&	0.888	\\
2	&	0.036	&	 0.964 / 0.963 	&	0.891	\\
3	&	0.030	&	 0.949 / 0.950	&	0.811	\\
4	&	0.029	&	 0.949 / 0.950	&	0.811 	\\
5	&	0.054	&	 0.948 / 0.967 	&	0.846	\\
6	&	0.052 	& 	 0.947 / 0.967	&	0.844 	\\
\vspace{-0.5mm}
7	&	0.000	&	 0.880 / 0.889	&	0.637	\\
8\tablenotemark{*}	&	0.107	&	 0.941 / 0.933	&	0.861	\\  
9	&	0.000	&	 0.771 / 0.738	&	0.401	\\ 
10	&	0.001	&	 0.762 / 0.728	&	0.408	\\
11	&	0.001	&	 0.643 / 0.617 	&	0.177 	\\
12	&	0.001	&	 0.644 / 0.620	&	0.181	\\
13	&	0.000	&	 0.882 / 0.890	&	0.637	\\
\vspace{-2.25mm}
\\
\tableline
\vspace{-2mm}
\\
Sim.	&	MgSiO$_3$	&	SiO$_2$	&	MgO		\vspace{0.75mm}
\\
\tableline
\vspace{-1.5mm} 
\\
1	&	0.423	&	 0.932 / 0.916	&	0.838	\\
2	&	0.428	&	 0.934 / 0.919	&	0.842	\\
3	&	0.334	&	 0.875 / 0.839	&	0.774	\\
4	&	0.333 	&	 0.875 / 0.840	&	0.775	\\
5	&	0.394	&	 0.799 / 0.870	&	0.811	\\
6	&	0.392	&	 0.798 / 0.869	&	0.809 	\\
\vspace{-0.5mm}
7	&	0.139	&	 0.676 / 0.674	&	0.615	\\
8\tablenotemark{*}	&	0.452	&	 0.930 / 0.893	&	0.813	\\  
9	&	0.044	&	 0.593 / 0.378	&	0.445	\\ 
10	&	0.054	&	 0.598 / 0.453	&	0.382	\\
11	&	0.014 	&	 0.240 / 0.197	&	0.181 	\\
12	&	0.015	&	 0.237 / 0.202	&	0.185	\\
13	&	0.140	&	 0.676 / 0.675	&	0.615	\\
\vspace{-2.25mm}
\\
\tableline
\vspace{-2mm}
\\
Sim.	&	 Si 	&	 FeS 		&	 Fe 	\vspace{0.75mm}
\\
\tableline
\vspace{-1.5mm} 
\\

1	&	0.998	&	0.626	&	 1.000 / 1.000	\\
2	&	0.998	&	0.631	&	 1.000 / 1.000	\\
3	&	0.998	&	0.520 	&	 1.000 / 1.000	\\
4	&	0.998	&	0.519	&	 1.000 / 1.000	\\
5	&	0.998 	&	0.540 	&	 1.000 / 1.000 	\\
6	&	0.998	&	0.537	&	 1.000 / 1.000	\\
\vspace{-0.5mm}
7	&	0.997 	&	0.314	&	 1.000 / 0.999	\\
8\tablenotemark{*}	&	0.996	&	0.674	&	 1.000 / 0.998	\\  
9	&	0.971	&	0.171	&	 0.999 / 0.962	\\ 
10	&	0.965	&	0.198	&	 0.997 / 0.961	\\
11	&	0.853 	&	0.046 	&	 0.932 / 0.789	\\
12	&	0.854	&	0.047	&	 0.928 / 0.790	\\
13	&	0.998	&	0.315	&	 1.000 / 0.999	
\enddata
\tablecomments{For species with two values, they show results from sputtering rates of \cite{Tielens:1994kx} and \cite{Nozawa:2006ve}, respectively.  Species with only one value show just the \cite{Nozawa:2006ve} sputtering rates.  In the context of the \cite{Tielens:1994kx} sputtering, SiO$_2$ represents ``silicates".}
\tablenotetext{*}{The high final dust mass values in Simulation 8 are a result of the analysis being truncated prior to ten full cloud-crushing times, as mentioned in the text. \vspace{1.0mm}}
\end{deluxetable}

\section{Summary and Future Work}\label{future}
We summarize the above the results into a few key points: 

\begin{itemize}
\item{Owing to linear density dependence in the sputtering rates, the high density clouds destroy significantly higher percentages of dust.  In some cases, as much as 70\% more dust is destroyed, with total destruction fractions as high as 80-100\%.}
\item{The relative velocity between the reverse shock and the ejecta clump strongly influences the total dust destruction.  For high-density ejecta clumps, the highest velocity shock results in an additional $\sim$50\% mass loss compared to the lowest velocity shock for some grain species.}
\item{The initial size distribution of dust grains has a considerable impact on the survival rate.  Species with appreciable mass in grains above 0.1 $\mu$m have the highest survival rate, losing only 30\% or less of their mass.  Grains that start out at less than 0.1 $\mu$m are often completely destroyed.}
\item{Total dust destruction varies widely across grain species and can be high for some portions of parameter space.  For high-density ejecta clumps and high-velocity shocks, the often studied species of C, SiO$\subscript{2}$, and Fe show 62\%, 20\%, and 80\% survival, respectively.}
\end{itemize}

From an observational perspective, these results may have interesting implications for the IR emission seen in supernova remnants, which is usually attributed to dust that has been formed and processed by the supernova explosion.  \cite{Rho:2008qf} modeled detailed mid-IR spectra of Cas A to determine the populations of dust that are likely present in the remnant.  They find a reasonably high abundance of Si dust, which agrees with the high survival rate of those grains.  Additionally, \cite{Sandstrom:2009fk} analyzed mid-IR spectra for SNR 1E0102-7219 and found both a lack of Al$_2$O$_3$ grains as well as a large population of amorphous carbon grains with radii of ~0.1 $\mu$m, in good agreement with our results.  Such observations are excellent indicators that the results from our simulations could be used to predict emission features in future remnant observations.

Since the plasma in these shock-cloud interactions will become enriched with high-mass ions as a result of dust-grain sputtering, we will investigate, in future work, the additional effects that this might have.  We have preliminary findings that indicate that for cooling simulations with enhanced metallicity ($Z \gtrsim$ 10 $Z\subscript{\odot}$), the results diverge further from the non-cooling simulations.  Specifically, the simulations with increased cooling owing to higher metal abundances tend to show a greater amount of dust destruction.  This likely stems from the previously mentioned effect in which the shorter cooling time leads to the formation and collapse of localized cloud over-densities that persist for a great fraction of the simulation time.  We also predict that, with accurate sputtering rates for enhanced metal abundances, we should see greater dust destruction by high mass ions, since the sputtering yields for these ions should will be factors of 10-100 higher.

While our current method of post-processing the tracer particles is sufficient as a first step in tracking dust populations in an ejecta clump, we would like to employ more active means of following dust erosion.  Therefore, to improve upon current simulations, we plan to sputter the dust grains ``on-the-fly" so that dust destruction occurs during simulation run time.  This should provide something closer to an instantaneous sputtering rate rather than the current low time resolution $da/dt$, which is applied only at each discrete data dump.  In addition, it will allow for a means of accounting for feedback effects that dust sputtering may have on the system.  As the dust grains are eroded, they reintroduce metals into the gas phase which would allow them to act as additional coolants and ions for grain sputtering.

Computing the feedback of metals by grain sputtering also motivates us to improve our cooling method by tracking the exact abundances of a variety of metals and then using Cloudy to get cooling rates that directly scale with the metal abundance pattern.  If we accurately track the abundances of the metals, we can increase the metal abundances in the hot plasma as the grains are sputtered.  Providing the link between grain sputtering and gas cooling could lead to a possible equilibrium between grain sputtering and plasma cooling.  At this point, collapsing gas drives up the sputtering rate, which leads to increased metal abundances that cool the gas even further, perhaps to low enough temperatures to shut off the thermal sputtering.

Finally, given the nature of the tracer particles used to track dust populations, we have only been able to model the dust as being directly coupled to the cloud material.  Therefore, we have only applied thermal sputtering rates and neglect non-thermal sputtering by grain motion through the plasma.  Dust coupling to the cloud may not be an unreasonable assumption if the grains are charged and tied to magnetic fields lines that might be present in the ejecta material.  It would be ideal to decouple the grains from the fluid and account for the effects of non-thermal sputtering as well as possible grain-grain collisions.  Additionally, if magnetic field lines are present, it would be important to investigate the effects of betatron acceleration \citep{Shull:1977lr, Shull:1978fk} as it might enhance sputtering of large grains.  Methods for including these factors, as well as those mentioned in the preceding paragraphs, are being considered for inclusion in a second paper on this topic.

\acknowledgements
We thank the anonymous referee for comments that helped to clarify and strengthen this work.  We also thank E. Hallman, B. O'Shea, and S. Skillman for helpful discussions about the inner workings of Enzo; M. Turk for the use of, and assistance with, his analysis software package, \texttt{yt} (yt.enzotools.org); R. Fesen for useful discussion and comments; and T. Nozawa for providing tabulated sputtering rates from previous work. This project was funded by NSF Theory Grant AST07-07474, STScI Archive Theory Grant AR11774.01-A, and recently by the NSF Graduate Research Fellowship program (D.W.S).

\bibliography{references}

\end{document}